\documentclass[PHD]{macro/neu_msthesis}
\usepackage{cite}
\usepackage{amsmath,amssymb,amsfonts}
\usepackage{algorithmic}
\usepackage{listings}
\usepackage{graphicx}
\usepackage{textcomp}
\usepackage{xcolor}
\usepackage[hyphens]{url}
\usepackage{fancyhdr}
\usepackage{booktabs}
\usepackage{tikz}
\usepackage{wrapfig}

\usepackage{tabularx}
\usepackage{ragged2e}
\usepackage{soul}
\usepackage{tikz}
\usepackage{tabularx}
\usepackage{makecell}
\usepackage{multirow}
\usepackage{tabu}

\usepackage{cite}
\usepackage{amsmath,amssymb,amsfonts}
\usepackage{algorithmic}
\usepackage{graphicx}
\usepackage{textcomp}
\def\BibTeX{{\rm B\kern-.05em{\sc i\kern-.025em b}\kern-.08em
    T\kern-.1667em\lower.7ex\hbox{E}\kern-.125emX}}
\usepackage{graphicx}
\usepackage[caption=false]{subfig}
\usepackage{float}
\usepackage{colortbl}
\usepackage{flushend}
\usepackage{array}
\usepackage{pifont}
\usepackage{xspace}

\usepackage{booktabs}
\usepackage[utf8]{inputenc}
\usepackage{array}
\usepackage{makecell}

\usepackage{tabularx}
\usepackage{pifont}

\usepackage{array}
\newcolumntype{x}[1]{>{\centering\arraybackslash\hspace{0pt}}p{#1}}

\usepackage[acronym]{glossaries}

\makeglossaries

\newcommand{\thesis}{dissertation }
\newcommand{\Thesis}{Dissertation }

\newif\ifreason

\newcommand{\reason}[1]{%
    \ifreason
       \textcolor{blue}{\fbox{\textbf{#1}}}%
    \fi
}

\newcommand{\tick}{\ding{52}}
\newcommand{\cross}{\ding{56}}

\newcommand{\algoref}[1]{Algorithm~\ref{#1}}

\newcommand{\mmhf}{\texttt{MMH4}\xspace}
\newcommand{\hacc}{\texttt{HACC}\xspace}
\newcommand{\mtag}{$\mathrm{TAG}$\xspace}

\newcommand*\circled[1]{\tikz[baseline=(char.base)]{
            \node[shape=circle,fill,inner sep=1pt] (char) {\textcolor{white}{#1}};}}

\reasonfalse % For final mode

\title{Enabling Accelerators for Graph Computing}

\author{Kaustubh Shivdikar}

\dept{Electrical and Computer Engineering}

\degreename{Electrical and Computer Engineering}

\field{Electrical and Computer Engineering}

\submitdate{April 2023}

\numberofmembers{4}

\principaladviser{Dr. David Kaeli}
\firstreader{Dr. Devesh Tiwari}
\secondreader{Second Reader Name}

\thirdreader{Third Reader Name}
\fourthreader{Fourth Reader Name}
\fifthreader{Fifth Reader Name}

\chairman{Dr. Srinivas Tadigadapa}

\dean{Dr. Waleed Meleis}

\usepackage{amsfonts}			% special math characters
\usepackage{times}		% use Postscript fonts

\usepackage[linesnumbered,algoruled,boxed,lined]{algorithm2e}
\usepackage{setspace}
\usepackage{pbox}
\usepackage[super]{nth}
\usepackage{placeins}

\usepackage{epigraph}

\usepackage{amsmath,amssymb,amsfonts}
\usepackage{tabularx,ragged2e}
\usepackage{soul}
\usepackage{array}
\usepackage{colortbl}
\usepackage{flushend}
\usepackage[caption=false]{subfig}
\usepackage{float}
\usepackage{textcomp}
\usepackage{booktabs}
\usepackage[utf8]{inputenc}
\usepackage{array}
\usepackage{makecell}

\usepackage{tabularx}
\usepackage{makecell}
\usepackage{multirow}
\usepackage{tabu}

\usepackage{xcolor}
\usepackage{xspace}

\newcommand{\figref}[1]{Figure~\ref{#1}}
\newcommand{\tabref}[1]{Table~\ref{#1}}

\newcommand{\secref}[1]{Section~\ref{#1}}

\newcommand{\ifno}[1]{}

\usepackage{multirow}
\makeindex
\usepackage{makeidx}
\usepackage{acronym}
\usepackage{amsmath}

\usepackage{url}
\ifnum\pdfoutput>0
\usepackage[pdftex,%                    % hyper-references for ps2pdf
bookmarks=true,%                        % generate bookmarks ...
bookmarksnumbered=true,%                % ... with numbers
hypertexnames=false,%                   % needed for correct links to figures
breaklinks=true           %Allows link text to break across lines;
]{hyperref}
\else
\usepackage[hypertex,%                  % hyper-references for ps2pdf
bookmarks=true,%                        % generate bookmarks ...
bookmarksnumbered=true,%                % ... with numbers
hypertexnames=false,%                   % needed for correct links to figures
breaklinks=true           %Allows link text to break across lines;
]{hyperref}
\fi

\hypersetup{				% setup PDF information fields
pdfauthor = {\authorRef},
pdftitle = {\titleRef},
pdfsubject = {\expandafter{\degreeRef} thesis submitted to Northeastern University},
pdfkeywords = {add keywords here}
pdfcreator = {LaTeX with hyperref package},
pdfproducer = {dvips + ps2pdf}}

\usepackage{appendix}

\begin{document}

% add a pdf bookmark to the cover page
\pdfbookmark[1]{Cover}{cover}

% --- title page ---
\titlepage

% --- front matter ---
\begin{frontmatter}
% print signature page
% \signaturepage
% dedication
% dedication.tex:

\begin{dedication}

To my family,

for their endless love, support, and encouragement.

\end{dedication}

% table of content (add bookmark for convenience)
\pdfbookmark[1]{Table of Contents}{contents}
\tableofcontents
\listoffigures
\newpage\ssp
\listoftables

% include a list of Acronyms (comment out if no acronyms are specified)
% acronyms.tex

% \chapter*{List of Acronyms}
% \printglossary[type=\acronymtype]
\printglossary[type=\acronymtype,title=List of Acronyms]
\addcontentsline{toc}{chapter}{List of Acronyms}

\newacronym{GCD}{GCD}{Greatest Common Divisor}
\newacronym{ABM}{ABM}{Advanced Bit Manipulation}
\newacronym{ATT}{ATT}{Address Translation Tables}
\newacronym{ADX}{ADX}{Multiprecision Add-Carry}
\newacronym{AESI}{AESI}{Advanced Encryption Instructions}
\newacronym{AMD}{AMD}{Advanced Micro Devices}
\newacronym{DGAS}{DGAS}{Distributed Global Address Space}
\newacronym{AVX2}{AVX2}{Advanced Vector Instructions 2}
\newacronym{BFS}{BFS}{Breadth-First Search}
\newacronym{BLAS}{BLAS}{Basic Linear Algebra Subprograms}
\newacronym{BMI2}{BMI2}{Bit Manipulation Instruction Set 2}
\newacronym{CRAM}{C-RAM}{Computational RAM}
\newacronym{ELL}{ELL}{ELLPACK}
\newacronym{EMMX}{EMMX}{Extended MMX Extension}
\newacronym{CLMUL}{CLMUL}{Carry-less Multiplication Extension}
\newacronym{CPU}{CPU}{Central Processing Unit}
\newacronym{CSR}{CSR}{Compressed Sparse Row}
\newacronym{CSC}{CSC}{Compressed Sparse Column}
\newacronym{DARPA}{DARPA}{Defense Advanced Research Projects Agency}
\newacronym{DFFT}{DFFT}{Dense Fast Fourier Transform}
\newacronym{DMA}{DMA}{Direct Memory Access}
\newacronym{DRAM}{DRAM}{Dynamic Random Access Memory}
\newacronym{F16C}{F16C}{16-bit Floating-Point Conversion}
\newacronym{FMA}{FMA}{Floating-Point Multiply-Add}
\newacronym{FMA3}{FMA3}{3-Operand Fused-Multiply-Add}
\newacronym{GCN}{GCN}{Graph Convolutional Network}
\newacronym{GEMM}{GEMM}{General Matrix Multiplication}
\newacronym{GNN}{GNN}{Graph Neural Networks}
\newacronym{GPGPU}{GPGPU}{General-Purpose Graphics Processing Unit}
\newacronym{GPU}{GPU}{Graphics Processing Unit}
\newacronym{HBM}{HBM}{High Bandwidth Memory}
% \newacronym{HIVE}{HIVE}{Hierarchical Identify Verify Exploit}
\newacronym{IPC}{IPC}{Instructions Per Cycle}
\newacronym{ISA}{ISA}{Instruction Set Architecture}
\newacronym{MAPCSR}{MAP-CSR}{Memory Aligned Parallel - Compressed Sparse Row}
\newacronym{MLP}{MLP}{Multi-layer Perceptron}
\newacronym{MKL}{MKL}{Math Kernel Library}
% \newacronym{MTP}{MTP}{Multi-threaded Pipeline}
% \newacronym{PIUMA}{PIUMA}{Programmable Integrated Unified Memory Architecture}
\newacronym{RF}{RF}{Register File}
\newacronym{OMP}{OMP}{OpenMP}
\newacronym{OS}{OS}{Operating System}
\newacronym{PIM}{PIM}{Processor in Memory}
\newacronym{RMAT}{RMAT}{Recursive Matrix}
\newacronym{SFFT}{SFFT}{Sparse Fast Fourier Transform}
\newacronym{SIMD}{SIMD}{Single Instruction, Multiple Data}
\newacronym{SMASH}{SMASH}{Sparse Matrix Atomic Scratchpad Hashing}
\newacronym{SPAD}{SPAD}{Scratchpad}
\newacronym{SPMD}{SPMD}{Single Program, Multiple Data}
\newacronym{SPGEMM}{SpGEMM}{Sparse Matrix-Matrix Multiply}
\newacronym{SPMV}{SpMV}{Sparse Matrix-Vector}
\newacronym{SSE42}{SSE4.2}{Streaming SIMD Extensions 4.2}
% \newacronym{STP}{STP}{Single-threaded Pipeline}

% \end{acronym}

% include any of the front matter files that contain text
% attention the input does cause a page break, the include on 
% the other hand does not
\begin{acknowledgements}

I wish to extend my profound appreciation to my advisor, Dr. David Kaeli, for his pivotal role in my academic journey. His remarkable foresight and unwavering dedication have played a crucial role in shaping my path towards academic excellence. Dr. Kaeli's profound expertise and innovative spirit have served as a beacon of guidance through challenging periods, illuminating the path with his profound knowledge and strategic insight. His exceptional commitment to research and education has not only inspired me but also significantly influenced my professional development. The supportive and enriching environment he fostered has been instrumental in navigating the complexities of doctoral research, especially as an international student adapting to a new academic and cultural setting. Dr. Kaeli's mentorship extends beyond academic achievements, providing invaluable lessons in resilience and integrity. I am eternally grateful for the privilege of being mentored by a scientist of such high calibre and integrity, whose influence will undoubtedly echo throughout my career.

I extend my heartfelt gratitude to Prof. Ajay Joshi for his invaluable insights on Homomorphic Encryption, Prof. John Kim for his expertise on network topologies, Prof. Jos\'e Luis Abell\'an for his deep knowledge on Graph Neural Networks, and Prof. Devesh Tiwari for his guidance on compute micro-architectures. Their contributions have been instrumental in shaping my research and enhancing my understanding of these complex areas. Additionally, my sincere thanks are due to all faculty members and administrative staff of the College of Engineering at Northeastern University for their unwavering support throughout my doctoral journey.

My sincere and warm thanks go also to my lab colleagues: Amir Taherin, Derek Rodriguez, Julian Gutierrez, Malith Jayaweera, Matin Raayai, Michael Shen, Neal Livesay, Nicolas Agostini, Sana Anvari, Trinayan Baruah, Yifan Sun, Alexander Ingare, and Zlatan Feric, for their invaluable contribution to my academic and personal growth. Their collaboration, insight, and camaraderie have been instrumental in fostering a nurturing and productive research environment. It has been a privilege to work alongside such dedicated and talented individuals. Their diverse perspectives and expertise have greatly enriched my research experience. Furthermore, their support and encouragement have been a constant source of motivation, making my journey through the intricacies of academic research both rewarding and memorable.

A Ph.D. journey that lasted 6 years, 7 months, and 25 days (or 2429 days in total) was made possible by the unwavering support from my friends Ahan Kak and Ashwin Shirsat. Sharing the joys of highs and support in the times of lows.
Finally, last but by no means least, words will fail to express my gratitude to my family, who have been my rock through this journey, especially my parents Chandrakant and Anagha, and my brother Saumil Shivdikar. This achievement would have not been possible without their love, support, and encouragement.

\end{acknowledgements}

% abstract.tex:

\begin{abstract}

The advent of Graph Neural Networks (GNNs) has revolutionized the field of machine learning, offering a novel paradigm for learning on graph-structured data. Unlike traditional neural networks, GNNs are capable of capturing complex relationships and dependencies inherent in graph data, making them particularly suited for a wide range of applications including social network analysis, molecular chemistry, and network security. The impact of GNNs in these domains is profound, enabling more accurate models and predictions, and thereby contributing significantly to advances in these fields.

GNNs, with their unique structure and operation, present new computational challenges compared to conventional neural networks. This requires comprehensive benchmarking and a thorough characterization of GNNs to obtain insight into their computational requirements and to identify potential performance bottlenecks. In this thesis, we aim to develop a better understanding of how GNNs interact with the underlying hardware and will leverage this knowledge as we design specialized accelerators and develop new optimizations, leading to more efficient and faster GNN computations.

A pivotal component within GNNs is the Sparse General Matrix-Matrix Multiplication (SpGEMM) kernel, known for its computational intensity and irregular memory access patterns. In this thesis, we address the challenges posed by SpGEMM by implementing a highly optimized hashing-based SpGEMM kernel tailored for a custom accelerator. This optimization is crucial to enhancing the performance of GNN workloads, ensuring that the acceleration potential of custom hardware is fully realized.

Synthesizing these insights and optimizations, we design state-of-the-art hardware accelerators capable of efficiently handling various GNN workloads. Our accelerator architectures are built on our characterization of GNN computational demands, providing clear motivation for our approaches. Furthermore, we extend our exploration to emerging GNN workloads in the domain of graph neural networks. This exploration into novel models underlines our comprehensive approach, as we strive to enable accelerators that are not just performant, but also versatile, able to adapt to the evolving landscape of graph computing.

\end{abstract}

\end{frontmatter}

% --- body of the document --- 

%\pagestyle{plain}
\pagestyle{headings}

% include each chapter like below
%  ___       _             
% |_ _|_ __ | |_ _ __ ___  
%  | || '_ \| __| '__/ _ \ 
%  | || | | | |_| | | (_) |
% |___|_| |_|\__|_|  \___/ 

\chapter{Introduction}
\label{chp1}
\epigraph{In the vast landscape of computation, graphs stand as the bridges connecting isolated islands of data, creating a coherent world from chaos.}{Inspired by Donald Knuth}

\reason{The emergence of graph computing (a historical perspective).}

Graph computing traces its lineage back to some of the earliest pursuits of mathematics. Historically, graph theory took its first steps in the $18^{th}$ century with Leonhard Euler's formulation of the Seven Bridges of K\"{o}nigsberg problem, where he proved that it was impossible to traverse each of the city's seven bridges once and only once without retracing any step~\cite{euler1956seven}.
% igniting a discipline that sought to understand complex systems through vertices and edges.
This abstract representation allowed for the modeling of a wide variety of systems, from social interactions to intricate molecular structures.
However, for much of its history, graph theory remained primarily an academic endeavor with limited computational exploration, owing to the computational constraints of the era.

With the digital revolution of the late 20th century, computing power saw unprecedented growth. As industries started grappling with vast amounts of interconnected data—from the nascent Internet's web pages, to the massive social networks—there emerged, a pressing need for effective means to process and analyze this data. It was in this backdrop that graph computing began its ascent, evolving from theoretical speculations to practical, essential toolsets. The challenge shifted from simply understanding graph structures to efficiently processing them, leading to the exploration of specialized hardware and software solutions. 
This combination of data-centric challenges and available computational power provides the motivation for this thesis.

\section{Background and Motivation}

\reason{Relating graph computing to Graph Neural Networks}

Graphs have seen a growing role in modern computational domains. With the rise of vast amounts of complex, interconnected data, traditional data processing methods have often fallen short. Enter Graph Neural Networks (GNNs), a specialized neural network architecture designed to handle such data and extract insights from these intricate connections. GNNs have shown remarkable potential in various domains, offering solutions where conventional methods have struggled. Their capacity to model relational data naturally fits a wide range of applications, spanning from social network analysis to molecular chemistry.
% Recognizing the versatility and potential of GNNs in handling such complex data patterns is essential.
To delve deeper into their capabilities, here is a list of GNN applications.

\begin{figure}[htbp]
    \centering
    \includegraphics[width=1.0\textwidth]{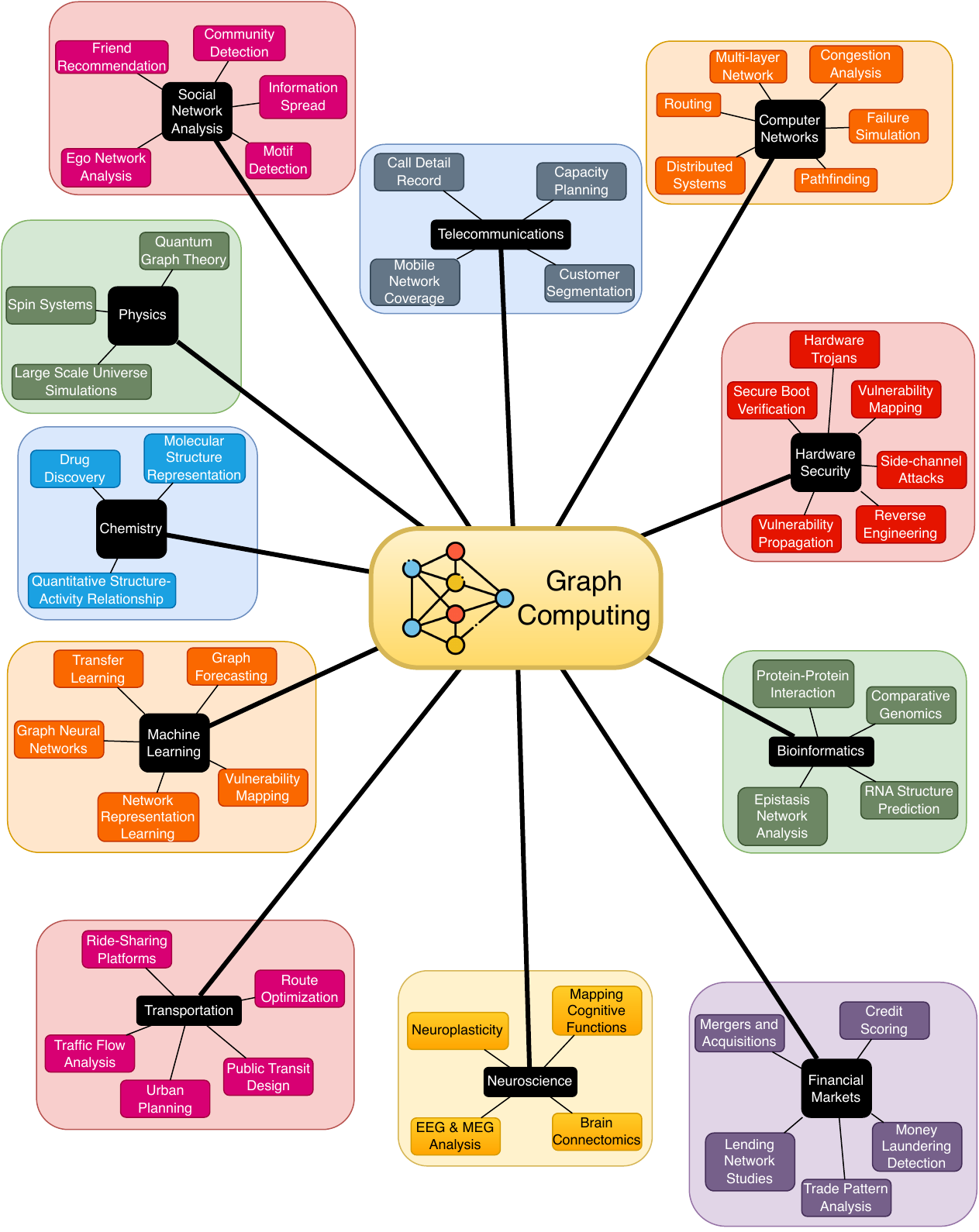}
    \caption{Graph Computing Applications}
    \label{chp1:fig:graph_applications}
\end{figure}

\reason{Applications of GNNs}

\begin{enumerate}

    \item \textbf{Social Network Analysis}: Graphs are essential in Social Network Analysis (SNA) to understand interactions within networks such as Facebook and Twitter. By examining these structures, SNA can detect key influencers, community structures, and predict potential trends or misinformation spread~\cite{campbell2013social,ugander2011anatomy,yamaguchi2010turank}.
    
    \item \textbf{Computer Networks}: Graphs represent devices and communication pathways in computer networks, aiding in tasks such as routing and fault detection. Graph-based algorithms optimize network design and manage potential vulnerabilities~\cite{hossmann2010know,goldberg2005computing,zhong2010graph}.

    \item \textbf{Hardware Security}: Circuits can be represented as graphs to detect anomalies, such as potential Hardware Trojans. Graph Neural Networks (GNNs) further enhance detection capabilities, pinpointing subtle irregularities~\cite{alrahis2022embracing,hashemi2022graph,yasaei2021gnn4tj}.

    \item \textbf{Bioinformatics}: In bioinformatics, graphs are used to model entities, for example, protein interactions and genetic sequences. They help identify conserved patterns in DNA or RNA, study metabolic pathways, and analyze genomic variations~\cite{bernardes2015evaluation,li2012comparison}.

    \item \textbf{Financial Markets}: Graphs illuminate interactions in financial markets. They help analysts identify correlated assets, assess systemic risks, and optimize algorithmic trading strategies~\cite{portfoliotheory,systemicrisk,algoTrading}.

    \item \textbf{Neuroscience}: Representing the brain's complex structure, graphs in neuroscience help analyze neural networks, understand (Functional magnetic resonance imaging) fMRI data, and study the brain's topological properties~\cite{sporns2011networks,bullmore2011brain,fmrigraphs,bassett2017network}.

    \item \textbf{Transportation and Logistics}: Graphs map transportation networks, aiding in solving optimization problems such as route planning and vehicle routing. They help ensure efficient deliveries, optimize public transit, and adapt to disruptions~\cite{golden2008vehicle,laporte1992traveling,dell2015public,yang2018dynamic}.

    \item \textbf{Chemistry}: Molecular structures in chemistry are represented using graphs, predicting chemical properties, identifying isomers, and aiding in drug design~\cite{trinajstic1983chemical,todeschini2016handbook,balaban1976applications,willett2006chemical}.

    \item \textbf{Physics}: Graphs in physics are used to model quantum states, lattice structures, and network systems. They aid in understanding quantum computing, material properties, and system dynamics~\cite{caldarelli2007scale,wu2004two,barabasi2016network,hein2006entanglement}.

    \item \textbf{Machine Learning and Data Mining}: In machine learning, graphs facilitate clustering, classification, and feature extraction. They model non-Euclidean data, aiding in semi-supervised learning and geometric deep learning~\cite{zhou2009graph,chapelle2006semi,bronstein2017geometric,wong2015frequent}.

    \item \textbf{Telecommunications}: Graphs underpin the telecommunications sector, assisting in routing, optimization, and network slicing for 5G technologies~\cite{gavrilovska2007graph,wang2003network,ahuja1993network,ordonez2017network}.

\end{enumerate}

This list showcases the wide impact of concepts from graph theory across a multitude of domains.
Figure~\ref{chp1:fig:graph_applications} illustrates specific applications of graphs across various domains.
As interconnected systems become increasingly central to understanding our world, the role of graph theory is likely to grow even more essential.

\reason{How the compute industry adapted to the growing graph computing needs.}

In the wake of the rising importance of graph-based computations, the hardware landscape within the compute industry began to undergo key shifts. Traditional Central Processing Units (CPUs), initially designed for sequential tasks, started incorporating SIMD-based graph extensions to enhance parallel processing capabilities~\cite{zheng2021efficient}.Graphics Processing Units (GPUs), with their inherent parallelism, were enhanced with kernel support tailored specifically for graph algorithms~\cite{tang2020high, rajamanickam2021kokkos}. Beyond these general-purpose processors, the industry also witnessed the advent of domain-specific accelerators~\cite{sarkar2023flowgnn, hwang2023grow, yao2022scalagraph, li2022lisa}, specifically crafted to speedup graph computations, addressing the unique challenges and demands that graph algorithms present.

On the software front, comprehensive software stacks have been designed from the ground up, specifically focusing on facilitating efficient graph processing. Libraries such as PyTorch-Geometric~\cite{fey2019fast} and Deep Graph Library (DGL)~\cite{wang2019deep} emerged, offering robust platforms for researchers and developers to implement and optimize graph algorithms. These software advances, in tandem with hardware innovations, have pushed the field of graph computing further forward. However, as with any rapidly evolving domain, while significant strides have been made, the journey is far from over. The vast potential of graph computing continues to present both challenges and opportunities that require further exploration and innovation.

\section{Challenges in Accelerating Graph Computing}

As promising as GNNs are, they are not without their computational challenges. Given the inherently recursive nature of GNNs, coupled with the irregular structure of many real-world graphs, we find significant bottlenecks in terms of their scalability and performance. Parallelizing GNN computations, which seems to be a logical solution given the abundance of task-level parallelism in graph processing, is riddled with data dependencies. The core challenges lie in the low spatial locality of data, which results in memory access inefficiencies and the computation stalls, making it difficult to achieve scalable parallel performance. Addressing these challenges is pivotal to harnessing the full potential of GNNs and making them a feasible solution for large-scale, real-world applications.

\subsection{Scalability Concerns in Graph Neural Networks}

GNNs present unique challenges in handling large graphs, which often involve billions to trillions of nodes and edges~\cite{fu2019exagraph}. The computational and memory demands of processing such exascale graphs escalate rapidly, often making computations infeasible on standard computational infrastructures. For instance, the Friendster social network graph comprises over 65 million nodes and 1.8 billion edges~\cite{yang2012defining}. When applying GNNs to this magnitude of data, the iterative and recursive nature of these networks requires massive memory bandwidth and computational power.  Even more complex graphs from biological and cosmological simulations are looming on the horizon, potentially reaching the exabyte scale~\cite{bader2019data}. Addressing the scalability challenges with GNNs on these graphs will be paramount to unlocking new scientific discoveries and advances in several fields.

\subsection{Task-Level Parallelism in Graph Computations}

Graph computations inherently possess a rich vein of task-level parallelism, given that many operations can theoretically be conducted simultaneously across different nodes or subgraphs~\cite{mccune2015thinking}. This natural parallelism emerges from the decentralized structure of graphs, where independent sub-tasks can be identified and processed in parallel, especially in sparse graph computations. However, this potential for parallelism is intermingled with intricate data dependencies among nodes and edges. These dependencies arise from the interconnections in the graph, leading to situations where the output of one task is contingent upon the result of another, thus necessitating synchronization and communication~\cite{beamer2017locality}. Consequently, fully capitalizing on the inherent task-level parallelism, while managing data dependencies, poses significant challenges, requiring advanced strategies to ensure efficiency and accuracy in graph computations.

\subsection{Data Spatial Locality and Computational Irregularity}

Graph computations, especially in the realm of GNNs and large-scale graph analytics, encounter two predominant challenges: 1) data spatial locality and 2) computational irregularity~\cite{chen2018graph}. First, the lack of data spatial locality implies that successive operations might access data dispersed across memory, leading to increased cache misses and degraded memory performance. Graphs, being inherently non-uniform, result in unpredictable memory access patterns, often unable to take advantage of cache hierarchies in modern processors~\cite{kiriansky2016avoiding}. Second, computational irregularity in graph algorithms surfaces due to the diverse node degrees and edge distributions, causing workload imbalance in parallel computing scenarios. This irregularity complicates efficient scheduling on parallel hardware and requires sophisticated load-balancing techniques to mitigate performance imbalance~\cite{khorasani2014cusha}. Both challenges pose barriers to fully realize the benefits of parallelism in graph computing.

\section{Objectives and Contributions}

The development of a graph accelerator tailored for GNN workloads requires following a systematic approach. Initially, our efforts are concentrated on conducting comprehensive benchmarks and characterizations of the GNN workloads, along with the various kernels targeted for optimization. Subsequently, our attention shifts to addressing a significant bottleneck in GNN workloads—the SpGEMM kernel—by devising and implementing an optimized version specifically for a custom accelerator. Finally, we integrate these insights and optimizations to architect a hardware accelerator, specifically designed to enhance the performance across a diverse range of GNN workloads.
A summary of these objectives and their associated contributions is presented in Table~\ref{tab:objectives_contributions}.

\begin{table}[ht]
\centering
\begin{tabular}{>{\raggedright\arraybackslash}p{6cm} >{\raggedright\arraybackslash}p{6cm}}
\toprule
\textbf{Objective} & \textbf{Contribution} \\
\midrule
Analyze the architectural impact of computing Graph Neural Networks. & Provide a benchmark suite for comprehensive evaluation of how different architectural components influence the performance and efficiency of GNN computations. \\
% \addlinespace
\midrule
Accelerate SpGEMM kernel on custom accelerator. & Propose and implement optimization strategies for the SpGEMM kernel, resulting in significant performance improvements on a custom hardware accelerator. \\
% \addlinespace
\midrule
Design a new accelerator to accelerate multiple GNN workloads. & Develop a novel hardware accelerator architecture tailored for various GNN workloads, demonstrating versatility and improved performance across sparse and dense compute kernels. \\
\bottomrule
\end{tabular}
\caption{Summary of Objectives and Contributions}
\label{tab:objectives_contributions}
\end{table}

\section{\Thesis Organization}

\begin{itemize}

    \item Chapter~\ref{chp1}: This chapter sets the stage for the thesis by highlighting the significance and impact of computations based on graph structures.

    \item Chapter~\ref{chp2}: This chapter provides background on the foundations of graph theory, providing a comprehensive overview of the various types of graphs, their properties and their representations. It also introduces machine learning on graphs, with a specific focus on Graph Neural Networks (GNNs), explaining their structure, functionality and applications.

    \item Chapter~\ref{chp3}: This chapter presents a thorough review of previous work in the domain, discussing existing approaches and solutions in workload characterization, GPU acceleration, Coarse-Grained Reconfigurable Arrays (CGRAs), and custom accelerators specifically designed for GNN workloads.

    \item Chapter~\ref{chp4}: We present a GNN benchmark suite in this chapter, offering a curated collection of workloads and tools to assess the performance of GNN computations. In addition, we conduct extensive analysis of the architectural requirements to efficiently run GNN workloads, examining how different components and configurations affect overall performance.

    %\item Chapter~\ref{chp5}: % Building on the insights gained from the previous chapters, we propose algorithmic optimizations for SpGEMM kernel designed to accelerate GNN workloads. 

    %\item Chapter~\ref{chp6}: % Incorporating insights from workload characterization, we deisgn a CGRA-based accelerator We present details of the design requirements, resulting architecture, and implementation of the solution, describing how we address the unique challenges posed by GNN computations.

    %\item Chapter~\ref{chp7}: % The final chapter of the thesis summarizes the key findings and contributions, discussing both the completed aspects of the research and outlining avenues for future work to be completed as a part of the thesis.

    \item Chapter~\ref{chp5}: We explore the SMASH (Sparse Matrix Atomic Scratchpad Hashing) algorithm, a novel SpGEMM kernel optimization aimed at enhancing GNN processing. We discuss the development and implementation of SMASH, including its various versions tailored to exploit distinct architectural features for improved efficiency in GNN workloads.
    
    \item Chapter~\ref{chp6}: In this chapter, we introduce NeuraChip, a custom CGRA-based accelerator designed to meet the unique demands of GNN computations. We provide detailed discussion on the architecture, including its heterogeneous processing approach, adaptive hash-based compute mapping, and mechanisms for rolling evictions, highlighting how NeuraChip addresses critical bottlenecks in GNN acceleration.
    
    \item Chapter~\ref{chp7}: The concluding chapter discusses the key insights and contributions of this work, ranging from the development of a GNN benchmark suite, to the introduction of algorithmic and hardware innovations such as SMASH and NeuraChip. We reflect on the impact of these contributions on the field of GNN acceleration.  We also cover potential directions for future GNN research to explore further advancements in graph computing.

\end{itemize}

% \section{Key Themes and Research Objectives}

% \subsection{The Power of Graph Neural Networks}

% \subsection{Architecting GNN Accelerators}

% TAlk about regular NN
% Talk about GNN

% Talk about accelerators
% Problems in GNNs running on accelerators
% This thesis will address it
% More Motivation
% Intro to this area of work
% Make it accessible to the non-specialist

% The story of how it is very easy to build custom accelerators now

\chapter{Fundamentals of Graph Computing and Accelerator Architectures}
\label{chp2}
% \section{Graph Theory}
% \subsection{Power Law Distribution}

% \section{Graph Neural Networks}

% \section{Accelerator Architecture}
% \subsection{Memory Hierarchy}
% \subsection{Compute Engines}
% \subsection{Interconnects}
% \subsection{ISA}
% \subsection{Compiler}

Before delving into the details of our accelerator proposal for graph computing, we explore the foundational concepts of graph computing and accelerator design. This chapter serves as a foundation, offering a systematic overview of the essential background, terminology and challenges that characterize graph computing, as well as its impact on hardware accelerators. By first reviewing these foundational elements, this chapter aims to provide the necessary background required to understand GNN accelerator design.

\section{Graph Theory Basics}
\begin{figure}[htbp]
    \centering
    \includegraphics[width=1.0\textwidth]{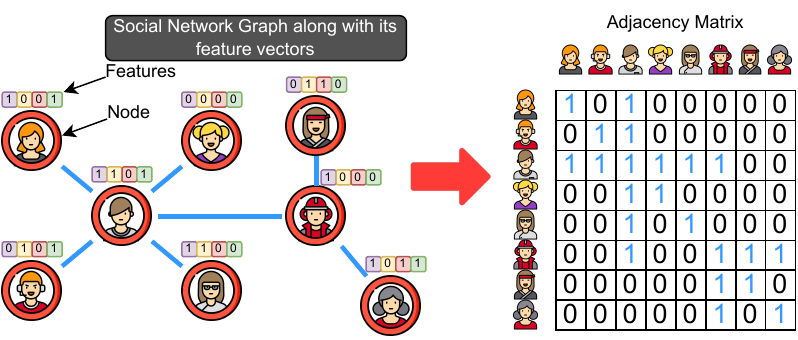}
    \caption{An example of a social network graph and its corresponding adjacency matrix. Each node in the graph is associated with a feature vector that contains the node's attributes.}
    \label{chp2:fig:social_network}
\end{figure}

Graph theory is a field within mathematics that explores the structure of interconnected nodes and edges. Graphs have gained widespread recognition for their ability to represent non-euclidean data. Graphs can represent many real-world problems, from network topologies and social networks to transportation systems and molecular structures. This section provides an overview of the fundamental elements of graphs, before diving into the use of graph-based applications in machine learning. Figure~\ref{chp2:fig:social_network} represents a social network graph, demonstrating how individuals are connected.  We also provide the associated adjacency matrix, which quantitatively captures these connections. Additionally, for each individual in the network, a feature vector is provided that captures various attributes or characteristics associated with that individual.

Mathematically, a graph $G$ is a pair of vertices (i.e.,  nodes) $V$ and edges $E$, that is represented as: 

\begin{equation}
G = (V, E)
\end{equation}

where $V$ is a set of vertices, and $E$ is an unordered set of pairs of vertices.
A singular edge $e$ within the set of edges $E$ is represented as $e = \{x, y\}$ or simply $e = xy$, where $x$ and $y$ are endpoints (nodes) of the edge.
$x$ and $y$ are said to be neighbors or adjacent nodes in the graph $G$.

\subsection{Graph Properties}
Graph properties provide insight into the structure, characteristics and behavior of graphs. Here are some of the fundamental properties and characteristics of graphs:

\begin{enumerate}

    \item \textbf{Degree}: The number of edges incident on a vertex. In directed graphs, we differentiate between in-degree (number of incoming edges) and out-degree (number of outgoing edges).

    \item \textbf{Order and Size}: The order of a graph refers to the number of  vertices and the size refers to the number of its edges.

    \item \textbf{Diameter}: The longest shortest path between any two vertices.

    \item \textbf{Radius}: The minimum eccentricity of any vertex in the graph. The eccentricity of a vertex is the greatest distance from the vertex to any other vertex.

    \item \textbf{Girth}: The length of the shortest cycle in the graph.

    \item \textbf{Adjacency}: Two vertices are said to be adjacent if they are connected by an edge.
    
    \item \textbf{Clique}: A set of vertices where each pair is adjacent.

    \item \textbf{Path}: A sequence of vertices where each adjacent pair is connected by an edge.

    \item \textbf{Cycle}: A path that starts and ends at the same vertex. A cyclic graph is a graph that contains at least one cycle, which is a sequence of vertices where the first and last vertices are the same, and each pair of consecutive vertices in the sequence is connected by an edge. An acyclic graph is a graph that contains no cycles.

    \item \textbf{Connectivity}: A graph is connected if there's a path between every pair of vertices. In directed graphs, if a graph is strongly connected, there is a directed path between any pair of vertices.

    \item \textbf{Connected Components}: 
    
    \textit{For undirected graphs}: A connected component is a subgraph in which any two vertices are connected to each other by paths, and which is connected to no additional vertices in the supergraph (i.e., the main graph). In simpler terms, a connected component is a ``piece" or ``part" of the graph where there's a route between any pair of nodes within that piece, but no connection to nodes outside of it.

    \textit{For directed graphs}: Connected components are further classified as Strongly Connected Components and Weakly Connected Components.

    \textit{\textbf{Strongly Connected Components}}: A strongly connected component of a directed graph is a maximal strongly connected subgraph. This means that for every pair of vertices $u$ and $v$ in the subgraph, there's a directed path from $u$ to $v$ and a directed path from $v$ to $u$.

    \textit{\textbf{Weakly Connected Components}}: If you were to ignore the directionality of the edges in a directed graph, and the graph becomes connected, then the graph is said to be weakly connected. The maximal subgraphs of this type are the weakly connected components.

    \item \textbf{Planarity}: A graph is said to be planar if it can be embedded (i.e., drawn) in the plane such that no edges intersect or cross each other except at their endpoints (vertices). In other words, a graph is planar if it can be drawn on a flat surface without any of its edges overlapping or crossing, except where they meet at nodes. This definition applies to both directed and undirected graphs.

    \item \textbf{Isomorphism}: Isomorphism refers to a one-to-one correspondence between the vertices of two graphs such that the adjacency relation between pairs of vertices is preserved. In simpler terms, two graphs are isomorphic if they are essentially the same in terms of structure, though they might look different in their graphical representation.
    
    Formally, two graphs $G_1$ and $G_2$ are said to be isomorphic if there exists a bijective function $f: V(G_1) \rightarrow V(G_2)$ such that for any two vertices $u$ and $v$ of $G_1$, there is an edge between $u$ and $v$ in $G_1$ if and only if there is an edge between $f(u)$ and $f(v)$ in $G_2$.

\end{enumerate}

\subsection{Types of Graphs}

Graphs can be categorized in various ways based on their properties, structures, and applications. Below is an overview of the primary types of graphs:

\begin{enumerate}
    \item \textbf{Undirected Graph}: An undirected graph is a simple structure consisting of nodes, also known as vertices, connected by edges. In this type of graph, the edges don't have a specific direction. That is, if vertex $A$ is connected to vertex $B$, then vertex $B$ is equivalently connected to vertex $A$. Such graphs are commonly used to represent symmetric relationships, for example, friendships in a social network.

    \item \textbf{Directed Graph (or Digraph)}: Directed graphs, often called digraphs, also comprise vertices and edges. However, the crucial difference is that the edges have a direction. An arrow from vertex $A$ to vertex $B$ signifies a one-way relationship. Digraphs are especially useful in representing prerequisites in a course structure or transitions in a finite automaton.

    \item \textbf{Weighted Graph}: A weighted graph assigns a specific weight or value to each of its edges. This weight can represent various characteristics such as distance, cost, or any measurable quantity relevant to the problem being addressed. For instance, in mapping out a city's road network, the weights could symbolize the distances or travel times between intersections.

    \item \textbf{Unweighted Graph}: Contrary to weighted graphs, unweighted graphs treat each edge equally, without any specific value or weight assigned. Such graphs are often utilized in scenarios where only the relationship or connection between nodes is of interest, without any quantitative measure on the edges.

    \item \textbf{Cyclic Graph}: A cyclic graph contains at least one cycle, which is a closed path in which a vertex is revisited without retracing any edge. Cyclic graphs can represent systems or networks where it's possible to return to a starting point via a unique route.

    \item \textbf{Acyclic Graph}: An acyclic graph is devoid of any cycles. This means it's impossible to start at one vertex and traverse the graph in such a way that you return to the starting vertex without backtracking. A classic example of an acyclic graph is a tree.

    \item \textbf{Connected Graph}: In a connected graph, there exists a path between every pair of vertices, ensuring that no vertex is isolated. Such graphs are particularly valuable in scenarios where continuous connectivity is essential (communication networks are one such example).

    \item \textbf{Disconnected Graph}: As the name suggests, a disconnected graph has one or more vertices that aren't connected to the rest of the graph. In other words, not all pairs of vertices are reachable from each other. Such graphs might represent fragmented or isolated systems.

    \item \textbf{Complete Graph}: A complete graph is a robust structure where every pair of distinct vertices is connected by a unique edge. In terms of social networks, a complete graph would mean every person knows every other person directly.

    \item \textbf{Bipartite Graph}: A graph $G$ is called bipartite if its vertex set can be partitioned into two disjoint sets $U$ and $V$ such that every edge connects a vertex in $U$ to one in $V$. This means that there are no edges that connect vertices within the same set $U$ or within the same set $V$.
    
    Formally, a graph $G = (V, E)$ is bipartite if there exists a partition $(U, V)$ of its vertex set $V$ such that for every edge $(x, y) \in E$, either $x \in U$ and $y \in V$ or $x \in V$ and $y \in U$.

    \item \textbf{Planar Graph}: A planar graph can be drawn on a plane without any edges crossing, except at their endpoints. Such graphs are beneficial in specific design and layout problems, ensuring no overlaps or intersections.

    \item \textbf{Tree}: A tree is a special kind of graph that's both connected and acyclic. Trees are hierarchical structures commonly used in computer science for data structures such as binary search trees and file systems.

    \item \textbf{Forest}: A forest is a collection of disjoint trees, meaning it's acyclic but not necessarily connected. Forests can represent multiple independent hierarchies or classifications within a system.

    \item \textbf{Multigraph}: Multigraphs allow for multiple edges, also termed parallel edges, between the same set of vertices. This type of graph can be useful in scenarios where multiple distinct relationships or connections exist between the same entities.

    \item \textbf{Simple Graph}: A simple graph is a basic structure where each pair of vertices shares at most one edge, and there are no loops. It's the foundational form of many other graph types and serves as a starting point in many graph theory discussions.

    \item \textbf{Hypergraph}: A hypergraph generalizes the traditional graph concept by allowing edges, often called hyperedges, to connect any number of vertices, not just two. Hypergraphs can represent complex relationships that don't fit neatly into pairwise associations.

    \item \textbf{Subgraph}: A subgraph is formed by selecting a subset of vertices and edges from a larger graph, without introducing any new ones. Subgraphs are crucial for analyzing specific portions or aspects of a larger network or system.

    \item \textbf{Regular Graph}: In a regular graph, every vertex has the same degree, meaning each node connects to an equal number of other nodes. This uniformity can simplify certain analyses and algorithms. Specifically in the context of graph computations, Regular graphs lead to close to uniform work distribution across the computing cores.

\end{enumerate}

\subsection{Graph Representation}
Graphs, as abstract mathematical structures, need to be represented in a tangible form, especially for computational processes. The choice of representation can significantly influence the efficiency of various graph algorithms. The most common forms of graph representation include:

\begin{enumerate}

    \item \textbf{Adjacency Matrix}: This is a 2D array of size $V \times V$ (where $V$ is the number of vertices in a graph). The entry $m_{ij}$ is either 1 (or the edge's weight) if there's an edge between vertices $i$ and $j$, and 0 otherwise. While this method provides a quick way to check the presence of a specific edge, it can be space-inefficient for sparse graphs as it requires $O(V^2)$ space. Figure~\ref{chp2:fig:social_network} illustrates a social network graph alongside its representation in adjacency matrix format.

    \item \textbf{Adjacency List}: For every vertex, a list of its adjacent vertices is maintained. This representation is more space-efficient for sparse graphs. In this method, an array of lists is used, with the size of the array being equal to the number of vertices. The $i$th position in the array holds a list of nodes to which node $i$ is connected.

    \item \textbf{Incidence Matrix}: This is a 2D array where rows represent vertices and columns represent edges. For example, for an undirected graph, the entry $m_{ij}$ is 1 if vertex $i$ is incident to edge $j$, -1 if $i$ is the edge's terminal vertex, and 0 otherwise. For a directed graph, the entry is -1 for the tail of the arrow (edge) and 1 for the head.

    \item \textbf{Edge List}: This is a list of pairs (or triples, if weights are present) that represent edges. For instance, an edge from vertex $A$ to vertex $B$ with weight $w$ can be represented as $ (A, B, w) $. This representation is particularly useful when the graph structure is more concerned with edges rather than vertices.

\end{enumerate}

The choice of representation often hinges on the specific operations that need to be optimized. For instance, adjacency lists are faster for traversal algorithms, while adjacency matrices can be more suitable for algorithms involving edge lookups or matrix operations.

\subsection{Graph Transformation}

Graph transformation is a powerful technique that focuses on the modification and manipulation of graph structures. It offers a systematic way to derive a new graph from an existing one, serving both as a computational tool and a conceptual methodology to analyze various properties and behaviors of graphs. Common types of graph transformations include:

\begin{enumerate}

    \item \textbf{Subgraph Extraction}: This involves creating a new graph by selecting a subset of vertices and edges from the original graph, usually based on certain criteria or conditions.

    \item \textbf{Graph Contraction}: This process combines multiple vertices into a single vertex, often to simplify a graph's structure while retaining its fundamental characteristics.

    \item \textbf{Graph Expansion (or Vertex Splitting)}: This is the reverse of contraction. A single vertex is expanded into multiple vertices, with edges adjusted accordingly.

    \item \textbf{Edge Contraction}: Two vertices connected by an edge are merged into a single vertex, and the edge is removed. The new vertex retains all edges that the original vertices had, except for the contracted edge.

    \item \textbf{Line Graph Transformation}: Given a graph, its line graph is another graph representing the relationship between the edges of the original graph. Each vertex in the line graph represents an edge in the original graph, and two vertices in the line graph are connected if their corresponding edges in the original graph are incident on a common vertex.

    \item \textbf{Dual Graph Transformation}: Applied typically to planar graphs, this creates a vertex in the dual graph for every face in the original graph, and two vertices in the dual graph are connected by an edge if their corresponding faces in the original graph are separated by an edge.

Graph transformations play a crucial role in a myriad of applications, including algorithm design, network analysis, and optimization problems, by enabling alternative perspectives and simplifications of the original structures.

\end{enumerate}

\section{Machine Learning on Graphs}
\label{chp2:sec:ml_on_graphs}
Building upon the principles of graph theory just reviewed, we now transition to the applications. Graphs are useful abstractions as then naturally map to a number of machine learning tasks. Graphs are used for non-Euclidean data structures that encapsulate relationships, hierarchies, and patterns, which are difficult to model in traditional data formats. In this section, we delve into the techniques, algorithms, and challenges associated with leveraging graph data for predictive and analytical tasks.
This emerging field promises important advancements in domains ranging from social network analysis and recommendation systems to bioinformatics and traffic routing.

\subsection{Graph Neural Networks}

Graph Neural Networks (GNNs) have emerged as powerful tools for learning and processing data structured as graphs. Unlike traditional neural networks that operate on fixed-size vectors, GNNs work directly with graphs, accommodating their non-euclidean nature and inherent irregularities. At the core of GNNs lies the principle of aggregating information from a node's neighbors to iteratively update the node's representation. This process captures both local structures and broader topological features of the graph. Through successive layers, GNNs can accumulate and transform information from increasingly larger neighborhoods around each node. This ability to learn meaningful representations of nodes, edges, or entire graphs has led to their successful application in diverse areas such as social network analysis, molecular chemistry, and recommendation systems, bridging the gap between the rich expressivity of graphs and the computational capabilities of deep learning.

\begin{figure}[htbp]
    \centering
    \includegraphics[width=1.0\textwidth]{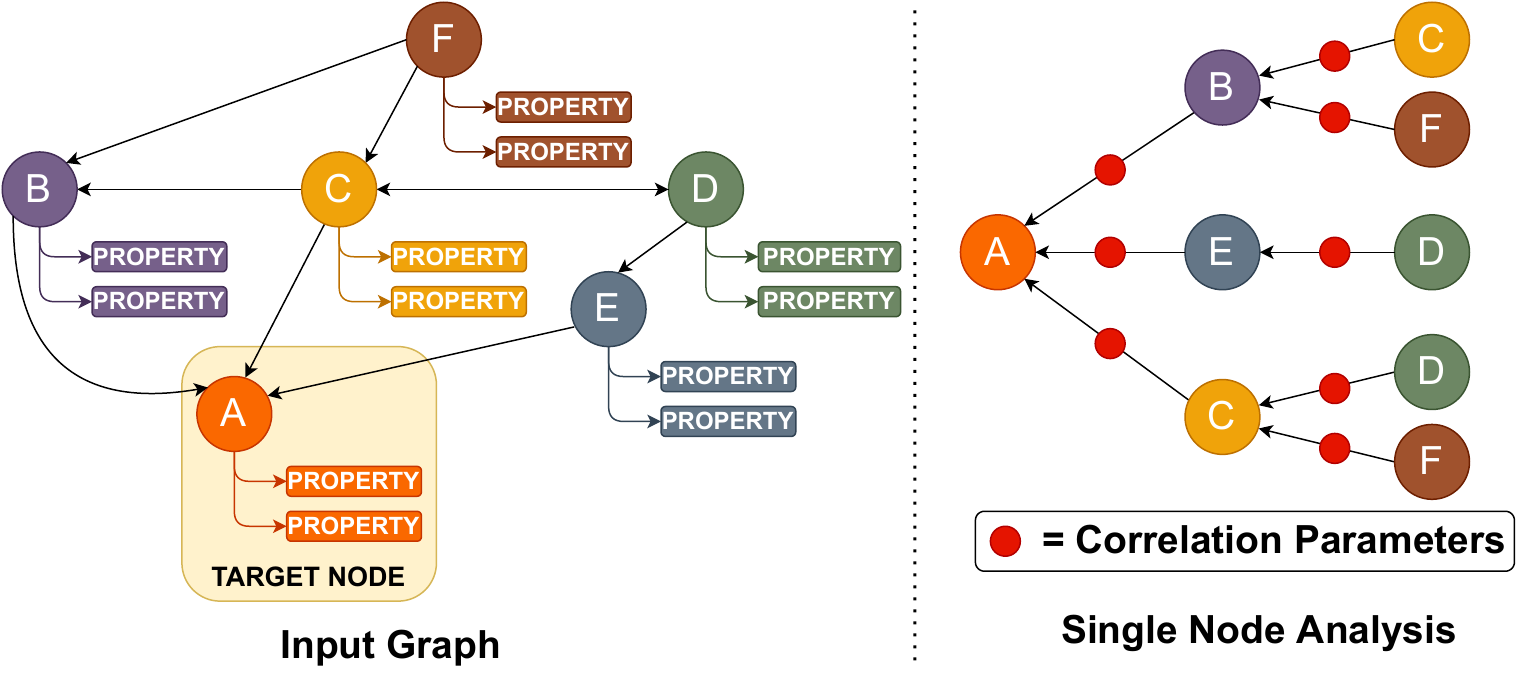}
    \caption{Analysis of Graph Neural Networks, demonstrating the propagation of node properties influenced by the graph's topology.}
    \label{chp2:fig:gnn_analysis}
\end{figure}

% A Graph Neural Network~(GNN) is a Machine Learning~(ML) model designed to work on non-euclidean data, originally proposed to solve node classification problems~\cite{scarselli2008graph}. The core idea behind using GNNs is to collect and aggregate information about a graph's structure to capture its inherent features and predict properties for specific nodes, connections, and generalizations to unseen graphs. 
% Figure~\ref{fig:intro_diag}~(left) shows an example input graph with several edges connecting six nodes (A--F), where each node is represented by a set of feature vectors (properties).

Before a GNN model can make predictions, the model must first be trained.  As shown in Figure~\ref{chp2:fig:gnn_analysis}, the goal of GNN training is to learn correlation parameters for each node, capturing its relation to the rest of the graph.
More specifically, a feature vector for node A relates node A's properties to its neighbors' properties, nodes B, C and E. Each of these neighbors, in turn, has their own feature vectors to relate to their own neighbors. Hence, the properties of node A can be associated with the properties of every other node in the graph. This feature of GNNs is also useful for finding missing properties of nodes in a graph. Similar to DNNs, a GNN can also have multiple layers, with each layer represented by two functions: i) an aggregation function and ii) an update function~(i.e., a combination function). As the name suggests, the aggregation function is responsible for collecting or pooling the features of the neighbors for a given node. On the other hand, the update function is responsible for updating each node's feature vectors using Multi-Layer Perceptrons~(MLPs). A GNN model can have layers with different aggregation and update functions. The deeper the GNN model, the more information a node has about other nodes that are distant from it in the graph. However, training deeper GNNs is difficult, primarily due to the vanishing gradient problem~\cite{hochreiter1998vanishing}. As GNNs grow deeper, the gradients become so small that the weights stop getting updated. This property makes it difficult to train the GNN further. To address these challenges, novel GNN architectures have been proposed to enable deeper GNN models~\cite{li2019deepgcns}.

\subsection{Graph Convolutional Networks}
Graph Convolutional Networks (GCNs) represent a key advancement in the domain of graph-based machine learning. These networks are designed to process data structured as graphs, allowing for the consideration of both node features and the graph's inherent structure. Traditional neural networks are ill-suited for graph data due to the irregular and non-Euclidean nature of graphs. In contrast, GCNs leverage the spatial relationship between nodes to propagate and aggregate information through the graph, thus learning a more comprehensive representation of the data.

The propagation rule for a GCN layer can be described as:
\begin{equation}
H_{(l+1)} = \sigma \left( \tilde{D}^{-\frac{1}{2}} \tilde{A} \tilde{D}^{-\frac{1}{2}} H_{(l)} W_{(l)} \right)
\end{equation}

Where:

\( H^{(l)} \) is the matrix of node features at layer \( l \).

\( \tilde{A} \) is the adjacency matrix of the graph with added self-connections 

\( \tilde{D} \) is the diagonal node degree matrix of \( \tilde{A} \).

\( W^{(l)} \) is the weight matrix for layer \( l \).

\( \sigma \) is an activation function, e.g., the ReLU function.

The key principle of a GCN is the neighborhood aggregation scheme. For each node in the graph, a GCN layer aggregates feature information from its neighbors and possibly itself. This aggregated information is then passed through a transformation (usually a linear transformation, followed by a non-linear activation function). The process can be iteratively run over multiple layers, enabling the aggregation of information from a larger neighborhood at each subsequent layer. The inclusion of this spatial-based information aggregation makes GCNs particularly adept at node classification, graph classification, and link prediction, especially when the structure of the graph plays a significant role in the underlying data distribution.

\subsection{Graph Isomorphism Networks}

Graph isomorphism is a central concept in the field of graph theory, revolving around the study of the structural equivalence between two graphs. Two graphs $G_1$ and $G_2$ are considered isomorphic if there exists a one-to-one correspondence (or bijective function) between their vertices, such that the adjacency relationship is preserved. In other words, the graphs are structurally identical, and one can be transformed into the other merely by relabeling the vertices without altering the underlying connectivity pattern. While the concept sounds straightforward, determining whether two large graphs are isomorphic in an efficient manner remains a challenging computational problem.

The update rule for the GIN can be formulated as:
\begin{equation}
h^{(l+1)}_v = \text{MLP}^{(l)} \left( h^{(l)}_v + \sum_{u \in \mathcal{N}(v)} h^{(l)}_u \right)
\end{equation}

Where:

\( h^{(l+1)}_v \) is the feature vector of node \( v \) at layer \( l+1 \).

\( \mathcal{N}(v) \) represents the neighbors of node \( v \).

\( \text{MLP}^{(l)} \) denotes a multi-layer perceptron used at layer \( l \).

The GIN introduces an additional learnable parameter to weigh the importance of self-features versus neighbor features. This ensures that the GIN can capture subtle structural details, making it a powerful tool for graph representation learning.

% The Weisfeiler-Lehman (WL) Isomorphism Test~\cite{shervashidze2011weisfeiler}, also known as the Weisfeiler-Lehman method of graph isomorphism, offers an iterative approach to distinguish non-isomorphic graphs. The method involves a color-refinement process wherein each vertex is assigned a multi-set of labels based on its neighborhood's structure. In every iteration, the method refines these labels based on previously assigned labels to the vertices and their neighbors. This process continues until either the labels become stable, or a set number of iterations are reached. If, at the end of this procedure, the two graphs have identical multi-set labels for their vertices, they are potentially isomorphic. However, the converse isn't necessarily true; the test can produce false positives, making it a necessary but not always sufficient condition for isomorphism.

Understanding and recognizing graph isomorphism has profound implications in numerous areas of science and technology. For example, in chemistry, graph isomorphism can be used to determine molecular similarity, as molecules can be represented as graphs where atoms are vertices and bonds are edges. Similarly, in computer science, isomorphic graphs might denote equivalent solutions or states in certain problems. Furthermore, in database search and pattern recognition, determining graph isomorphism efficiently can aid in retrieving or recognizing specific patterns amidst a large dataset. However, due to the complexity of the problem, especially with large graphs, much research has been invested in finding efficient algorithms and heuristic methods to tackle the isomorphism challenge.

%%%%%%%%%

%% Graph Attention Networks

\subsection{Graph Attention Networks}

Graph Attention Networks (GATs) mark a significant evolution in graph neural network technology by introducing an attention mechanism that allows nodes to dynamically assign importance to their neighbors' information. This attention-based approach enables the model to focus more on relevant features from a neighborhood, enhancing the adaptability and performance of the network on graph-structured data. GATs address the limitation of conventional graph neural networks, such as GCNs, which treat all neighbors equally during the aggregation process. By incorporating attention, GATs can assign a weight based on the influence of each neighbor based on the task at hand, leading to more effective learning outcomes.

The propagation rule for a GAT layer can be described as:
\begin{equation}
h^{(l+1)}_i = \sigma \left( \sum_{j \in \mathcal{N}(i) \cup \{i\}} \alpha_{ij}^{(l)} W^{(l)} h^{(l)}_j \right)
\end{equation}

Where:

\( h^{(l+1)}_i \) is the feature vector of node \( i \) at layer \( l+1 \).

\( \alpha_{ij}^{(l)} \) represents the attention coefficient between nodes \( i \) and \( j \) at layer \( l \), indicating the significance of node \( j \)'s features to the update of node \( i \)'s features.

\( W^{(l)} \) is the weight matrix for layer \( l \).

\( \sigma \) is an activation function, such as the ReLU function.

The computation of attention coefficients involves a self-attention mechanism where a shared attention function, applicable to all edges, computes the coefficients based on the features of the nodes at either end of the edge. This process allows GATs to perform feature extraction that is both context-aware and adaptive, leading to more nuanced representations of nodes based on their local graph topology.

GATs have demonstrated superior performance on various tasks, including node classification, graph classification, and link prediction, particularly in scenarios where the relevance of neighboring nodes varies significantly. The model's ability to selectively prioritize information makes it highly effective in capturing the complex dependencies characteristic of graph-structured data, thereby pushing the boundaries of what is achievable with graph neural networks.

%% Graph SAGE

\subsection{GraphSAGE}

GraphSAGE (Graph Sample and AggregatE) is a novel framework designed to efficiently generate node embeddings for large-scale graphs. Unlike traditional graph neural networks, such as GCNs, that require the entire graph to be processed simultaneously, GraphSAGE introduces a more scalable approach by sampling a fixed-size neighborhood around each node and aggregating their features. This methodology allows GraphSAGE to efficiently deal with graphs of varying sizes and topologies, including those that evolve over time, by learning a function that can generate embeddings for unseen nodes based on their local neighborhoods.

The propagation rule for GraphSAGE can be generalized as:
\begin{equation}
h^{(l+1)}_i = \sigma \left( W^{(l)} \cdot \text{AGGREGATE}^{(l)} \left( \{h^{(l)}_j | j \in \mathcal{N}(i) \} \right) \right)
\end{equation}

Where:

\( h^{(l+1)}_i \) is the feature vector of node \( i \) at layer \( l+1 \).

\( \mathcal{N}(i) \) denotes the set of neighbors for node \( i \).

\( \text{AGGREGATE}^{(l)} \) is a function that combines the feature vectors of the sampled neighborhood nodes at layer \( l \).

\( W^{(l)} \) is the weight matrix for layer \( l \).

\( \sigma \) represents an activation function, such as the ReLU function.

GraphSAGE's aggregation functions can vary, including mean, LSTM, and pooling aggregators, which allows the model to be tailored to specific types of graph data and applications. This flexibility, combined with the efficiency of sampling, makes GraphSAGE particularly suitable for dynamic graphs and scenarios where real-time embedding generation is crucial.

Moreover, by learning to aggregate information from a node's local neighborhood, GraphSAGE can leverage the structural information inherent in the graph, allowing for powerful representations that capture both the features of individual nodes and their relational context within the graph. This approach has proven effective across a range of tasks, including node classification, link prediction, and graph classification, particularly in domains where the graph structure is indicative of underlying patterns or relationships, such as social networks, recommendation systems, and knowledge graphs.

%% Principal Neighborhood Aggregation

\subsection{Principal Neighborhood Aggregation}

Principal Neighborhood Aggregation (PNA) addresses the need for more nuanced aggregation mechanisms capable of capturing the diverse structural properties within graphs. Diverging from traditional graph neural networks such as GCNs, which primarily utilize simplistic aggregation functions such as sum, mean, or max, PNA introduces a multifaceted approach by integrating several aggregation schemes and a degree-scaling component. This mixture of techniques enhances the model's ability to represent the intricate patterns and relationships inherent in graph-structured data.

The core idea of PNA is to leverage the strengths of multiple aggregation functions simultaneously, thereby improving the feature representation of each node by capturing various aspects of its neighborhood's structure and feature distribution. The inclusion of a degree-scaling component further refines this process by adjusting the influence of neighboring nodes based on their connectivity, thus providing a more balanced and informative aggregation outcome.

The general update rule for PNA can be encapsulated as:
\begin{equation}
h^{(l+1)}_i = \sigma \left( \sum_{\text{agg} \in \mathcal{A}} \delta_{\text{agg}} \cdot W^{(l)}_{\text{agg}} \cdot \text{agg}\left( \{h^{(l)}_j | j \in \mathcal{N}(i) \cup \{i\} \} \right) \right)
\end{equation}

\noindent
Where:

\noindent
- \( h^{(l+1)}_i \) denotes the feature vector of node \( i \) at layer \( l+1 \).

\noindent
- \( \mathcal{A} \) is a set of aggregation functions, such as sum, mean, and max.

\noindent
- \( \delta_{\text{agg}} \) represents a degree-scaling factor associated with each aggregation function, optimizing the impact of node degrees on the aggregation process.

\noindent
- \( W^{(l)}_{\text{agg}} \) is a weight matrix specific to each aggregation function at layer \( l \).

\noindent
- \( \sigma \) is an activation function, for instance, the ReLU function.

PNA's design is particularly effective in graphs where nodes exhibit significant variability in their degree distribution. By considering multiple aggregation perspectives and adjusting for node degree, PNA ensures an accurate representation of each node's neighborhood, significantly improving the performance on tasks such as node classification, graph classification, and link prediction.

\section{Graph Neural Network Frameworks}
Support for GNN primitives in popular ML frameworks is increasing. Today, researchers from the ML community are developing libraries in the form of extensions to frameworks such as PyTorch and TensorFlow. The two most popular libraries/extensions that implement customized GNN kernels, as well as provide programming support in the form of APIs, are PyTorch Geometric~(PyG)~\cite{fey2019fast} and the Deep Graph Library~(DGL)~\cite{wang2019deep}. PyG is an extension based on top of the PyTorch library and so only supports PyTorch. On the other hand, DGL provides support for PyTorch, TensorFlow, and MXNet. 

Spektral~\cite{grattarola2020graph} and Aligraph~\cite{zhu2019aligraph} are two librariesk built on top of TensorFlow, used for GNN training. GraphNets~\cite{battaglia2018relational} is a GNN framework from Google, supported using TensorFlow as the backend. As PyG and DGL bridge both the semantic and performance gaps when developing GNN models, they are the most widely used frameworks by both the ML community~\cite{zhang2020deep} and architecture community.~\cite{zhang2020architectural,yan2020characterizing,yan2020hygcn,liang2020engn}.
Table~\ref{tab:pyg_vs_dgl} presents a comprehensive comparison between PyTorch Geometric (PyG) and Deep Graph Library (DGL), two prominent libraries for implementing Graph Neural Networks (GNNs). Both libraries are actively maintained, with extensive documentation and a large user community, ensuring robust support for developers and researchers in the field of graph-based deep learning.

\begin{table}[h]
\centering
\caption{Comparison between PyTorch Geometric (PyG) and Deep Graph Library (DGL)}
\label{tab:pyg_vs_dgl}
\begin{tabularx}{\textwidth}{l X X}
\toprule
\textbf{Feature/Aspect} & \textbf{PyTorch Geometric (PyG)} & \textbf{Deep Graph Library (DGL)} \\
\midrule
\midrule
\textbf{Programming Language} & Primarily Python & Primarily Python \\
\midrule
\textbf{Deep Learning Framework} & Built on top of PyTorch & Supports both PyTorch and TensorFlow \\
\midrule
\textbf{Ease of Use} & High-level API, easy to use for beginners & Low-level C++ API available \\
\midrule
\textbf{Performance} & Efficient and scalable & Highly optimized for sparse operations \\
\midrule
\textbf{Models Available} & Extensive collection of pre-implemented GNN models & Newer models only optimized for PyG \\
\midrule
\textbf{Community and Support} & Large community, active development & Relatively new library with community beginning to grow \\
\midrule
\textbf{Documentation} & Extensive documentation with examples & Comprehensive documentation \\
\midrule
\textbf{Graph Types Supported} & Supports heterogeneous and temporal graphs & Supports heterogeneous graphs \\
\midrule
\textbf{Scalability} & Optimized for single-machine, multi-GPU setups & Designed for distributed training on multiple machines and large graphs \\
\midrule
\textbf{Extensibility} & Easy to extend and contribute & Low-level C++ API relatively difficult to extend \\
\bottomrule
\end{tabularx}
\end{table}

\section{Accelerators: An Overview}

Accelerators are specialized hardware components designed to perform specific computational tasks more efficiently than general-purpose processors. The primary motivation behind their design lies in addressing the performance and efficiency bottlenecks encountered in conventional computing systems, particularly for workloads that require intensive data processing or have unique computational patterns. Accelerators are engineered to offload specific tasks from the central processing unit (CPU), thereby enhancing the overall system performance and energy efficiency.

\subsection{Unique Advantages of Accelerators}
The distinct advantages of accelerators stem from their specialized architecture, which is tailored to execute a specific set of operations. This specialization enables several benefits:

\begin{itemize}
    \item \textbf{Enhanced Performance:} Accelerators can execute certain tasks or algorithms much faster than general-purpose CPUs due to their optimized hardware design for those tasks.
    \item \textbf{Energy Efficiency:} By offloading intensive tasks from CPUs, accelerators can reduce overall power consumption, making them ideal for energy-sensitive applications.
    \item \textbf{Parallel Processing Capabilities:} Many accelerators, just like GPUs, are capable of handling multiple operations concurrently, effectively leveraging the parallel hardware of the accelerator.
    \item \textbf{Customizability:} Accelerators can be customized for the specific needs of an application, allowing for greater flexibility in handling diverse computational requirements.
\end{itemize}

\subsection{Advantages and Disadvantages of Designing Accelerators}
While accelerators offer considerable advantages, there are also trade-offs involved in their design and deployment.

\subsubsection{Advantages}
\begin{itemize}
    \item \textbf{Highly Efficient for Targeted Tasks:} Accelerators provide optimized performance for specific applications, such as graphics processing, machine learning, and now, graph computing.
    \item \textbf{Scalability:} They can be scaled to handle larger workloads more effectively than general-purpose processors.
    \item \textbf{Innovation:} The development of accelerators drives technological innovation, particularly in fields that require high computing power.
\end{itemize}

\subsubsection{Disadvantages}
\begin{itemize}
    \item \textbf{Limited Flexibility:} Being specialized, accelerators are not as versatile as CPUs for general computing tasks.
    \item \textbf{Development Complexity:} Designing and implementing accelerators can be complex and resource-intensive.
    \item \textbf{Integration Challenges:} Integrating accelerators into existing systems may require significant architectural changes and software support.
    \item \textbf{Cost:} The development and deployment of accelerators can be costly, especially for cutting-edge designs.
\end{itemize}

Overall, accelerators represent a critical advancement in computing technology, offering specialized solutions for a range of emerging applications. In particular, their role in graph computing has opened new avenues for handling complex data structures and algorithms efficiently. However, the decision to utilize accelerators must consider the balance between their specialized capabilities and the associated design and integration challenges.

% \section{Compiler Design}

%  ____      _       _           _  __        __         _
% |  _ \ ___| | __ _| |_ ___  __| | \ \      / /__  _ __| | __
% | |_) / _ \ |/ _` | __/ _ \/ _` |  \ \ /\ / / _ \| '__| |/ /
% |  _ <  __/ | (_| | ||  __/ (_| |   \ V  V / (_) | |  |   <
% |_| \_\___|_|\__,_|\__\___|\__,_|    \_/\_/ \___/|_|  |_|\_\
\chapter{Related Work}
\label{chp3}

In this chapter, we review the existing body of work on graph computing. Our initial focus is on the examination of prior research concerning the benchmarking of graph-based workloads. Next, we discuss previous work aimed at accelerating graph neural networks.

\section{Graph Computing Benchmark Suites}

Past GPU benchmark suites have provided guidance to GPU architects. To date, GPU benchmarks fall into one of two categories. They either evaluate general-purpose GPU computing capabilities~\cite{rodinia, parboil, che2013pannotia, shoc, polybench, nupar}, or target assessment of the performance of a specific class of workloads~\cite{lonestar, heteromark, tartan}. With the growing popularity of DNN workloads, a new wave of DNN benchmarks have been developed. 

\textbf{Benchmarking Deep Learning Workloads and Workload Characterization:} Early DNN benchmark suites explored the execution performance of low-level primitives~\cite{deepbench, dnnmark}, as well as end-to-end inference and training~\cite{adolf2016fathom, coleman2017dawnbench}. 
Later efforts included a more diverse set of DNN algorithms, including a broader range of network models and commercial efforts.  TBD~\cite{zhu2018benchmarking} is a DNN benchmark suite proposed by Zhu et al. to study DNN training performance on GPUs. AIBench~\cite{aibench} is an industry-initiated benchmark suite that is focused on industrial AI services. Mattson et al.~\cite{mattson2020mlperf} proposed the MLPerf training and MLPerf inference suites. MLPerf adopts ideas from prior DNN benchmark suites to develop an industry-standard DNN-focused benchmark suite, designed so that new hardware and software optimizations can be evaluated fairly. To date, MLPerf has primarily focused on DNNs.  In terms of architectural characterization, Dong et al.~\cite{dong2018characterizing} looked at the architectural implications of CNN training on a GPU. Mojumder et al.~\cite{mojumder2018profiling} profiled DNN models trained on an NVIDIA DGX-1 system. However, all these prior workload studies were limited to DNNs that operate on euclidean data (e.g., images, video and speech). In this thesis we develop GNNMark.  We specifically aim to bridge this gap, providing the architecture community with an appropriate benchmark suite to study GNN training behavior. We also plan to work with the MLPerf consortium to integrate our GNN models into their training suite in the future. 

\textbf{Workload Characterization for GNNs:}
GNNs have recently attracted attention from the computer architecture community due to their growing popularity in the machine learning domain. Yan et al.~\cite{yan2020characterizing} have characterized Graph Convolutional Network~(GCN) inference performance, focusing on aggregation and model update phases. Zhang et al.~\cite{zhang2020architectural} have also characterized the inference performance of GNNs. Their work decomposes GNN inference execution into a Scatter-ApplyEdge-Gather-ApplyVertex~(SAGA) pipeline and then analyzes the behavior of each phase. They also present insights on how to efficiently design a GNN accelerator for inference. While a benchmark suite is also created as a part of their study, it is designed primarily for inference and is not available publicly. Most prior GNN studies focused on inference, ignoring the training process that tends to consume a large number of GPU hours. Also, the models evaluated are only designed to process homogeneous graphs. Other related work focused on characterizing GNN inference and designing customized accelerators for that purpose~\cite{yan2020hygcn,liang2020engn,kiningham2020grip}. In contrast, GNNMark includes GNN models that work across a wide range of graph data, including spatio-temporal graphs and heterogeneous graphs. GNNMark also includes multi-GPU implementations of GNNs,  making it suitable for research on GNN training behavior targeting GPUs.
The applications included in GNNMark can also be used to drive inference studies by first training the models to a target accuracy and then using the pre-trained models to characterize inference. We plan to extend the suite to support inference studies by providing a set of pre-trained models in the future. 

% Dwivedi et al.~\cite{dwivedi2020benchmarking} propose a benchmark suite targeting GNN models for the machine learning community. In contrast to GNNMark, which is designed to drive architectural studies, their benchmark suite is more focused on comparing the accuracy and performance of different popular GNN models in the literature, evaluating the models against a standard collection of graph datasets.

% \section{SpGEMM Accelerators}

% \section{Overview of Existing Architectures}

\section{Prior GNN Accelerators}

The cause for a critical computational bottleneck within GNN workloads is the presence of irregular memory access patterns. Accelerating GNN workloads for large input graphs, especially those with skewed sparsity patterns, is particularly challenging.
Previous accelerator designs have attempted to hide memory bottlenecks by performing high-level and low-level pipelining~\cite{kiningham2022grip}, row-remapping~\cite{geng2020awb}, generating two distinct implementations for the aggregation and combination phases~\cite{yan2020hygcn}, and leveraging flexible network topologies~\cite{li2021gcnax}. 
% Previous accelerator designs have attempted to bypass this memory bottleneck through strategies such as high-level and low-level pipelining, row-remapping, developing separate implementations for the aggregation and combination phases, and flexible network topology, among others.
While these enhancements improve performance for specific GNN workloads on predefined datasets, they fall short in terms of general applicability across a range of GNN workloads~\cite{abadal2021computing}.
% Moreover, these accelerators suffer from inefficient hardware utilization due to workload imbalance when processing new graphs or when executing different GNN workloads~\cite{abadal2021computing}.
Table~\ref{chp3:tab:prior_compare_workloads} illustrates support proposed in prior studies for various GNN workloads. Table~\ref{chp3:tab:prior_compare_optimization} further extends Table~\ref{chp3:tab:prior_compare_workloads} by summarizing the optimization techniques incorporated by prior GNN accelerators.

% \subsection{Existing Accelerators: GPUs}

% \subsection{Novel Custom Accelerators}

\begin{table*}
\centering
\caption{Comparison of prior state-of-the-art GNN accelerators' support for different graph neural network models. The table indicates the support provided by accelerator for GCN, GIN, GAT, SAGE, PNA, and DGN models}
% \vspace*{-2mm}
\label{chp3:tab:prior_compare_workloads}

\begin{tabular}{l | c c c c c c } 
    % \hline
    % \hline
    % \toprule[1pt]
    \arrayrulecolor{black}\toprule
   \textbf{Accelerator} & GCN & GIN & GAT & SAGE & PNA & DGN \\
   \arrayrulecolor{black}\toprule
   \arrayrulecolor{black}\toprule
   % \arrayrulecolor{black!20}\bottomrule
    % \\[-1.5ex]
   I-GCN~\cite{geng2021gcn} & \tick & \tick & \cross & \tick & \cross & \cross \\
   \arrayrulecolor{black!20}\midrule
   AWB-GCN~\cite{geng2020awb} & \tick & \cross & \cross & \cross & \cross & \cross \\
   \arrayrulecolor{black!20}\midrule
   HyGCN~\cite{yan2020hygcn} & \tick & \tick & \cross & \tick & \cross & \cross \\
   \arrayrulecolor{black!20}\midrule
   EnGN~\cite{liang2020engn} & \tick & \cross & \cross & \tick & \cross & \cross \\
   \arrayrulecolor{black!20}\midrule
   GraphPE~\cite{auten2020hardware} & \tick & \cross & \tick & \cross & \cross & \cross \\
   \arrayrulecolor{black!20}\midrule
   GNNerator~\cite{stevens2021gnnerator} & \tick & \cross & \cross & \tick & \cross & \cross \\
   \arrayrulecolor{black!20}\midrule
   GCoD~\cite{you2022gcod} & \tick & \tick & \tick & \tick & \cross & \cross \\
   \arrayrulecolor{black!20}\midrule
   ReGNN~\cite{chen2022regnn} & \tick & \tick & \cross & \tick & \cross & \cross \\
   \arrayrulecolor{black!20}\midrule
   ReFlip~\cite{huang2022accelerating} & \tick & \tick & \tick & \tick & \cross & \cross \\
   \arrayrulecolor{black!20}\midrule
   GROW~\cite{hwang2023grow} & \tick & \tick & \tick & \tick & \cross & \cross \\
   \arrayrulecolor{black!20}\midrule
   FlowGNN~\cite{sarkar2023flowgnn} & \tick & \tick & \tick & \tick & \tick & \tick \\
   \arrayrulecolor{black!20}\midrule
   % \arrayrulecolor{black}\bottomrule
   % \textbf{NeuraChip} & \tick & \tick & \tick & \tick & \tick & \tick & \tick & \tick & \tick & \tick & \tick & \tick \\
    % \hline
   \arrayrulecolor{black}\bottomrule

\end{tabular}

% \scriptsize

% \centering

% \diamd \  \todo{Address diamonds}

% \vspace*{-6mm}

\end{table*}

\begin{table*}
\centering
\caption{Comparison of various Graph Neural Network (GNN) Accelerators on optimization techniques incorporated}
% \vspace*{-2mm}
\label{chp3:tab:prior_compare_optimization}

\begin{tabular}{l | x{0.54in} x{0.7in} x{0.6in} x{0.7in} x{0.7in} } 
    % \hline
    % \hline
    % \toprule[1pt]
    \arrayrulecolor{black}\toprule
   \textbf{Accelerator} & Kernel Fusion & Loop Reordering & Pruning & Bank Mapping & Load Balancing \\
   \arrayrulecolor{black}\toprule
   \arrayrulecolor{black}\toprule
   % \arrayrulecolor{black!20}\bottomrule
    % \\[-1.5ex]
   I-GCN~\cite{geng2021gcn} & \cross & \cross & \tick & \tick & \tick \\
   \arrayrulecolor{black!20}\midrule
   AWB-GCN~\cite{geng2020awb} & \tick & \tick & \cross & \tick & \tick \\
   \arrayrulecolor{black!20}\midrule
   HyGCN~\cite{yan2020hygcn} & \tick & \tick & \cross & \tick & \tick \\
   \arrayrulecolor{black!20}\midrule
   EnGN~\cite{liang2020engn} & \cross & \tick & \cross & \tick & \tick \\
   \arrayrulecolor{black!20}\midrule
   GraphPE~\cite{auten2020hardware} & \cross & \cross & \cross & \tick & \cross \\
   \arrayrulecolor{black!20}\midrule
   GNNerator~\cite{stevens2021gnnerator} & \cross & \cross & \cross & \cross & \cross \\
   \arrayrulecolor{black!20}\midrule
   GCoD~\cite{you2022gcod} & \cross & \tick & \tick & \cross & \tick \\
   \arrayrulecolor{black!20}\midrule
   ReGNN~\cite{chen2022regnn} & \tick & \cross & \cross & \cross & \tick \\
   \arrayrulecolor{black!20}\midrule
   ReFlip~\cite{huang2022accelerating} & \cross & \cross & \cross & \cross & \tick \\
   \arrayrulecolor{black!20}\midrule
   GROW~\cite{hwang2023grow} & \cross & \tick & \cross & \cross & \tick \\
   \arrayrulecolor{black!20}\midrule
   FlowGNN~\cite{sarkar2023flowgnn} & \tick & \cross & \cross & \tick & \tick \\
   \arrayrulecolor{black!20}\midrule
   % \arrayrulecolor{black}\bottomrule
   % \textbf{NeuraChip} & \tick & \tick & \tick & \tick & \tick & \tick & \tick & \tick & \tick & \tick & \tick & \tick \\
    % \hline
   \arrayrulecolor{black}\bottomrule

\end{tabular}

% \scriptsize

% \centering

% \diamd \  \todo{Address diamonds}

% \vspace*{-6mm}

\end{table*}

% \textbf{CGRAs}

\textbf{I-GCN}~\cite{geng2021gcn}: I-GCN is a hardware accelerator designed to improve the performance of GCN inference. One of the primary challenges in accelerating GCNs is the poor data locality and redundant computation arising from the large size, high sparsity, and irregular non-zero distribution of real-world graphs. To tackle these issues, I-GCN employs an online graph restructuring algorithm known as islandization. This algorithm identifies clusters of nodes with strong internal connections, but weak external ones, which improves on-chip data reuse and minimizes off-chip memory accesses.

\textbf{AWB-GCN}~\cite{geng2020awb}: Autotuning-Workload-Balancing GCN (AWB-GCN) is a hardware accelerator specifically designed for speeding up GCN inference. Addressing the challenges of processing large and unbalanced real-world graphs, AWB-GCN employs three hardware-based autotuning techniques: dynamic distribution smoothing, remote switching, and row remapping. These techniques enable the system to dynamically adjust the workload distribution across a large number of processing elements.

\textbf{HyGCN}~\cite{yan2020hygcn}: HyGCN is an accelerator designed to address the unique computational challenges arising from the hybrid execution patterns of GCNs. These patterns comprise a dynamic and irregular aggregation phase, and a static and regular combination phase. The design of HyGCN is motivated by a characterization of GCN execution patterns on an Intel Xeon CPU. The accelerator employs a new programming model to exploit fine-grained parallelism and features two efficient processing engines tailored to handle the irregularity in the aggregation phase and the regularity in the combination phase. These engines are optimized for various levels of parallelism and data reusability.

\textbf{EnGN}~\cite{liang2020engn}: EnGN is an accelerator architecture that addresses the substantial computational and memory overhead present within GNN workloads. EnGN focuses on accelerating the three key stages of GNN propagation by abstracting them as common computing patterns. The architecture employs a ring-edge-reduce (RER) dataflow and corresponding RER PE-array to manage the poor locality associated with sparsely and randomly connected vertices. Additionally, EnGN utilizes a graph tiling strategy to fit large graphs into the accelerator's memory.

\textbf{GraphPE}~\cite{auten2020hardware}: The GraphPE architecture incorporates dedicated hardware units engineered to manage the irregular data movement typical of graph computations while also ensuring high computational throughput for GNN models.

\textbf{GNNerator}~\cite{stevens2021gnnerator}: GNNerator is an accelerator for GNNs, designed to address the computational challenges arising from the dual nature of GNN operations: dense and regular computations for feature extraction, and sparse and irregular computations for message passing between nodes. GNNerator employs heterogeneous compute engines specifically optimized for these contrasting computational patterns. The paper also introduces the concept of feature-blocking, a dataflow technique that adjusts the trade-off between irregular and regular memory accesses, increasing computational efficiency during both feature extraction and aggregation stages.

\textbf{GCoD}~\cite{you2022gcod}: GCoD is a hardware-software co-designed framework aimed at addressing the computational inefficiencies associated with GCNs when applied to large, sparse and irregular real-world graphs. In terms of algorithmic design, GCoD employs a "split and conquer" training strategy that locally polarizes graph densities, resulting in adjacency matrices with enhanced regularity and thus, achieves good acceleration. On the hardware side, a specialized two-pronged accelerator is developed, featuring separate engines to process denser and sparser graph workloads, thereby further improving utilization and acceleration efficiency.

\textbf{ReGNN}~\cite{chen2022regnn}: ReGNN is a GNN accelerator aimed at eliminating computational and communication redundancy inherent in traditional GNNs. ReGNN is built on a hardware-software co-design approach incorporating a dynamic redundancy-eliminated neighborhood message-passing algorithm. ReGNN is a configurable, pipelined architecture adaptable to various GNN variants without compromising accuracy.

\textbf{ReFlip}~\cite{huang2022accelerating}: ReFlip is a GCN accelerator that aims to improve the overall efficiency of both regular neural network computations and irregular graph analytics. ReFlip employs a unified architecture based on Processing-in-Memory (PIM)~\cite{jonatan2024scalability} and features a crossbar structure. This unified architecture is augmented by novel algorithm mappings that maximize performance by leveraging the inherent parallelism of crossbar structures.

\textbf{GROW}~\cite{hwang2023grow}: GROW is an accelerator for Graph Convolutional Neural Networks (GCNs), designed to optimize the two primary stages of GCNs—aggregation and combination—that have distinct dataflow. GROW utilizes Gustavson's algorithm to implement a row-wise product-based sparse-dense GEMM accelerator. By co-designing software and hardware, GROW claims to achieve a balance between data locality and parallelism.

\textbf{FlowGNN}~\cite{sarkar2023flowgnn}: FlowGNN introduces a dataflow architecture tailored for the acceleration of GNNs that utilize message-passing mechanisms. The FlowGNN architecture is scalable and supports a broad spectrum of GNN models, featuring a configurable dataflow that simultaneously computes node and edge embeddings as well as facilitates message passing, making it universally applicable across different models. A significant advantage of FlowGNN is its ability to perform GNN inference without any prior graph processing.

\textbf{LISA}~\cite{li2022lisa}: LISA is a compiler-oriented approach to map computations of GNNs on Coarse-Grained Reconfigurable Arrays (CGRA) spatial accelerators. CGRAs, known for their potential to enhance computational performance and energy efficiency, require sophisticated compiler designs to unlock their full capabilities. LISA introduces a solution by leveraging GNNs to analyze and interpret the structural characteristics of dataflow graphs (DFGs), which represent application-specific computations. This analysis facilitates the automatic identification of near optimal mappings for DFGs onto new accelerator architectures, considering both node placement and dependency routing. The integration of a simulated annealing-based mapping strategy, informed by GNN-generated insights, ensures that the mapping process is both efficient and effective. LISA dramatically reduces the time required to generate high-quality mappings for spatial accelerators, thereby accelerating the development cycle and enhancing the performance of computing systems.

% \section{Bottlenecks and Limitations}
% \section{Opportunities for Optimization}

%   ____ _                          _            _          _   _
%  / ___| |__   __ _ _ __ __ _  ___| |_ ___ _ __(_)______ _| |_(_) ___  _ __
% | |   | '_ \ / _` | '__/ _` |/ __| __/ _ \ '__| |_  / _` | __| |/ _ \| '_ \
% | |___| | | | (_| | | | (_| | (__| ||  __/ |  | |/ / (_| | |_| | (_) | | | |
%  \____|_| |_|\__,_|_|  \__,_|\___|\__\___|_|  |_/___\__,_|\__|_|\___/|_| |_|

\chapter{GNN Workload Characterization}
\label{chp4}
% \section{Irregular Memory Access Patterns}
% \section{GNNMark: Architectural Impact of GNNs on GPU}

\reason{Introduction}

Owing to the energy efficiency and high-performance capabilities of GPUs, GPUs are a natural choice for accelerating the training of GNNs. This forms the core motivation to understand the architectural and system-level implications of training GNNs on GPUs. Previously to our work, no benchmark suite existed to examine the architectural implications of GNN training workloads.

\reason{An overview of what is GNNMark}

In this \thesis, we address this need by presenting GNNMark~\cite{baruah2021gnnmark}, a feature-rich benchmark suite that encompasses the diversity present in GNN training workloads, datasets and GNN frameworks. Our benchmark suite consists of GNN workloads that utilize various graph-based data structures, including homogeneous graphs, dynamic graphs, and heterogeneous graphs commonly used in a number of application domains that we mentioned in Section~\ref{chp2:sec:ml_on_graphs}. We use this benchmark suite to explore and characterize GNN training behavior on GPUs. We study a variety of aspects of GNN execution, including both compute and memory behavior, highlighting major bottlenecks observed during GNN training. At the system level, we evaluate multiple metric, including the scalability of training GNNs across a multi-GPU system, as well as the sparsity of data encountered during training. The insights derived from our work can be leveraged by both hardware and software developers to improve both the hardware and software performance of GNN training on GPUs.
The contributions of this part of the thesis include:

\begin{enumerate}

    \item \textbf{GNN training-focused benchmark suite:} We deliver an open-source benchmark suite named GNNMark~(\url{https://gitlab.com/GNNMark/gnnmark}), designed to characterize the training behavior of GNNs on GPUs. Our suite includes a diverse set of popular GNN models that the machine learning community has developed. The workloads span seven different application domains and three different types of graph-based data types.

    \item \textbf{Architecture-level characterization of GNNMark:} We characterize the workloads in GNNMark, considering their architectural implications during the training process on a GPU. We are the first to provide a detailed execution time breakdown of different operations executed during GNN training and identify the major bottlenecks.  We find that these workloads are much more diverse than typical DNN training workloads. GNN execution is highly input data and model dependent. We find that integer operations play a critical role, a factor that has been relatively ignored in DNN training studies on GPUs. We also observe significant sparsity during GNN training. This can potentially be leveraged to train larger graphs on a single GPU. We also consider multi-GPU support in the suite, enabling scaling studies of GNNs across multi-GPU systems.

    \item \textbf{Recommendations to improve GPU architectures:} We present insights drawn from our detailed characterization and suggest changes to improve GPU architectures and system design so that GNNs can be trained efficiently.

\end{enumerate}

\section{Motivation for Characterizing GNN Workloads}

% \reason{DNNs are great, but GNNs are better for non-Euclidean data}

Deep Neural Networks (DNNs) have revolutionized numerous areas, such as image classification~\cite{druzhkov2016survey, shivdikar2015automatic}, speech recognition~\cite{padmanabhan2015machine, ding2022audio} and autonomous systems~\cite{mao2021one, mnih2013playing}. Notable DNN architectures such as Convolutional Neural Networks (CNNs)~\cite{fukushima1980neocognitron} and Transformers~\cite{vaswani2017attention} primarily operate on Euclidean data. This type of data, inherently 1D or 2D, includes images and speech datasets~\cite{jayaweera2021jaxed}. Yet, much of the data we encounter in the real world is non-Euclidean in nature~\cite{bronstein2017geometric}, encompassing structures of molecules, social networks, sensor systems, and manifolds. Traditional DNNs, designed for Euclidean data, often fall short in efficiently processing non-Euclidean data due to challenges in directly applying operations, such as convolutions~\cite{bronstein2017geometric, shivdikarspeeding}.

% \reason{Example applications of GNNs in real life}

To bridge this deficiency, GNNs~\cite{kipf2016semi,wu2020comprehensive} have been developed, specializing in non-Euclidean data training. For instance, Pinterest employs a GNN model, PinSAGE~\cite{ying2018graph}, for its recommendation algorithms, while Twitter researchers utilize GNN models with temporal graph data~\cite{rossi2020temporal}. Similarly, the Drug Repurposing Knowledge Graph (DRKG)~\cite{ioannidis2020drkg} adopts GNN models to research the applications of existing drugs for novel diseases.

% \reason{We need a study of the architectural impact of GNNs on GPUs}

With GPUs establishing themselves as the go-to platform for DNN and GNN training due to their advanced capabilities, many leading GNN frameworks, for example, the Deep Graph Library~\cite{zheng2020learning} and PyTorch Geometric~\cite{fey2019fast}, have integrated GPU training support. As GNNs continue to surge in popularity, there's a pressing need for optimizing GPU platforms to train them. Thoroughly analyzing GPU behavior during GNN training is paramount. By dissecting the myriad of GNN operations and their execution during training, GPU architects can pinpoint and address performance bottlenecks. Aspects such as GNN training scalability on multi-GPU setups and the presence of data sparsity during training can be tapped into for training large graphs, especially those surpassing a single GPU's memory capacity~\cite{rhu2018compressing, shivdikar2022accelerating, livesay2023accelerating}. A thorough analysis of GNN training workflows will enhance our understanding of the computational and memory constraints associated with running GNN workloads on GPUs.

% In this work, we address this need by presenting GNNMark, a feature-rich benchmark suite that covers the diversity present in GNN training workloads, datasets, and GNN frameworks. Our benchmark suite consists of GNN workloads that utilize a variety of different graph-based data structures, including homogeneous graphs, dynamic graphs, and heterogeneous graphs commonly used in a number of application domains that we mentioned above. We use this benchmark suite to explore and characterize GNN training behavior on GPUs. We study a variety of aspects of GNN execution, including both compute and memory behavior, highlighting major bottlenecks observed during GNN training. At the system level, we study various aspects, including the scalability of training GNNs across a multi-GPU system, as well as the sparsity of data, encountered during training. The insights derived from our work can be leveraged by both hardware and software developers to improve both the hardware and software performance of GNN training on GPUs.

\section{Prior Work on GNN Characterization}

Prior work on characterizing GNNs has focused primarily on the inference behavior for \acrshort{GCN}s~\cite{yan2020characterizing} or targeted a limited set of \acrshort{GNN} models~\cite{zhang2020architectural}. Both model and dataset diversity~\cite{zhang2020architectural} have not been considered by these studies. By dataset diversity, we mean different types of graphs, including homogeneous, heterogeneous, knowledge, and dynamic graphs~(explored in detail in Section~\ref{chp4:sec:graph_types}). Model diversity implies different types of GNN models, such as Graph Transformers, Spatio-Temporal GNNs, and LSTM based GNNs. In previous benchmarking efforts, GNN inference has been the primary target for characterization studies~\cite{yan2020characterizing, zhang2020architectural}.
Popular benchmark suites for DNN training, such as the MLPerf Training Suite~\cite{mattson2019mlperf}, Training Benchmarks for DNN~(TBD)~\cite{zhu2018benchmarking}, DNNMark~\cite{dong2017dnnmark}, Fathom~\cite{adolf2016fathom}, and DawnBench~\cite{coleman2017dawnbench}, do not consider GNNs as part of their workloads and deal exclusively with DNNs that deal with euclidean data. To comprehensively characterize the execution behavior of GNN training on GPUs, we need a benchmark suite that includes diverse GNN models that are trained on diverse datasets. Currently, no such benchmark suite exists. To fill this gap, we develop \textbf{GNNMark}, a collection of representative workloads that can be used by the computer architecture community to study the execution of GNNs on GPUs. We then analyze the workloads in the GNNMark benchmark suite, specifically focusing on their behavior during GNN training on a GPU. Apart from the fact that prior GNN workload characterization studies primarily focused on inference, we chose to focus on training, given that GPUs remain the best platform in terms of performance for GNN training.
% In contrast, customized accelerators have been shown to outperform GPUs when executing GNN inference~\cite{yan2020hygcn}

\section{Input Graph Types}
\label{chp4:sec:graph_types}
% GNNs have been explored in many different fields. Each type of GNN has inherently different types of graph data~\cite{wu2020comprehensive}. In our analysis of GNNs, we observe that there are three such major categories of graph data:
GNNs have evolved over time, and we find that each variation of GNN is typically associated with a distinct form of graph data~\cite{wu2020comprehensive}. In our examination of GNNs, we identify three primary classifications of graph data:

\begin{enumerate}

    \item \textbf{Homogeneous Graphs:} A homogeneous graph contains nodes and edges of a single type. For example, social network graphs are typically homogeneous, where each node represents a user, and an edge can represent if that one user follows another. Homogeneous graphs can be directed~(e.g., following a user on Twitter) or undirected~(e.g., adding a friend on Facebook). Another notable collection of homogeneous graph datasets that are used to evaluate GNN models are citation datasets~(e.g., Cora, PubMed, Citeseer)~\cite{kipf2016semi}.

    \item \textbf{Heterogeneous Graphs:} A heterogeneous graph contains nodes and edges of multiple types. A widely used form of a heterogeneous graph is found in recommendation generation scenarios. For example, in a dataset designed to recommend music to users, the graph will consist of two types of nodes: i) music nodes and ii) user nodes. The edges will correspond to different interactions between the user and a music piece. In addition, edges may contain additional information such as ratings or like/dislike attributes. Knowledge graphs that are used to model relations between an object and entities are another form of a heterogeneous graph -- e.g. when users search for a famous celebrity on Google~(an object). 

    \item \textbf{Dynamic Graphs:} A dynamic graph is a special type of graph where the graph itself, as well as its properties, can evolve over time. Many real-world graphs, such as social-network graphs~\cite{rossi2020temporal}, traffic data graph~\cite{yu2018spatio} and communication-network graphs, are dynamic~\cite{eppstein1999dynamic}. Note that dynamic graphs can be either homogeneous or heterogeneous. For example, if we take a homogeneous social network graph, where nodes represent people and edges represent whether there is a relationship, the number of relations a person has or the relations between two people can change over time. Another common use case of dynamic graphs is to model traffic data as a dynamic graph and use it for traffic forecasting and prediction~\cite{diao2019dynamic}.

\end{enumerate}

\section{Benchmark Suite Design}

% \reason{Requirements of Benchmark suite}

Characterizing the behavior of GNN training on a GPU requires a set of representative workloads to cover the wide variety of GNNs~\cite{wu2020comprehensive,zhang2020deep}. The variants should include GNNs used across multiple application domains, including recommendation systems, classification of molecules, traffic forecasting, etc. The representative suite should also include models that consider different classes of real-world graphs, including knowledge graphs, heterogeneous graphs, and dynamic graphs. In addition, multi-GPU GNN training should be supported to evaluate the efficacy of training GNNs on multi-GPU systems.

% \reason{Added support for PyTorch Geometric and Deep Graph Library}

To satisfy all the above-mentioned criteria, we offer GNNMark, a benchmark suite designed for studying the behavior of GNN training on GPUs. Similar to benchmark suites that target DNN training, such as TBD~\cite{zhu2018benchmarking} and MLPerf~\cite{mattson2019mlperf}, we curate our benchmark suite from open-source publicly available implementations of GNN models. As PyTorch Geometric~(PyG) and Deep Graph Library~(DGL) are the main frameworks employed for developing GNN models by the ML community, we use models developed using these frameworks. Since both of these frameworks support PyTorch, we have chosen models developed in PyTorch. The specific models chosen for this suite, along with their associated application domains and datasets, are summarized in \tabref{tab:benchmarks}. Below, we provide more details about each GNN model.

\begin{table*}[t]
\centering
\begin{tabularx}{\textwidth}{@{}l>{\RaggedRight}p{4.5cm}>{\RaggedRight}p{4.5cm}>{\RaggedRight}p{4.5cm}@{}}
  \toprule
  \thead{Abbv} & \thead{GNN Model} & \thead{Application Domain} & \thead{Graph Input Type} \\
  \arrayrulecolor{black!20}\midrule
  
  \textbf{PSAGE} & PinSAGE & Recommendation & Heterogeneous Graph \\
  \arrayrulecolor{black!20}\midrule
  
  \textbf{STGCN} & Spatio Temporal GCN & Traffic Forecasting & Dynamic Graph \\
  \arrayrulecolor{black!20}\midrule
  
  \textbf{DGCN} & Deep GCN & Molecular Property Prediction & Homogeneous Graph \\
  \arrayrulecolor{black!20}\midrule
  
  \textbf{GW} & GraphWriter & Text Generation & Heterogeneous Graph \\
  \arrayrulecolor{black!20}\midrule
  
  \textbf{KGNN} & k Graph Neural Networks & Protein Classification & Homogeneous Graph \\
  \arrayrulecolor{black!20}\midrule
  
  \textbf{ARGA} & Adverserially Regularized Graph Autoencoder & Node Clustering & Homogeneous Graph \\
  \arrayrulecolor{black!20}\midrule
  
  \textbf{TLSTM} & Tree Long Short-Term Memory Networks & Sentiment Classification & Homogeneous Graph \\

  \arrayrulecolor{black}\bottomrule
\end{tabularx}
\caption{Workloads in GNNMark Benchmark Suite.}
\label{tab:benchmarks}
\vspace{-0.5cm}
\end{table*}

\begin{table*}[htbp]
\centering
\begin{tabularx}{\textwidth}{@{}l>{\RaggedRight}p{9.6cm}rr@{}}
  \toprule
  \thead{Abbv} & \thead{Datasets} & \thead{\# Node} & \thead{\# Edge} \\
  \arrayrulecolor{black!20}\midrule
  
  \multirow{2}{*}{\textbf{PSAGE}} &
      Nowplaying~(NWP) \cite{zangerle2014nowplaying} & 22.9M & 1.9M \\
      
      \arrayrulecolor{black!10}\cmidrule{2-4}
      &
      Movielens~(MVL)~\cite{harper2015movielens} & 1.9M & 9.7K \\
     
  \arrayrulecolor{black!20}\midrule
  
  \multirow{2}{*}{\textbf{STGCN}} &
      LA \cite{yu2018spatio} & 207 & 325 \\
      
      \arrayrulecolor{black!10}\cmidrule{2-4}
      &
      PEMS\_Bay~(PEMS) \cite{yu2018spatio} & 1722 & 2694\\
      
  \arrayrulecolor{black!20}\midrule
  
  \multirow{2}{*}{\textbf{DGCN}} &
      MOLHIV \cite{hu2020open} & 1.04M & 1.1M \\
      \arrayrulecolor{black!10}\cmidrule{2-4}
      &
      MOLTOX \cite{hu2020open} & 145K & 151K \\
      
  \arrayrulecolor{black!20}\midrule
  
  \textbf{GW} &
      AGENDA ~\cite{koncel2019text} & 885K & 2.57M \\
      
  \arrayrulecolor{black!20}\midrule
  
  \textbf{KGNN} &
      Proteins~(PROT)~\cite{KKMMN2016} & 43K & 162K \\
      
  \arrayrulecolor{black!20}\midrule
  
  \multirow{3}{*}{\textbf{ARGA}} &
      Cora \cite{yang2016revisiting} & 2K & 10.5K \\
      \arrayrulecolor{black!10}\cmidrule{2-4}
      &
      CiteSeer~(CSEER) \cite{yang2016revisiting} & 3.3K & 9.2K\\
      \arrayrulecolor{black!10}\cmidrule{2-4}
      &
      PubMed~(CSEER) \cite{yang2016revisiting} & 19.7K & 88.6K\\
      
  \arrayrulecolor{black!20}\midrule
  
  \textbf{TLSTM} &
      Stanford Sentiment Treebank~(SNTM)~\cite{socher2013recursive} & 318K & 310K \\
      
  \arrayrulecolor{black}\bottomrule
\end{tabularx}
\caption{Workloads in GNNMark Benchmark Suite.}
\label{tab:benchmarks_values}
\vspace{-0.5cm}
\end{table*}

% \reason{Models incorporated into GNNMark}

\textbf{PinSAGE:} GNNs that operate on heterogeneous knowledge graphs can be used for recommendation tasks. These are commonly used in social networks. PinSAGE~\cite{ying2018graph} is one such GNN model that has been developed at Pinterest. Since the original PinSAGE model is not publicly available, we use the implementation that has been published by the developers of DGL. PinSAGE is an improvement upon the GraphSAGE model~\cite{hamilton2017inductive} for training on large graphs. It uses a {\em random walk} mechanism~\cite{wei2004towards} during aggregation to identify the importance of a node in the graph without the need to process the entire graph. This effectively allows a user to train a model on graphs that do not fit in GPU memory. 

\textbf{Spatio-Temporal Graph Convolutional Network:} Traffic forecasting is an important problem that falls into the domain of time-series prediction and uses dynamic graphs. This task is highly relevant for use in urban areas where traffic control and guidance are required. Solving this problem using conventional Euclidean-based DNNs is challenging because of the nonlinearity involved in traffic data~\cite{nair2001non}. One approach to deal with nonlinearity is to represent the problem as a graph and then apply depth-wise convolutions on the graph. Spatio-Temporal graph Convolutional Networks~(STGCN)~\cite{yu2018spatio} represent one such model that has been proposed to solve the problem of traffic forecasting. We include an STGCN to represent a GNN model that deals with dynamic graphs.

\textbf{DeepGCNs:} One of the key challenges with the original GCN models, such as the one proposed by Kipf and Welling~\cite{kipf2017semi}, is that increasing the depth of the model does not improve the accuracy of the model. This is due to the vanishing gradient problem~\cite{hochreiter1998vanishing}, which has made implementing deep GCNs challenging. Therefore, researchers have developed mechanisms to train deeper GCNs~\cite{li2019deepgcns}, using ideas borrowed from DNN research, such as residual layers and skip-connections used in models such as ResNet~\cite{he2016deep}. DeepGCN is a novel GCN architecture that allows GCNs to have more layers.  Additional layers in a GCN can significantly improve training accuracy~\cite{li2019deepgcns}, so we include it in our study.  Specifically, we use a DeepGCN model and train it to perform graph property prediction, a common task in molecular property prediction.

\textbf{GraphWriter:} Automated generation of text from a knowledge graph to form meaningful and coherent sentences is an open and challenging problem~\cite{ji2020survey}. Text encoding models, such as the popular Transformer model~\cite{vaswani2017attention}, cannot be directly applied to a knowledge graph as they do not work with non-euclidean data. Therefore, ML researchers have developed GNN-based Transformer models for this task. Graphwriter~\cite{koncel2019text} is one such novel GNN-based Transformer model designed to operate on knowledge-graphs for text generation.

\textbf{k-GNNs:} Most GNN models are one-dimensional in nature and cannot effectively capture any higher-order information, such as the properties of subgraphs, within the graph. As a result, they fail the graph isomorphism test proposed by Weisfeiler and Lehman~\cite{weisfeiler1968reduction}~(WL algorithm). The WL algorithm is a test used to determine the expressiveness power of a GNN by testing if an algorithm is able to distinguish whether two graphs are isomorphic or not. Two graphs are said to be isomorphic if they have the same number of vertices, edges, and connectivity. Therefore, researchers have developed higher-dimensional hierarchical GNNs, called k-GNNs~(where the k stands for the dimension), which can capture properties of subgraphs~\cite{morris2019weisfeiler}. This enables GNNs to perform close to the k-WL graph isomorphism test~\cite{morris2019weisfeiler}. We include two variants of k-GNNs, (KGNNL and KGNNH to denote a lower and higher dimensional version of k-GNN, respectively) and use them to perform classification of protein molecules. The primary reason we include this workload in our suite is to study how application characteristics and behavior change as we move towards higher-dimension GNNs.

\textbf{Adversarially Regularized Graph Autoencoder:} Generative Adversarial Networks~(GANs) are gaining popularity due to their ability to learn with limited amounts of data~\cite{pan2019recent}. Due to this property, GAN-based architectures are also being explored for GNNs. An Adversarially Regularized Graph Autoencode~(ARGA)~\cite{panadversarially} is one such GNN-based GAN model that is proposed for graph embedding. ARGA has an encoder-decoder architecture where the encoder is trained to form a compact representation of a graph, and the decoder is trained to generate the graph structure. The model is designed to perform node clustering, which is an unsupervised learning task, on real-world graphs. ARGA employs this encoder-decoder architecture within a GAN framework, so that it can successfully learn the low-dimensional features of the graph from the high-dimensional graph features. This process is referred to as graph embedding~\cite{cai2018comprehensive}. We include ARGA as a representative GAN-based GCN to further increase the diversity of our benchmark suite. We train ARGA to perform node clustering on real-world homogeneous graphs, such as Cora, PubMed, and CiteSeer~\cite{kipf2017semi}.

\textbf{Tree-LSTM:} Sentiment classification is an important task in the Natural Language Processing~(NLP) domain. Tree Long Short-Term Memory Networks~(Tree-LSTMs)~\cite{tai2015improved} are one group of models developed for this task. In contrast to the linear model used in an LSTM, Tree-LSTMs use a tree-structured network topology and can outperform linear LSTMs in the sentiment classification task~\cite{tai2015improved, intrator2017missing}. The Tree-LSTM method implemented in DGL uses the idea of batching. The basic idea of batching is to collect smaller graphs that are part of the dataset and convert them into a batched larger graph. We include the Tree-LSTM model in GNNMark to study how batching multiple small graphs to a larger graph impacts the behavior of an application.

\section{Profiling Methodology}

\subsection{Experimental Platform}

% \reason{Hardware used}

To demonstrate the utility of GNNMark, we use an NVIDIA V100~\cite{nvidia2017v100}, a commonly used GPU for running neural network training. V100 is part of the NVIDIA Volta family of GPUs. Our test system is equipped with an Intel(R) Xeon(R) CPU E5-2630 CPU that operates at a frequency of 2.4GHz. The GPU has 80 Streaming Multiprocessors~(SMs) and is rated to deliver 14 TFLOPS  of single-precision performance. The GPU  memory uses HBM2 with 16 GB capacity and bandwidth of 900 GB/s. The combined L1 cache/shared memory/texture cache has a capacity of 128 KB and is private to each Streaming Multiprocessor (SM). The L1 memory is backed by a 6.14 MB L2 cache, which is banked and shared across all SMs. 

% \reason{Multi-GPU setup}

For our multi-GPU experiments, we use 4 V100 GPUs on a node equipped with Intel(R) Xeon(R)  E5-2686 v4 2.4GHz CPUs, hosted on Amazon AWS EC2. Each GPU is interconnected using NVIDIA NVLink technology, providing a total of six links, for an aggregate bandwidth of 300 GB/s. Both the single-GPU and multi-GPU systems used in our experiments run CUDA 10.2, cuDNN 7.6.5, and PyTorch 1.5.0. The workloads included in GNNMark use either DGL version 0.5.2 or PyTorch Geometric 1.6.1.

Since multi-GPU training has been shown to improve the performance of DNN training~\cite{mojumder2018profiling}, we also look at how well GNN training can scale across multiple GPUs. GNN training can be sometimes be limited by GPU memory capacity, especially given the continual growth in the size of the input graph~\cite{jia2020improving}. One approach to counter this problem is to compress the data transferred from the CPU to the GPU and store the compressed data in GPU main memory. This is only possible if the data transferred is highly sparse~\cite{rhu2018compressing}. Therefore, we also characterize sparsity levels of the data transferred between the CPU and GPU during GNN training in our suite.

\subsection{Profiling Tools}
We use several tools for collecting the metrics of interest. For the kernel-level characteristics, such as cache statistics and comparisons between compute versus memory behavior, we use the NVIDIA nvprof profiler~(version 10.2)~\cite{bradley2012gpu}. Similar to DNNs, GNNs typically launch the same kernel many times during training. Therefore, when profiling and collecting hardware performance counters using nvprof, we profile the same kernel for either fifty kernel invocations or for one epoch, whichever is shorter. However, nvprof does not provide any mechanism to collect the memory divergence behavior of a workload. Therefore, we use the NVBit framework~\cite{villa2019nvbit}~(version 1.4), which is a binary instrumentation tool to collect the memory divergence information at a kernel level. To collect the sparsity of the data transferred from the host to the device during GNN training, we modified the PyTorch source code to collect this information.

\subsection{Metrics of Interest}
Characterizing the behavior of GNNs requires an understanding at both the architectural level and the system level. In this work, we profile and collect the following metrics:

\begin{enumerate}

    \item \textbf{Ratio of time spent in different operations:} Prior work on classifying the phases of GNN execution have categorized GNN execution into two phases: i) aggregation and ii) update phases~\cite{yan2020characterizing}. While classifying into these two phases is beneficial for machine learning purposes, we believe that architectural studies can be guided by a lower level of abstraction (i.e., operations), which has been proposed by Adolf et al.~\cite{adolf2016fathom}. In our profiling experiments for GNN, we observe commonly used operations across various GNN workloads, such as sparse matrix GEMM operations~(spgemm), scatter and gather operations, reduce operations, embedding operations, index selection operations, sorting operations, and element-wise operations. These operations may be embedded within one or multiple kernels within GNN training workloads. Understanding the time spent in these different operations and how they vary across different datasets for the same model can shed insight into where the majority of the execution time is actually spent during GNN training on a GPU.

    \item \textbf{FLOPS and Arithmetic Intensity Analysis:}  To understand how well the GPU can handle GNN training, it is important to analyze the arithmetic intensity and count the number of floating point operations (FLOPS). Arithmetic intensity~(AI) is defined as the ratio of the total number of floating point operations performed to the ratio of data transferred~(in bytes). AI can be used to gauge the data reuse of an algorithm. A higher AI is better since it implies more computations are performed for every byte of data. Analyzing the FLOPS vs AI shows whether a workload is mainly compute or memory bound.
    % , which is essential to understand where GNN training bottlenecks.
    % A similar analysis for different operations that arise in GNN training is helpful to pinpoint operation is mainly compute vs memory bound.

    \item \textbf{Stall Analysis:} To improve the performance of GNNs on GPUs, GPU application developers need to have an understanding of major stalls incurred during GNN training of different GNN workloads. Such an understanding of stalls at an operation level can be helpful to understand the performance of each aforementioned operation.

    \item \textbf{Cache behavior:} While not as important as they are for CPUs, caching can still benefit GPU applications with high spatial and temporal locality. Therefore, having a basic understanding of the hit rates of different levels of the cache is important. Another key characteristic relevant to caches is memory divergence. Memory divergence demonstrates the scattered memory access pattern of a given operation. The memory divergence of a single transaction is calculated by the number of unique cache lines that are touched by a warp. For example, if each of the 32 threads in a warp accesses a different cache line, then the divergence is 32. A scattered memory access pattern i.e., where threads in a warp end up accessing different cache lines, is detrimental on GPUs as the memory transactions cannot be coalesced. This, in turn, leads to serialization and can hurt performance. It is well known that memory divergence hurts the performance of typical graph workloads such as Breadth First Search and PageRank~\cite{che2013pannotia}. Therefore, it is essential to understand the level of memory divergence within a broader class of GNN workloads.

    \item \textbf{Sparsity during GNN training:} Sparsity of the data that is transferred between the CPU and the GPU during training can be leveraged to use optimizations such as DMA compression as proposed by Rhu et al.~\cite{rhu2018compressing}. Therefore, in this work, we also aim to understand the sparsity and compressibility during GNN training.

    % \item \textbf{Scaling with Multiple GPUs:} Multiple-GPUs can also be deployed to train GNNs in a large environment. This can theoretically provide performance benefits if the communication between GPUs is kept low such that each GPU can work in its own piece of data. Understanding of how different GNN models scale across multiple GPUs is therefore important to understand the scalability of these workloads as well as the major bottlenecks that can arise in scaling GNNs 

\end{enumerate}

\subsection{Multi-GPU Implementations} 
We also include multi-GPU versions of each workload in GNNMark to enable users to study the scalability of GNN training on multi-GPU systems.
% , as well as understand how well high-level support for multi-GPU DNN training in frameworks such as PyTorch scale in practice for GNNs.
The multi-GPU implementations are built on the PyTorch \texttt{Distributed Data Parallel}~(DDP) method to train GNNs across multiple GPUs, exploiting data-level parallelism. In practice, DDP has been shown to scale well on up to 256 GPU nodes~\cite{li2020pytorch} for DNN training.

\begin{figure*}[!t] \centering
\includegraphics[width=\linewidth]{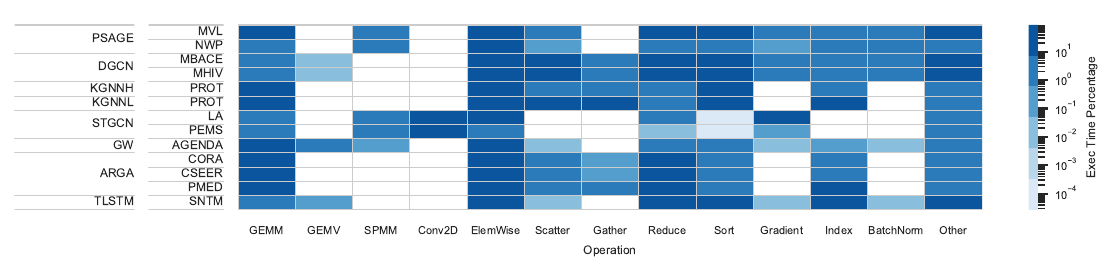} 
\caption{Execution breakdown, reported as the percent of total execution time, for individual operations across the different workloads of GNNMark.\label{fig:time_split}} 
\end{figure*}

\section{Benchmarking Results}

\subsection{Execution Time Breakdown}
We start our analysis by breaking down the time spent in the different GNN operations across the different workloads in our suite. Similar to DNNs~\cite{adolf2016fathom}, GNN training can be broken down into layers or operations. Prior work divided GNN training into two phases: i) an aggregation phase, and ii) an update phase~\cite{yan2020characterizing}. While this division is appropriate when looking at GNNs from an machine learning perspective, we believe that deeper insights are needed to fully characterize their behavior. Therefore, we work at the abstraction level of individual operations~\cite{adolf2016fathom}.

We identify a common set of operations performed during GNNMark execution. These operations include general matrix multiply~(GEMM), sparse matrix-matrix multiplication~(SpMM), convolutions, scatters, gathers, reductions, index selection, sorting, and element-wise operations. Element-wise operations are operations that operate on individual elements of a tensor and perform operations such as multiplication of all elements in the tensor by a scalar, changing the sign of all elements in the tensor, or adding two tensors of similar dimensions.

\figref{fig:time_split} shows a breakdown of the percentage of time spent in individual operations across the different workloads of GNNMark. \figref{fig:time_split} illustrates the percentage breakdown of operations varies significantly across workloads. For instance, STGCN, a spatio-temporal GNN, is dominated by 2D convolution operations~(60\% on average), while DGCN is dominated by element-wise operations~(31\% on average). 

The execution time breakdown across operations in a GNN differs greatly from the mix in a typical DNN. Across all the workloads, we observe that only 25\% of the execution time is spent executing GEMM and SpMM operations. This is in stark contrast to the mix of operations commonly found in DNN workloads, where GEMM~(convolutional layers and fully-connected layers) dominate the execution~\cite{dnnmark}. We find that GNN training also differs significantly from GNN inference workloads~\cite{yan2020characterizing}, where GEMM operations are reported to consume more than 50\% of the execution time. 

Other common operations, such as sorting, index selection, reductions, and scatter-gather operations, account for 20.8\% of the total execution on average. These operations are primarily used in the graph's aggregation phase, where the nodes exchange information with one another before updating the feature vectors. 

PSAGE, when trained on the MVL dataset, spends 20.7\% of its execution on sorting and only 7.0\% of the time on reductions, whereas ARGA~(using the Cora data) spends 23\% on reductions and only 6.1\% on sorting. This great diversity and variety of tasks in GNN training present challenges to architects designing customized accelerators for GNN training, given that accelerators are typically designed to optimize only for a single set of operations.

% The dimension of the GNN plays a key role in the execution time distribution. When comparing high and low dimensional GNNs using kGNNs, while training on the (\texttt{PROT}) dataset, we observe that the high-dimensional GNNs spend more execution time in the scatter and gather phase (16\% on average) as compared to the low-dimensional models (4\% on average). This is because high-dimensional GNNs are hierarchical in nature, meaning that they need to collect and pool features across the subgraphs of the graph, leading to an increase in scatter/gather operations. 

%TODO: Maybe another app here is better if we comment out the last two para
In contrast to typical DNN workloads, GNN workloads tend to be more input data-dependent. For PSAGE, the percentage of time spent in element-wise operations is much higher when training on the (\texttt{NWP}) dataset~(78\%), versus training on the (\texttt{MVL}) dataset~(36\%). This is because, when training on the NWP dataset, the feature vectors are 10$\times$ larger than when training on the MVL dataset. As element-wise operations operate on each value of the input feature vector, the time spent executing these operations becomes more dominant when graphs with larger input features are used. 

\noindent\rule{\columnwidth}{0.4pt}
\textbf{GNN Execution Characteristics}: Our performance analysis shows that GNN training workloads exhibit more diverse behavior as compared to DNN training workloads. Each model's characteristics can differ vastly from others. Even the same GNN model can exhibit different characteristics depending on the input graph type. In addition, execution hot spots are no longer limited to convolution and GEMM operations. We find operations such as reductions, scatter, gather and sorting also need to be optimized. The solution of attaching a single-purpose accelerator to primarily accelerate GEMM operations~\cite{qin2020sigma} during DNN training may not work well for GNN training.\newline
\noindent\rule{\columnwidth}{0.4pt}

\subsection{Instruction Mix and GFLOPS/GIOPS Analysis}
\begin{figure}[!t] \centering
\includegraphics[width=\linewidth]{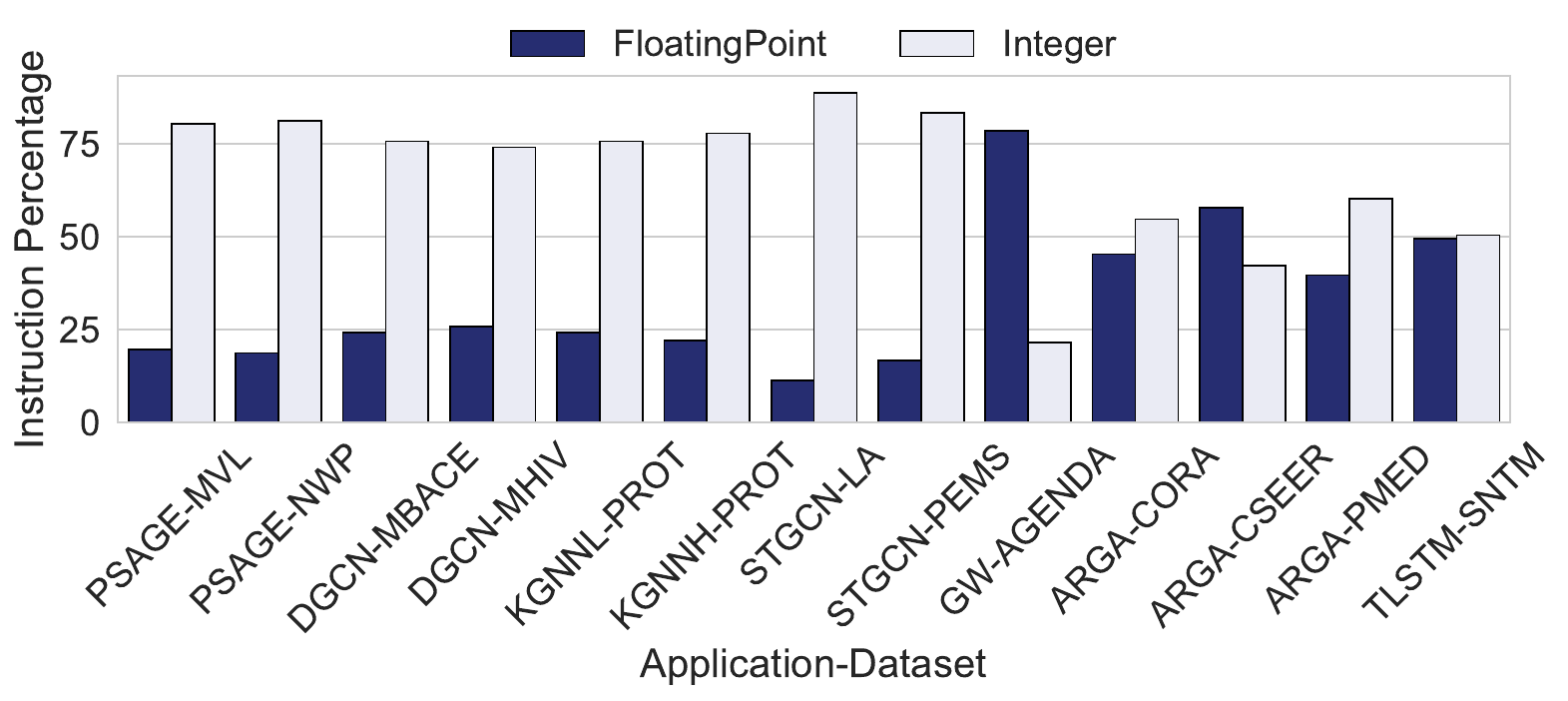} \caption{Breakdown of fp32 vs. int32 instructions across the different workloads in GNNMark.\label{fig:inst_breakdown}} 
\end{figure}

Another aspect of GNN training behavior is the dynamic instruction mix present in different workloads. As shown in \figref{fig:inst_breakdown}, integer instructions play a larger role than floating-point instructions across all workloads. On average, 64\% of the executed instructions are integer~(int32) instructions, whereas only 28.7\% are single-precision floating point~(fp32) instructions. The only workload where this trend is reversed is in GraphWriter~(\texttt{GW}). This is because, in \texttt{GW}, a majority of the time is spent on GEMM and SpMM operations~(as seen in \figref{fig:time_split}), which work on fp32 data. While improving the performance of fp32 instructions has received much attention, int32 instructions have not received the same. Given that int32 instructions dominate GNN execution during training, improving the performance of integer math on a GPU is a critical factor when trying to accelerate GNN training.

\begin{figure}[!t] \centering

\includegraphics[width=\linewidth]{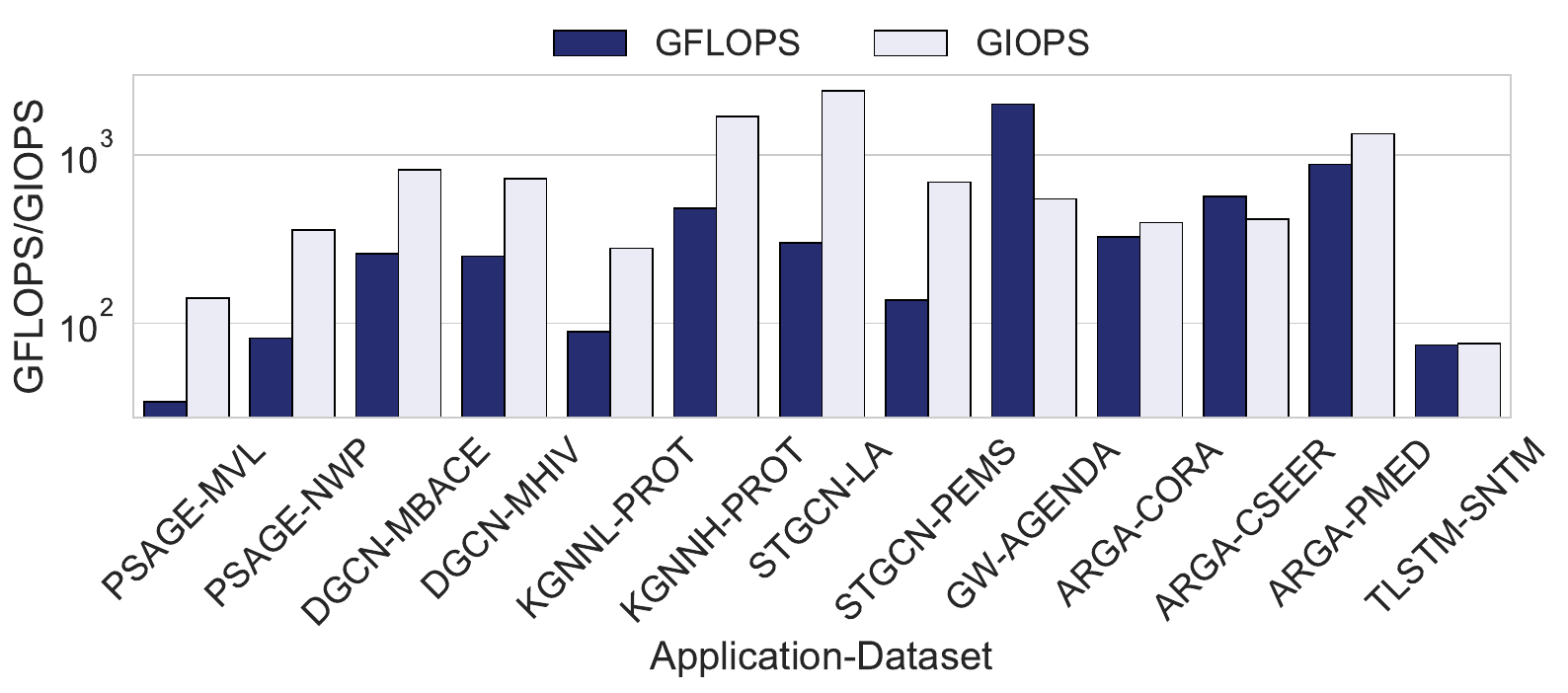} \caption{GFLOPS and GIOPS across the different workloads in GNNMark. \label{fig:flops_app}} 
\vspace{-0.5cm}
\end{figure}

 \figref{fig:flops_app} presents the number of GFLOPS and GIOPS executed by our workloads in GNNMark. We observe that the average GFLOPS rate is 214 GFLOPS, and the average GIOPS rate is 705 GIOPS. The observed average GFLOPS rate is much lower than the theoretical max GFLOPS of the V100, which is 14 TFLOPS for fp32 arithmetic~\cite{nvidia2017v100} (the V100 specs do not mention the peak theoretical GIOPS. We believe it to be close to the peak theoretical GFLOPS). \texttt{GW} has the highest fp32 performance of 1.99 TFLOPS. Being a transformer-based ML model, \texttt{GW} can effectively use most of the parallel resources on a GPU~\cite{vaswani2017attention}. 
We also observe that, while graph batching has been proposed to improve performance in DGL, \texttt{TLSTM} is still able to achieve only 74 GFLOPS.

The average IPC measured across all the workloads was found to be 0.55, which reflects the memory-bound nature of the workloads. When comparing the GFLOPS and GIOPS of different operations, we observe that the GEMM operations typically have a higher GFLOPS~(in the mid 300s) as opposed to other operations, such as reductions, scatters, and gathers that have lower rates~(in the 100 GFLOPS/GIOPS range) suggesting a very low overall GPU utilization. Given that these operations can dominate the execution time, it is important for both hardware and software developers to focus on improving the performance of these operations.

\noindent\rule{\columnwidth}{0.4pt}
\textbf{Instruction Set Usage Summary}: Our analysis reveals that, during GNN training, execution is dominated by integer operations. Thus, to accelerate GNN training on either GPUs or accelerators, int32 arithmetic performance will be key. The overall performance in terms of GFLOPS/GIOPS for GNNs is relatively low compared to the peak performance of the hardware. This suggests that GNN training is primarily memory-bound. Given that operations such as reductions, scatters, gathers, and sorting can occupy a good chunk of the execution time during GNN training, it is important for both hardware and software developers to focus on improving the performance of these operations. \newline
\noindent\rule{\columnwidth}{0.4pt}

\subsection{Stalls and Cache Analysis}
\begin{figure}[!t] \centering
\includegraphics[width=\linewidth]{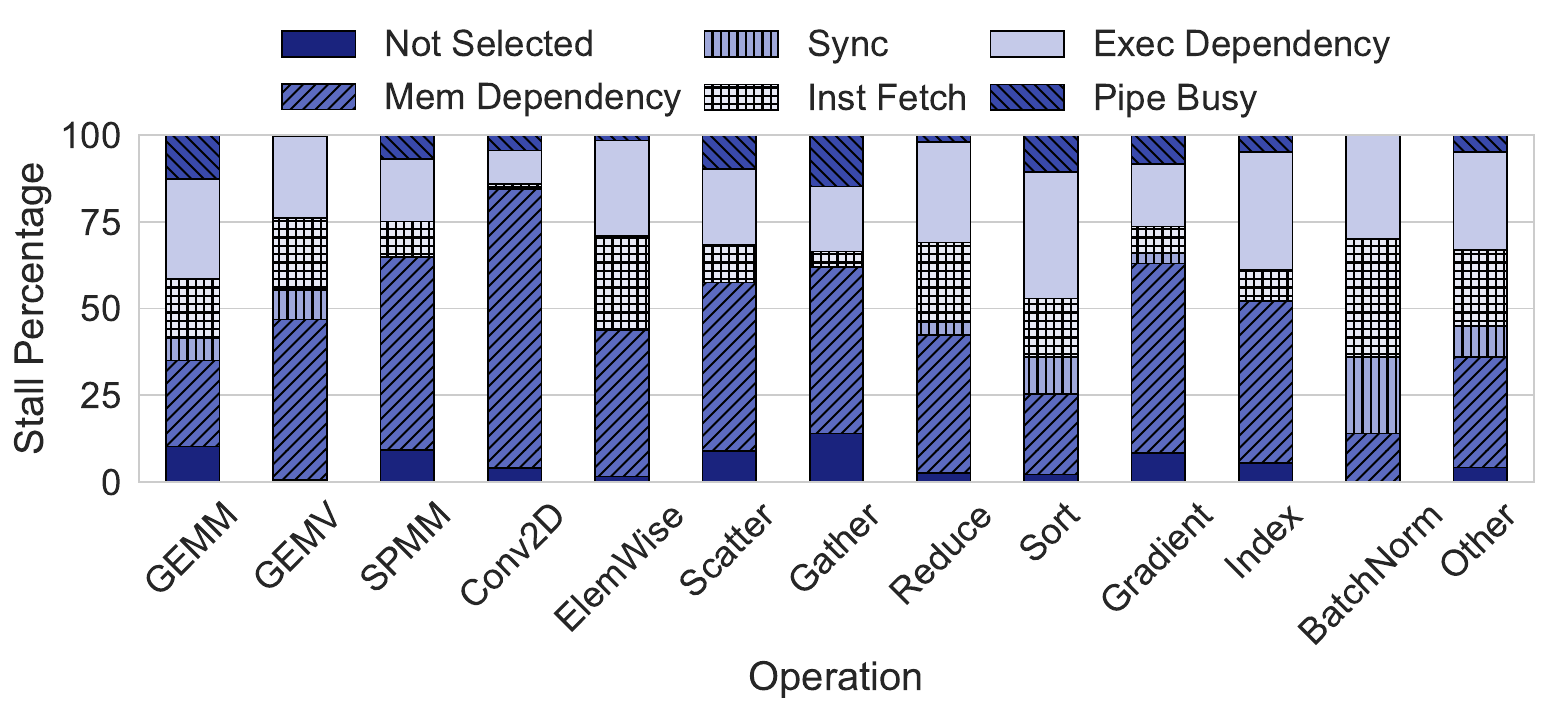} 
\caption{Stall breakdown across operations in GNNMark. \label{fig:stall_operations}} 
\vspace{-0.5cm}
\end{figure}

\begin{figure}[!t] \centering
\includegraphics[width=\linewidth]{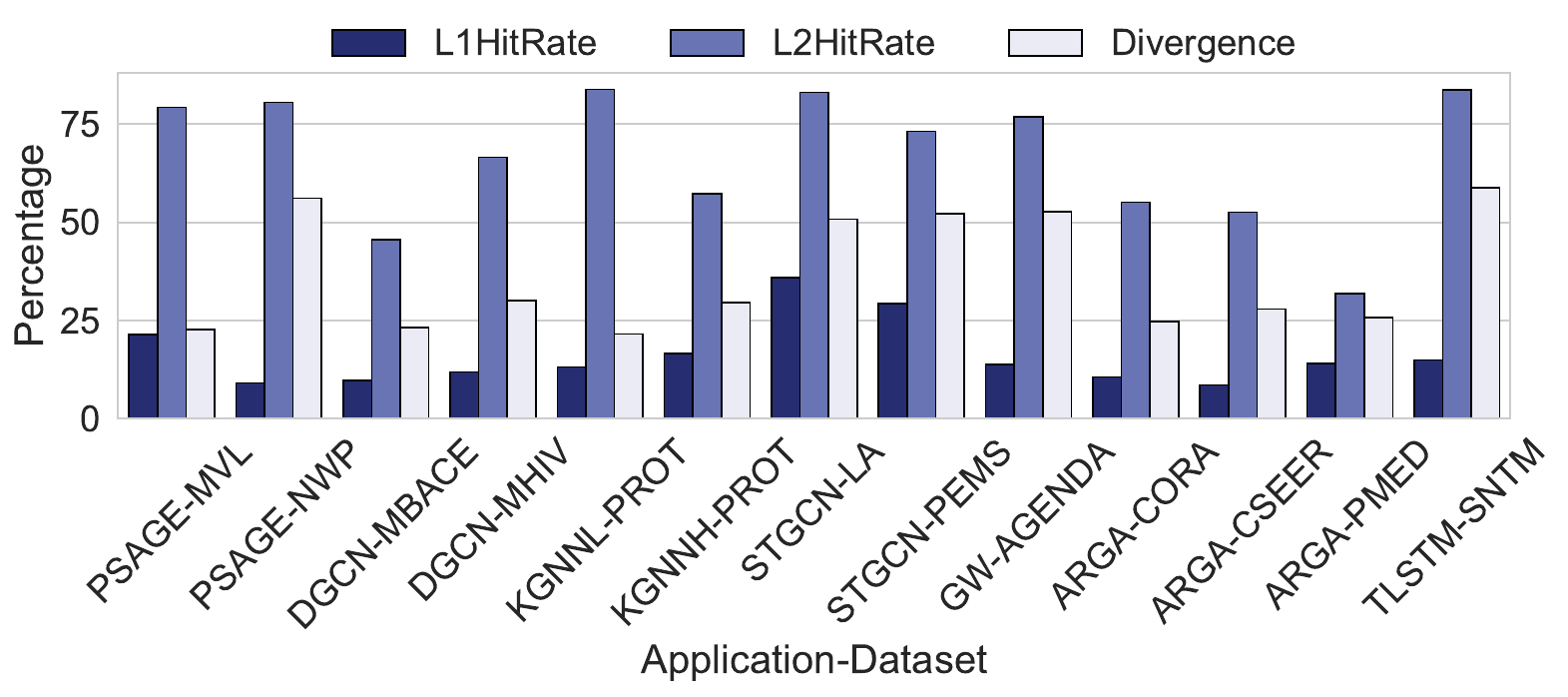} \caption{L1-D and L2-cache hit ratios, and divergent load ratios for GNNMark workloads.\label{fig:cache_app}} 
\vspace{-0.5cm}
\end{figure}

Developing a comprehensive understanding of major stalls in the GPU hardware during GNN training can help guide architectural design decisions when tuning the performance of these workloads. Given that caches can greatly improve the performance of GPU applications, it is also important to look at their efficiency in the context of GNN training. In \figref{fig:stall_operations}, we provide a distribution of different types of stalls observed in GNN training. We find that execution is stalled primarily due to \emph{Memory Dependency}, \emph{Execution Dependency}, and \emph{Instruction Fetch}. The high percentage of \emph{Memory Dependency} stalls~(34.3\% on an average) suggests that the memory subsystem is inefficient in serving data read requests to the GPU cores. From~\figref{fig:cache_app}, we observe that GNN workloads have an extremely low L1 D-cache hit rate on the V100~(a mere 15\%, on average), which is the primary reason for these stalls.

We also analyze the impact of divergent load instructions. The load instructions associated with a warp are considered divergent if they access more than one cache line~(a line is 128B on the V100). Memory divergence can impact the performance of typical graph workloads, such as Breadth First Search and PageRank~\cite{che2013pannotia}. Therefore, it is important to characterize the degree of memory divergence present during GNN training.

Of all the load instructions, we observe $32.5\%$ of load instructions exhibit divergence across different GNN training workloads. This percentage is large and is highly correlated with the resulting low L1 D-cache hit rates. While the larger L2-cache~(6MB) on the V100 fares significantly better with a $70\%$ hit rate on average, the inability of the L1 D-cache to effectively hold the working set can put pressure on the L2 cache to satisfy the memory requirements. Across the different operations, we observe that GEMM, SpMM, and GEMV have poor locality (i.e., a low L1 D-cache hit rate, less than $10\%$ on average). The L1 D-cache hit rates of other operations, such as indexing, scatters, gathers and sorting, are also low~(below $15\%$, on average).


The high percentage of \emph{Execution Dependency} stalls (i.e., 29.5\% of the stall cycles on an average) points to the fact that, across the entire set of workloads, there are many dependencies between instructions in a warp, which results in low instruction-level parallelism. Microarchitectural enhancements to support out-of-order execution in the GPU pipeline~\cite{gong2019haws} can potentially accelerate GNN training.

Surprisingly, \emph{Instruction Fetch} stalls are also significant~(21.6\% on average). This is due to two reasons. The first is that the instruction cache is ineffective in caching all the instructions. Although the V100 architecture has a new 12KB L0 I-cache that is backed by a larger 128KB L1 I-cache, it seems to not be highly effective in caching all instructions during kernel execution. The second reason is loop unrolling techniques~\cite{murthy2010optimal}, which are used to improve the performance of a GPU kernel, can negatively impact the instruction cache hit rate and increase the stalls due to instruction fetching~\cite{davidson1995aggressive}.

% As observed in~\figref{fig:stall_operations}, for commonly used GNN operations~(Conv2D is used only for STGCN and BatchNorm for DeepGCN), the scatter and gather operations, index selection operations have a higher rate of stalls when compared to GEMM due to memory dependencies. This is primarily because both scatter and gather, as well as index selection operations, exhibit an irregular memory access pattern.

In~\figref{fig:stall_operations}, it is evident that scatter and gather operations, along with index selection operations, exhibit a higher frequency of stalls in comparison to GEMM, particularly for commonly utilized GNN operations (notably, Conv2D for STGCN and BatchNorm for DeepGCN). The primary reason for this is the irregular memory access patterns demonstrated by both scatter and gather, as well as index selection operations, which lead to memory dependencies.

\noindent\rule{\columnwidth}{0.4pt}
\textbf{Takeaways}: 
Our analysis of the stalls during GNN training shows that stalls due to \emph{Memory Dependency}, \emph{Execution Dependency}, and \emph{Instruction Fetch} can be significant. While GPU architecture research has focused on removing the first two types of stalls, improving the performance of instruction fetching has been neglected. Therefore, architects and compiler developers should focus on developing techniques to improve instruction fetch to optimize the performance of GNN training.

GNN training also suffers from a high degree of L1 D-cache misses and a significant number of divergent load instructions across all operations. The extremely high L1 D-cache miss rates suggest that caching is not effective for GNN workloads. We envision two potential solutions to alleviate this problem. The first is to employ half-precision-training for GNN training, which uses only 16-bit data instead of 32-bit data, thus can significantly reduce the L1 D-cache miss rates. Alternatively, L1 cache bypassing solutions~\cite{xie2013efficient,xie2015coordinated} can be explored to alleviate this problem.
% The average load divergent instructions account for $32.5\%$ across various GNN workloads.
Among all the load instructions, $32.5\%$ exhibit divergence across various GNN workloads.
\newline
\noindent\rule{\columnwidth}{0.4pt}

\subsection{Sparsity during GNN training}
\begin{figure}[!t] \centering
\includegraphics[width=\linewidth]{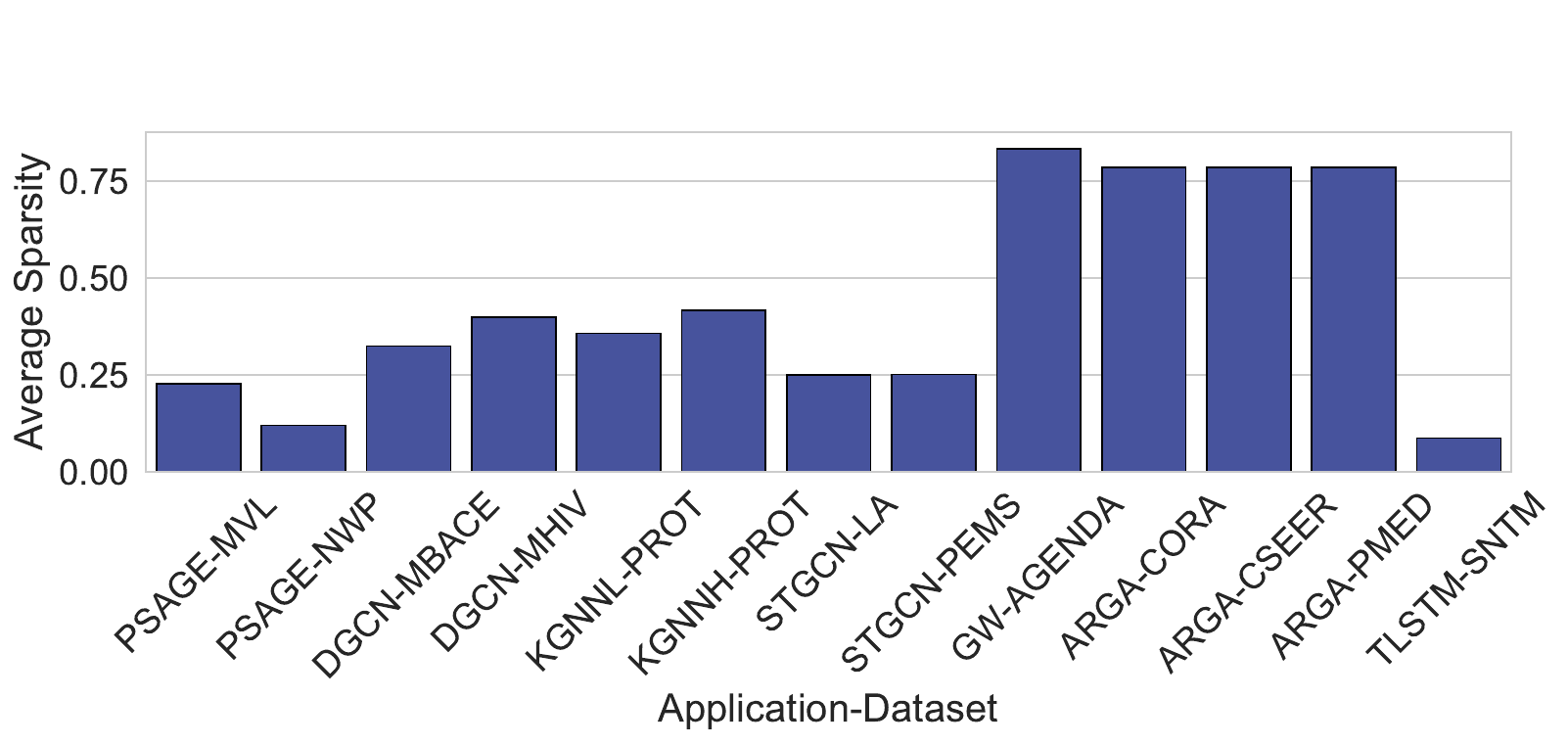} \caption{Average sparsity in the data transferred from CPU-to-GPU during GNN training in GNNMark workloads.\label{fig:sparsity}} 
\end{figure}

Training sparsity refers to the zero values (as a percentage of all values)  that are transferred during CPU-to-GPU memory copies during the GNN training process. For characterizing the average sparsity, we report the percentage of zero values observed in CPU-to-GPU data transfers during GNN training. From~\figref{fig:sparsity}, an average sparsity of 43.2\% was observed during GNN training. 
This suggests that compression techniques could be employed. Rhu et al.~\cite{rhu2018compressing} proposed using compression to alleviate the problem of training large DNN models on a GPU. While GNN models are smaller than conventional DNN models~(e.g., Resnet-50 is 50-layers deep, whereas most GNNs today have fewer than 10 layers), the input graph can occupy a significant portion of GPU memory~(up to 90\% in our experiments). While the machine learning community has proposed sampling the graph to address this problem~\cite{ying2018graph}, there are situations where training on the whole graph has been shown to provide better accuracy~\cite{jia2020improving}. We suggest compressing the data in GPU memory to facilitate training on large graphs.

\begin{figure}[!t] \centering
\includegraphics[width=\linewidth]{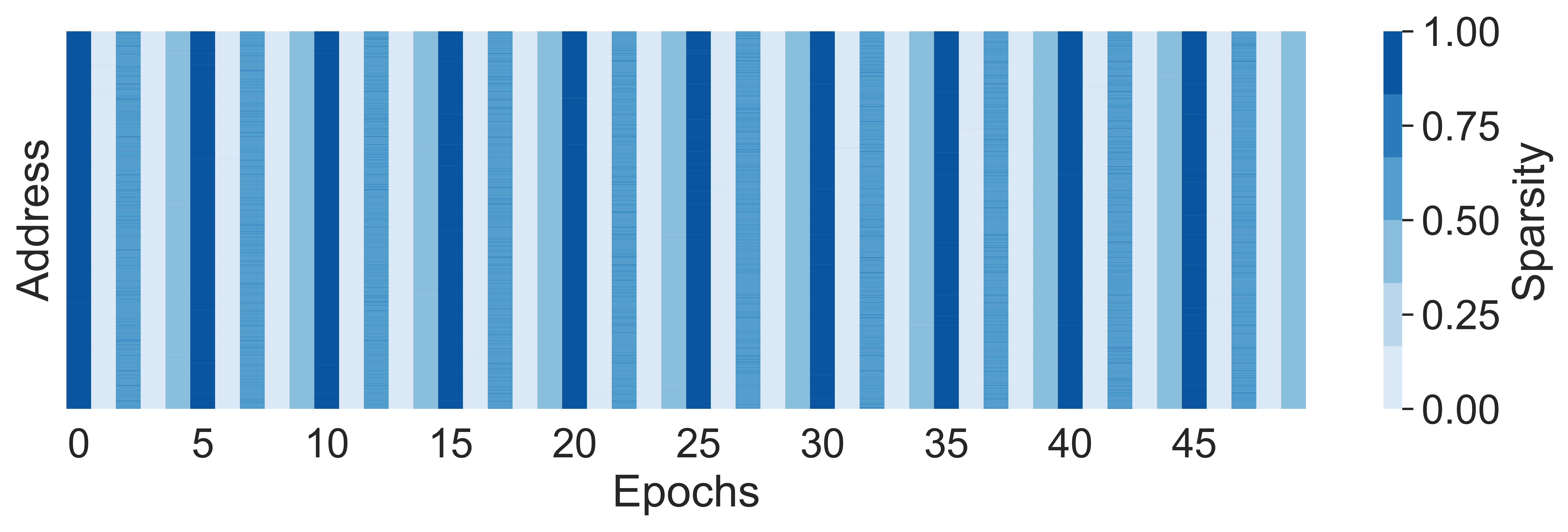} \caption{Sparsity heat map for DeepGCN when running on the MOLHIV dataset.
\label{fig:sparsity_dgcn}} 
\end{figure}

We can also observe a  predictable pattern in the data sparsity~(from ~\figref{fig:sparsity_dgcn}), providing opportunities to apply adaptive compression algorithms~\cite{tavana2019exploiting}. As the sparsity values change during training, the GNN training framework may need to exploit different compression solutions and formats that work the best for a specific sparsity level.

Looking at the average sparsity for \texttt{PSAGE} in~\figref{fig:sparsity}, we can conclude that training sparsity is a function of both the model and the input graph. When using the MVL dataset, the average sparsity is 22\%, but it reduces to 11\% when training on the NWP dataset.

In terms of models, since many GNNs such as GraphTransformer, DeepGCNs and ARGA use activation functions such as ReLU and PReLU in their layers, they produce highly sparse data. We suggest applying compression to take advantage of this sparsity. The result will be that we can train larger graphs on a single GPU. We plan to pursue this path in later  work in this thesis.

\noindent\rule{\columnwidth}{0.4pt}
\textbf{Takeaways}: Training on graphs that are larger than the size of GPU memory is a challenging problem. Thus, exploiting the high degree of sparsity present in GNN workloads by using compression techniques can begin to address this problem.
\newline
\noindent\rule{\columnwidth}{0.4pt}

\subsection{Scalability of GNN training using multi-GPU systems}
\begin{figure}[!t] \centering
\includegraphics[width=\linewidth]{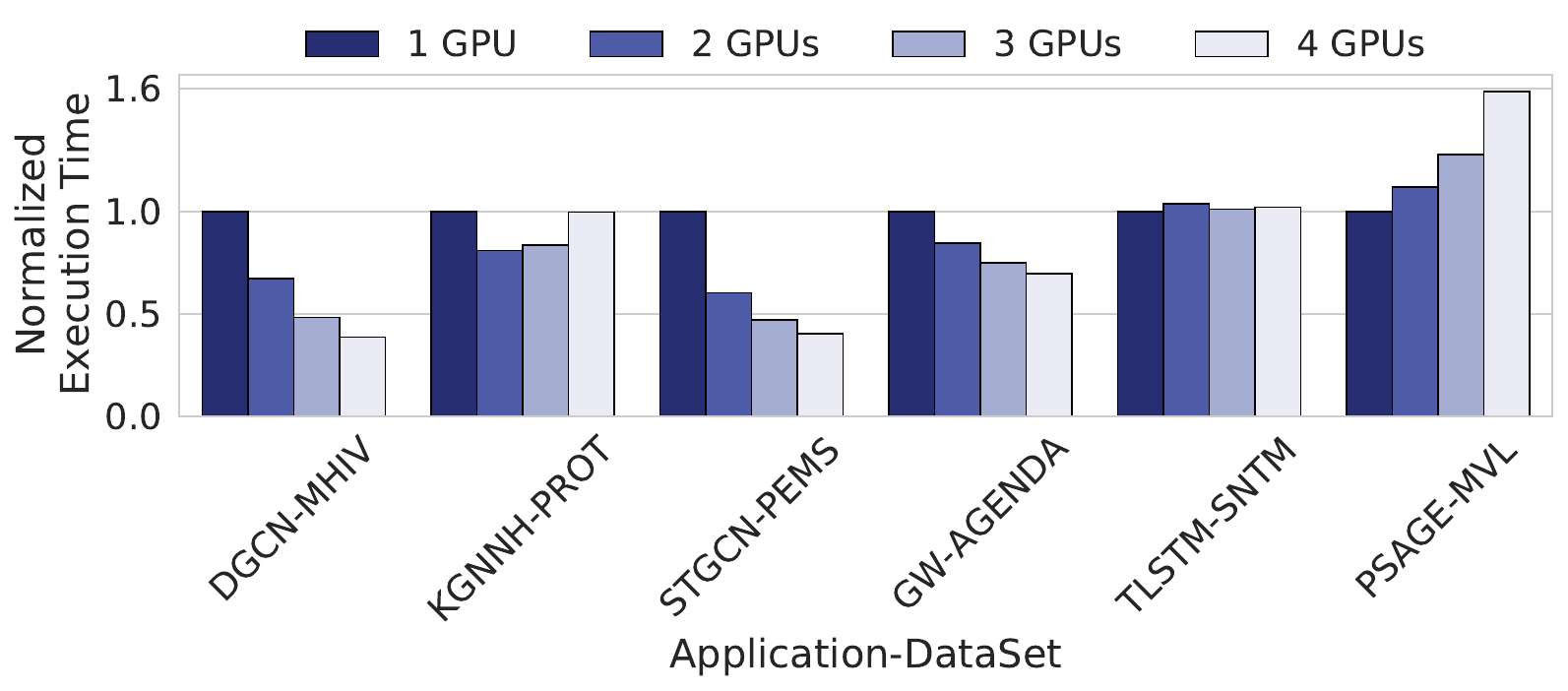} \caption{Multi-GPU performance scaling.
\label{fig:multi_gpu}} 
\vspace{-0.5cm}
\end{figure}

Using the multi-GPU implementations that we developed for the GNNMark workloads using PyTorch DDP, we evaluate the strong scaling characteristics of the workloads in GNNMark. We train all our models for five epochs~(we observe similar performance across all epochs) and report the average \texttt{time-per-epoch}, an approach used in previous work~\cite{mojumder2018profiling}, to understand the performance of DNN workloads on multi-GPU systems. We do not evaluate ARGA, as the application inherently sends the entire graph to the GPU as a part of its training process, and therefore, distributing the same graph across multiple GPUs does not help. The first thing we can clearly observe from~\figref{fig:multi_gpu} is that not all workloads benefit from multi-GPU training. While DGCN, STGCN, and GW show considerable performance gains, the same does not hold true for the other applications. TLSTM does not benefit from multi-GPU training. Given that this is an LSTM-based GNN model with low computational GFLOPS/GIOPS intensity, the application is unable to take advantage of the additional computing power offered by multi-GPU systems. For PSAGE, we observe performance degradation when scaling across multiple GPUs. This is primarily because the PSAGE implementation in DGL uses a batch sampling mechanism, which is not compatible with PyTorch DDP. As a result, the training data gets replicated across multiple devices, and this replication results in redundant computation and unnecessary communication, which in turn hurts performance.

\noindent\rule{\columnwidth}{0.4pt}
\textbf{Takeaways}: Multi-GPU systems do not always benefit GNN training. Therefore, ideas such as topology-aware scheduling and fine-grained graph partitioning that have been proposed by researchers in graph-centric GNN frameworks, such as ROC~\cite{jia2020improving} and NeuGraph~\cite{ma2019neugraph}, should be adopted by high-level frameworks, such as PyG and DGL, to enable more efficient GNN training. Currently, these frameworks are not open source, and hence, we cannot evaluate them for the GNNMark workloads.
\newline
\noindent\rule{\columnwidth}{0.4pt}

\section{GNNMark Summary}
In this \thesis, we present GNNMark, a diverse benchmark suite of GNN workloads designed for the characterization of GPU performance. To the best of our knowledge, we are the first to propose a GNN training focused benchmark suite for the architecture community. We use GNNMark to perform a detailed characterization of GNNs to understand the architectural implications of training on GPU systems. Our work provides novel insights that show the major architectural bottlenecks in GNN training and suggests how they can be potentially addressed.  

A single GNN model can exhibit different characteristics based on the input graph. We observe that unlike DNNs, GEMM and convolution operations are less dominant in GNN execution. Instead, integer operations required for graph processing can dominate execution, suggesting that improving the performance of integer math is paramount. A high degree of instruction fetch stalls shows that the instruction cache on the GPU can limit GNN performance. Finally, we also report on the training sparsity and strong scaling characteristics of GNN training using our suite.

% For future work, given the high sparsity found in these workloads, we plan to explore using compression to accelerate GNN training on large graphs, as well as understand the weak scaling characteristics of GNN training. We also plan to update our suite using the \texttt{time-to-train} metric proposed by the developers of MLPerf~\cite{mattson2020mlperf} and support half-precision training in GNNMark.  We also plan to add more models, such as those in the realm of Reinforcement Learning based GNNs~\cite{wang2018nervenet}, to GNNMark.

% \input{chapters/05_a_accelerator}

% \input{chapters/06_completed}

\chapter{Algorithmic Strategies for GNN Acceleration}
\label{chp5}
% \section{SMASH: Sparse Matrix Atomic Scratchpad Hashing}

% Following the prior chapter's focus on the characterization of GNN workloads, this chapter applies the knowledge gained to improve the performance of a critical kernel in GNN computations: Sparse matrix-matrix multiplication (SpGEMM).

Graph Neural Networks (GNNs) are characterized by their computational structure, which involves dense matrix operations in the combination phase and sparse matrix operations in the aggregation phase. This chapter focuses on the acceleration of the combination phase through the development of an accelerator specifically designed for Sparse General Matrix-Matrix Multiplication (SpGEMM). Subsequently, in the following chapter, we introduce an accelerator designed to efficiently manage both the sparse (combination) and dense (aggregation) computational phases of GNN workloads, thereby establishing a comprehensive and versatile GNN accelerator.

% Sparse matrices and sparse matrix multiplication (SpGEMM) kernels are commonly found in a wide range of applications spanning graph traversal algorithms to machine learning workloads.
% Why SpGEMM is difficult?
Optimizing the performance of multiplications that operate on sparse matrices is challenging, especially given the associated irregular memory access patterns, resulting in load imbalance on today's parallel architectures.  Given that the input matrix data possesses low temporal and spatial locality, this leads to inefficient cache usage and pipeline stalls. 
% thus creating redundant input fetches from DRAM. %, consuming the precious DRAM memory bandwidth.
%% Why we need hardware accelerators for SpGEMM
Modern-day CPUs and GPUs struggle to produce scalable performance when executing sparse matrix multiplication workloads. CPUs fail to monetize on the parallelism present in such workloads, while GPUs struggle to balance tasks across their thousands of hardware threads.
While a number of sparse matrix formats have been proposed to better handle the sparsity, we lack a single format that works over a range of different sparsity patterns.  Given the growing popularity of sparse datasets in emerging applications, we need to explore how we can leverage a novel architecture to accelerate SpGEMM workloads.

%% Contribution of the paper

As part of this
\thesis, we explore a novel distributed memory SpGEMM implementation.  We specifically target this work for a custom accelerator.
Our approach improves performance by mapping the computation of row-wise products, hashtables, and on-chip accumulation to the accelerator's scratchpads.
We provide a new set of performance metrics for this class of workload and demonstrate their utility using a suite of micro-benchmarks run using synthetic, as well as real-world, datasets.
We also introduce a new Memory-aware Aligned Parallel Compressed Sparse Row matrix storage format called MAP-CSR to further accelerate local memory accesses. Running on a custom graph-based accelerator, we are able to achieve consistent speedup over MKL and provide insights on the scalability of our implementation.

% an average speedup of $19.27\times$ over MKL, XX$\times$ over cuSPARSE and XX$\times$ over InnerSP~\cite{baek2021innersp} a previous state of the art SpGEMM accelerator.

%  ___       _                 _            _   _             
% |_ _|_ __ | |_ _ __ ___   __| |_   _  ___| |_(_) ___  _ __  
%  | || '_ \| __| '__/ _ \ / _` | | | |/ __| __| |/ _ \| '_ \ 
%  | || | | | |_| | | (_) | (_| | |_| | (__| |_| | (_) | | | |
% |___|_| |_|\__|_|  \___/ \__,_|\__,_|\___|\__|_|\___/|_| |_|

%% INTRODUCTION DRAFTS %%
% 1. 
% 2. What is row-wise? and Why row-wise application is better?
% 3. Flowchart of papers
% 4. Clearly define objectives of your paper

\section{Motivation for SpGEMM kernel acceleration}
\label{sec:intro}
%% Why is SpGEMM gaining importance?

%% Talk about the degraded performance of SpGEMMs on GPUs and x86. 

%% Talk how data is getting sparse (Example of sparse graphs) but the architectures of the past are not designed to handle such sparse data.

%% SpGEMM is notoriously difficult to optimize, the following section talks about some of the challenges faced in implementing SpGEMM.

%% SpGEMM reuses data, so new format is reused.

Multiplication of two sparse matrices (i.e., a SpGEMM kernel) is commonly found in many emerging workloads.  Some examples of popular algorithms that need to process sparse matrices include:
\begin{itemize}
    \item Scientific computations: algebraic multigrid solvers (AMG)~\cite{baker2011challenges}, volumetric mesh analysis~\cite{mueller2017ternary} and linear-scaling electronic structure computations~\cite{bowler2001parallel};
    \item Graph-based computations: triangle counting~\cite{davis2018graph}, path planning~\cite{niu2016efficient}, community detection~\cite{ediger2011tracking}, breadth-first-search~\cite{bulucc2011parallel}, recommendation systems~\cite{naumov2019deep}, graph neural networks~\cite{baruah2021gnnmark}, label propagation, network packet routing~\cite{thakkar2017video}, graph centrality measures, and graph contractions~\cite{arrigo2017sparse}.
\end{itemize}
\begin{figure}[htbp]
	\centering
	\includegraphics[width=0.93 \textwidth]{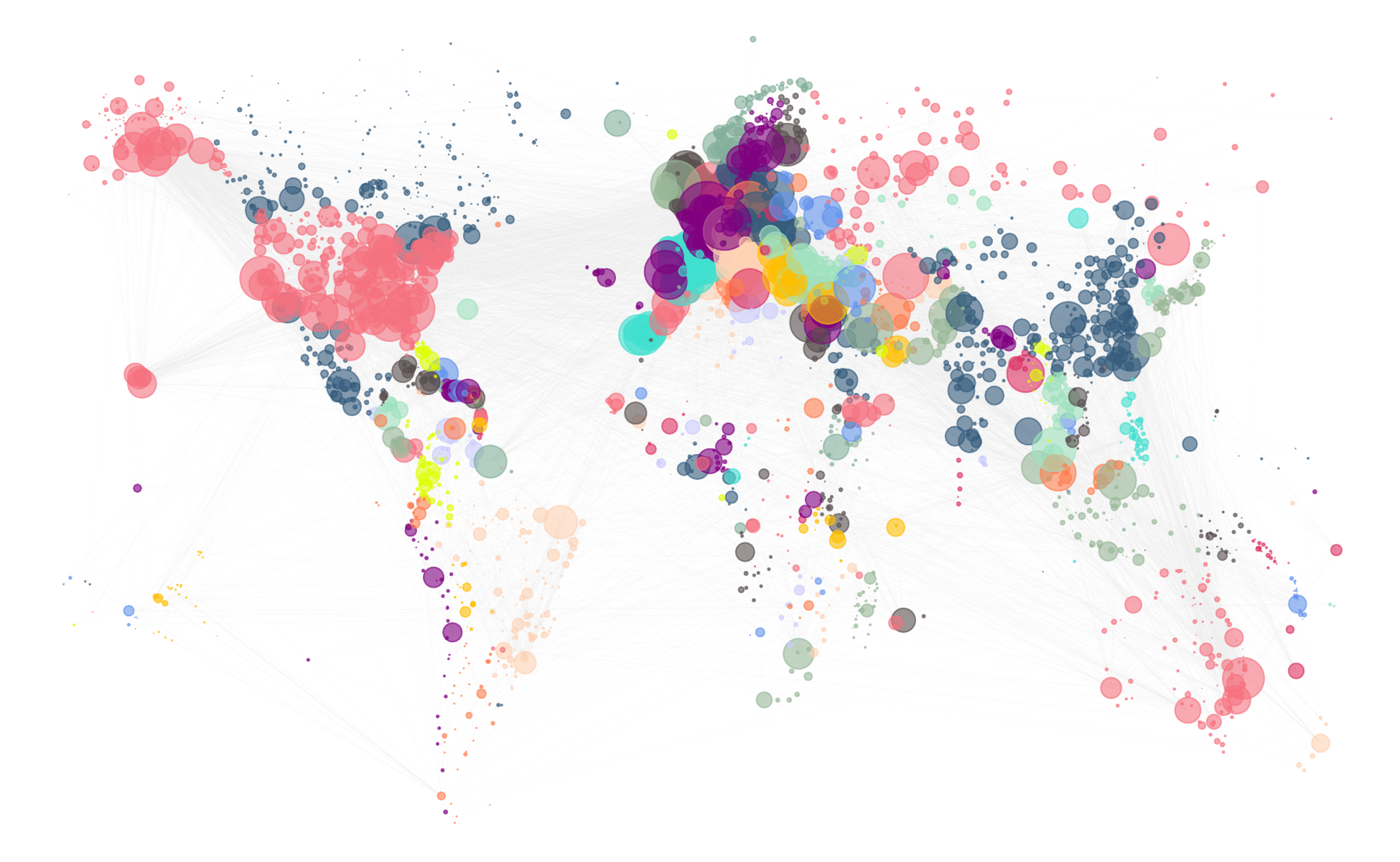}
	\caption{World exposure graph centrality, as generated using an SpGEMM kernel (with nodes as cities, node size proportional to Maximal Frontier Betweenness Centrality (MFBC), edges as air travel corridors, and colors representing countries)}
	\label{fig:epidemic}
\end{figure}

% SpGEMM has been heavily used in epidemic control applications. SpGEMM is a key kernel used in computing centrality measures for airports~\cite{solomonik2017scaling, arrigo2017sparse}, and providing valuable insights in routing vaccinations to prime cities~\cite{hou2020effectiveness}. Figure~\ref{fig:epidemic} plots the world exposure graph computed recursively using airport network dataset, a real-world data set that possesses a $99.63\%$ sparsity, simulated on a single node of our targeted accelerator~\cite{opsahl2011anchorage}.

SpGEMM plays a pivotal role in applications for controlling epidemics. It serves as an essential kernel in calculating centrality measures for airports~\cite{solomonik2017scaling, arrigo2017sparse}, and offers crucial guidance for directing vaccine distribution to key cities~\cite{hou2020effectiveness}. The world exposure graph, plotted in Figure~\ref{fig:epidemic}, is recursively generated using the airport network dataset.
We incorporate this dataset (with a sparsity of $99.63\%$) in the analysis of our SpGEMM kernel implementation on the custom accelerator.
% This real-world dataset, featuring a sparsity of $99.63\%$, is simulated on a single node of the accelerator under consideration~\cite{opsahl2011anchorage}.

Modern trends in Big Data have witnessed an increase in data sparsity, along with an increase in data set size. In 2021, Facebook claimed they had $2.89$ billion monthly active users, with studies suggesting an average of 338 friends per user~\cite{smith_2020}. The resulting adjacency matrix of the Facebook user graph would approach $99.99\%$ sparsity.
Graph analytics of such highly sparse datasets push the limits of the current computing infrastructure and expose innate problems exhibited by traditional architectures. Sparse graph workloads are dominated by highly irregular and uncoalesced memory access patterns.

Current multi-core CPU architectures, given their limited number of compute units (i.e., cores), fail to capitalize on the fine-grained parallelism present in these workloads. \acrfull{SIMD} style GPU architectures struggle to evenly distribute tasks among their threads, leading to under-utilization of hardware. In this work, we present an SpGEMM algorithmm called \acrfull{SMASH}, tailored to a custom multi-threaded accelerator.  This accelerator provides a novel MIMD-style architecture based on simple in-order cores.
{\em SMASH}, a state-of-the-art SpGEMM kernel implementation, provides $1.6\times$ average speedup over MKL with synthetic datasets, a $1.29\times$ average speedup over real world datasets and a $1.04\times$ average speedup over real world datasets as compared to an A100 GPU.
% Have propagated us away from the traditional systolic SIMD architectures requiring uniform workload to a MIMD architecture versatile to skewed work distribution.
% Sparse matrix multiplication is one such workload that leads to performance degradation from skewed sparsity patterns.

% Sparse matrix multiplication (SpGEMM) is a memory-intensive kernel that is bound by memory bandwidth~\cite{srivastava2020matraptor}. This is because of the skewed sparsity patterns that cause irregular memory accesses, leading to poor data locality and inefficient cache usage.

% Challenges of SpGEMM..
% Scaling...
% The scaling of massively parallel SpGEMM kernel is heavily bound by communication~\cite{azad2016exploiting}
% Some previous work...
% Breakthroughs in PIUMA that help SpGEMM...

To summarize, the key contributions of this part of the thesis work include:
\begin{enumerate}
    \item We characterize the challenges faced while developing efficient SpGEMM kernels. To this end, we perform an analysis of various sparse matrix multiplication methods and evaluate the advantages and limitations of each..
    \item We present a new sparse matrix storage format called \acrfull{MAPCSR}, which allows us to compute each row of the sparse matrix in parallel. \acrshort{MAPCSR} improves the efficiency of memory accesses, ensuring memory-aligned storage of each row. Our \acrshort{MAPCSR} implementation is able to improve the performance of SpGEMM by $1.58\times$.
    \item Finally, we present \acrfull{SMASH}, an efficient SpGEMM kernel implementation that leverages distributed memory on a custom accelerator. We provide three different versions of \acrshort{SMASH}, with iterative improvements, each capitalizing on a different feature of the underlying architecture.
\end{enumerate}

\section{Background on SpGEMM}
\label{sec:background}

The SpGEMM kernel operation generally consists of two distinct phases, each with its own computational requirements and challenges:
\begin{enumerate}
    \item \textbf{The Multiplication Phase}: In this initial phase, the algorithm performs element-wise multiplication between corresponding non-zero elements of the sparse matrices involved. Given the sparse nature of the matrices, the algorithm needs to identify matching elements efficiently. This often involves complex data structures like compressed sparse row (CSR) or compressed sparse column (CSC) to store only the non-zero elements along with their indices. The computational complexity in this phase is primarily determined by the number of non-zero elements in the matrices.
    \item \textbf{The Accumulation Phase}: Following the multiplication of elements, this phase focuses on summing up the products to generate the final sparse matrix. This involves aggregating values that are multiplied with the same index, essentially condensing them into a single entry in the resulting matrix. The challenge here is to perform this aggregation in an efficient manner, especially when the product matrix has fewer zeros, i.e., is less sparse than the input matrices. Optimizations often target reducing memory access latency and improving data locality in this phase.

\end{enumerate}

\begin{figure}[htbp]
	\centering
	\includegraphics[width=0.98\textwidth]{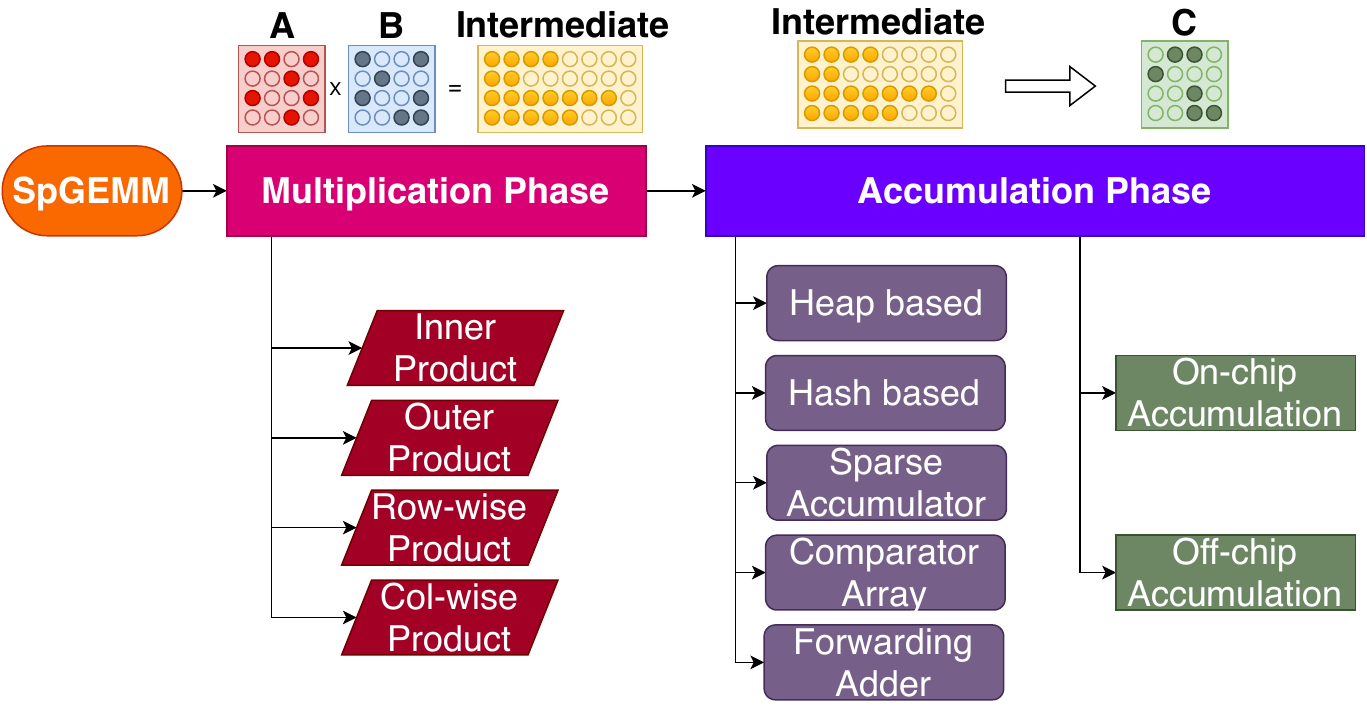}
	\caption{Methods of implementing the two distinct phases of SpGEMM kernel}
	\label{fig:spgemm_types}
\end{figure}

\begin{figure}[htbp]
	\centering
	\includegraphics[width=0.98 \textwidth]{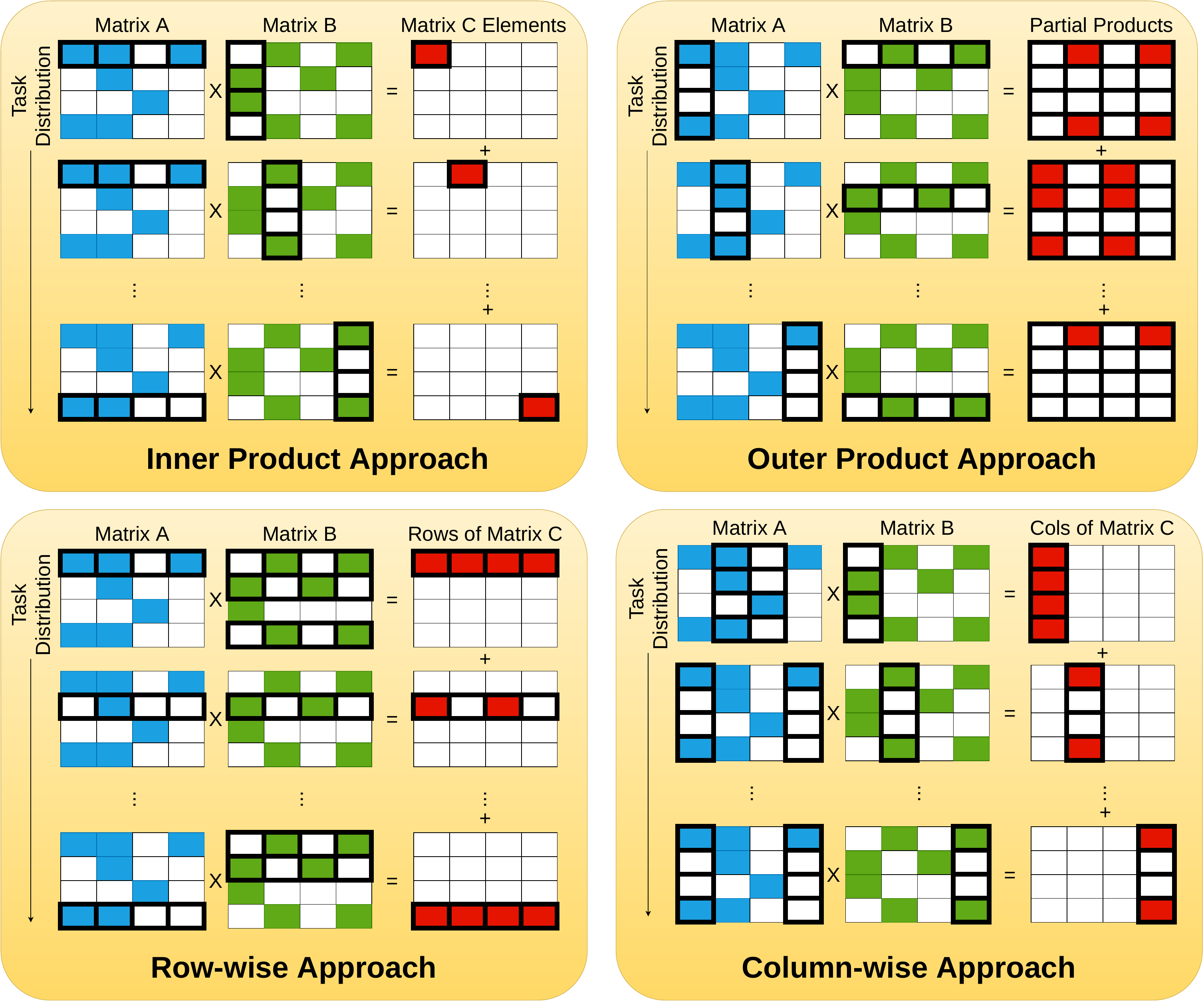}
	\caption{Matrix Multiplication Methods}
	\label{fig:matmul_types}
\end{figure}

Variations in the implementations of these two phases give rise to various SpGEMM algorithms.  There are four methods to compute the first multiplication phase, as shown in Figure~\ref{fig:matmul_types} and Figure~\ref{fig:spgemm_types}.  Each method exhibits different memory access patterns and provides varying degrees of parallelism.

While the inner product multiplication computes output matrix elements directly, its performance is crippled by poor input reuse. On the other hand, the outer product multiplication suffers from poor output locality arising from the endless batches of partial product matrices  generated~\cite{zhang2020sparch}. In our work, we incorporate row-wise multiplication, owing to the massive parallelism exposed by this method. Row-wise multiplication also does not suffer from the memory bloat problem when dealing with a large number of intermediate partial products~\cite{baek2021innersp}.
% Table~\ref{table:parallelism} provides an overview comparing these four types of multiplication phases.

The next phase, called the accumulation phase, can be distinguished based on the underlying data structures.
Examples of accumulation techniques include heap-based~\cite{azad2016exploiting}, hash-based~\cite{nagasaka2017high}, sparse accumulator (SPA) based~\cite{gilbert1992sparse}, comparator array based~\cite{zhang2020sparch}, and Forwarding Adder Network (FAN) based~\cite{qin2020sigma} to name a few.
Depending on the memory hierarchy used for accumulation, this phase can be further classified into on-chip and off-chip accumulation.

This work presents \acrshort{SMASH}, a scalable sparse matrix multiplication kernel based on the row-wise multiplication method. SMASH incorporates on-chip, hash-based accumulation to lower redundant memory accesses. We implement three different versions of this kernel on a custom graph accelerator.

%  __  __    _    ____        ____ ____  ____  
% |  \/  |  / \  |  _ \      / ___/ ___||  _ \ 
% | |\/| | / _ \ | |_) |____| |   \___ \| |_) |
% | |  | |/ ___ \|  __/_____| |___ ___) |  _ < 
% |_|  |_/_/   \_\_|         \____|____/|_| \_\

\section{MAP - CSR Storage Format}
\label{sec:mapcsr}

\begin{figure}[htbp]
	\centering
	\includegraphics[width=0.93 \textwidth]{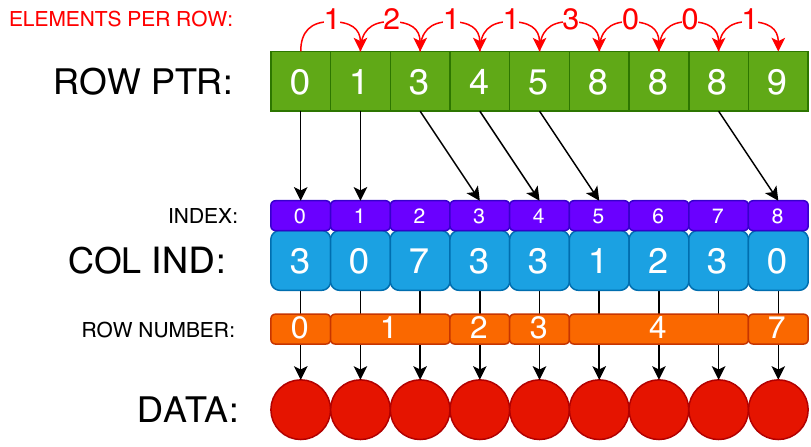}
	\caption[Conventional CSR Format]{\label{fig:conv_csr} Conventional CSR Format}
\end{figure}

The traditional method of storing sparse matrices is \acrshort{CSR}~\cite{smailbegovic2005sparse} (see Figure~\ref{fig:conv_csr}), which is memory efficient as it only stores $n + nnz$ elements instead of $n^2$ (where $n$ is the dimension of a square matrix and $nnz$ is the total number of non-zeros). But what it gains in memory efficiency, it lacks in exposing parallelism. For example, while writing to a \acrshort{CSR} matrix, the rows are required to be written sequentually. If the data needs to be written in parallel, using the \acrshort{CSR} format requires knowledge of the number of non-zeros in all rows in advance to allocate memory preemptively.

The Conventional \acrshort{CSR} format allows only sequential write operations, which poses a considerable challenge to implement \acrshort{SPGEMM} kernels that scale on multi-node systems. It also introduces many synchronization operations, leading to performance degradation.
We introduce a novel matrix storage format called \acrshort{MAPCSR}~\cite{shivdikar2021smash}, that is designed to scale well on large-scale distributed systems.

\subsection{MAP-CSR Implementation}
\begin{figure}[htbp]
	\centering
	\includegraphics[width=0.98 \textwidth]{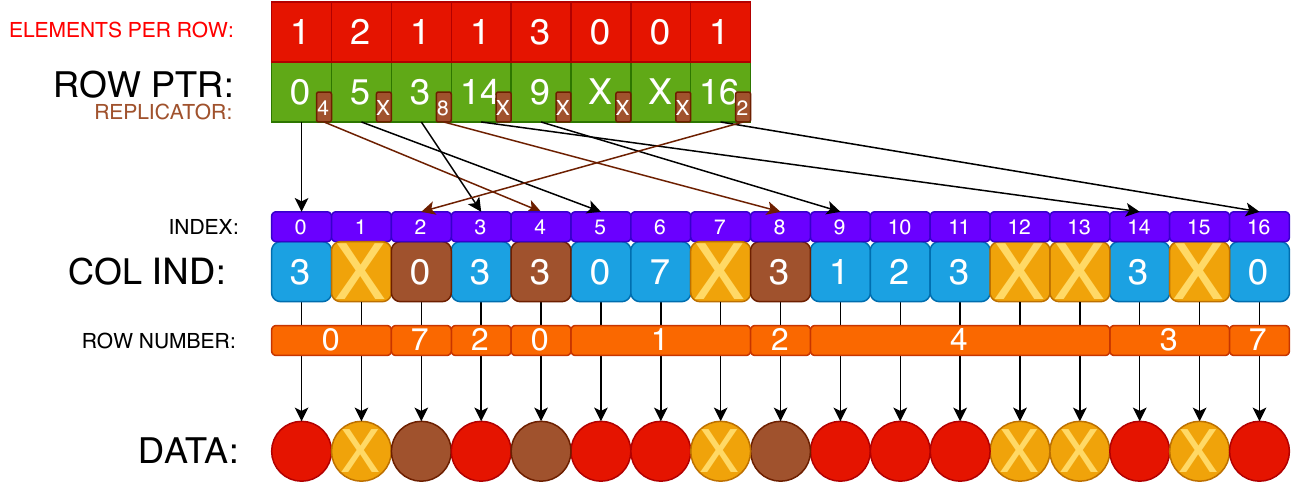}
	\caption[The MAP-CSR Format]{\label{fig:map_csr} MAP-CSR Format}
\end{figure}

Instead of the traditional $3$-array storage format used conventional \acrshort{CSR} (refer to Figure~\ref{fig:conv_csr}), \acrshort{MAPCSR} utilizes a $5$-array storage (see Figure~\ref{fig:map_csr}) as follows:
\begin{enumerate}
    \item Elements per row array: Stores the number of non-zero elements present in each row.
    \item Row pointer array: Points to the start of each row (stores the offset to the start of each row in the column pointer and the data array).
    \item Replicator array: Similar to the row pointer array, but points to the replica of rows in the column pointer array and the data array.
    \item Column pointer array: Stores the column indices of elements in each row.
    \item Data Array: Stores the value of each element. 
\end{enumerate}

Using a $5$-array storage format allows us to write the rows of the matrix in a random order, as compared to the sequential order imposed by Conventional \acrshort{CSR}.
% MAP-CSR can be viewed as a simple memory allocator implementation where each row of a matrix is a call to allocate memory, which is responded with a pointer to the memory location (ROW PTR) and the size of the memory allocated (ELEMENTS PER ROW).
In addition to storing rows in a random order, our approach also allows padding rows with zeros. This enables us to store rows in specific memory banks in main memory. Data accesses to different memory banks can have different latencies depending on which core is making the request.
The ability to select a memory bank for storing specific rows allows for the optimization of memory access latency, particularly for rows that are accessed frequently.

A similar approach of storing sparse matrices was incorporated by Buluc et al.~\cite{bulucc2009parallel}, where they stored rows in random order.
With \acrshort{MAPCSR}, we add another feature called the \textit{Replicator array}. This array, as the name suggests, allows rows to be replicated multiple times in the column pointer and data array. For example, row $7$ is replicated in Figure~\ref{fig:map_csr}. The row pointer array points to offset $16$, the location where one copy of the row is located, and the replicator array points to offset $2$, where a second copy of the row is present. There might even exist more than two copies of each row, in which case, the replicator array for each core points to their respective offsets with low latency. To compare memory requirements of the \acrshort{MAPCSR} format with the memory consumed by the Conventional \acrshort{CSR} format, we can compute the replication ratio $\Re$ as follows:
\begin{equation}
\Re \approx \frac{nnz + nnz^{\prime} + nz_{pad}}{nnz}
\label{eq:replication_ratio}
\end{equation}
where $nnz$ represents number of non-zeros, $nnz^{\prime}$ represents the number of non-zeros from the replicated rows, $nz_{pad}$ denotes the number of zeros used for padding and $\Re$ is the replication ratio, where $1 \leq \Re < \infty$.

\begin{figure*}[htbp]
	\centering
	\includegraphics[width=1.0\textwidth]{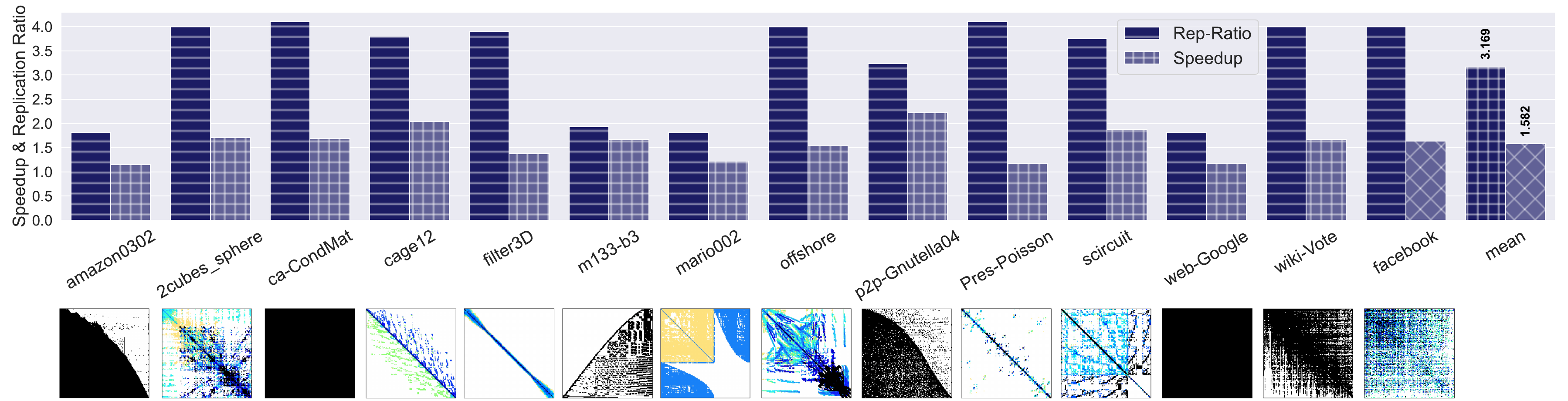}
	\caption[MAP-CSR Speedup]{\label{fig:map_csr_speedup} Replication Ratio (lower is better) and Speedup (higher is better) of SMASH using MAP-CSR v/s CSR storage format on real world datasets.}
\end{figure*}

We benchmark our SpGEMM implementation using the MAP-CSR storage format and compare it to the performance of a vanilla CSR (traditional CSR) storage format. Figure~\ref{fig:map_csr_speedup} provides information on the replication ratio (Equation~\ref{eq:replication_ratio}) for each dataset, as well as the speedup obtained by using MAP-CSR as compared to the CSR storage format.
A higher ratio denotes a larger memory footprint, hence we aim to lower the replication ratio. On average, we obtain a $1.582\times$ speedup by utilizing MAP-CSR, as compared to using a CSR storage format, with an average replication ratio value of $3.169$.

\subsection{MAP-CSR Advantages}
\acrshort{MAPCSR} offers many advantages over conventional \acrshort{CSR}, namely:
\begin{enumerate}
    \item Allows rows to be stored in a random order.
    \item Allows zero-padding to be aligned on memory banks.
    \item Allows for rows to be replicated for faster reads.
    \item Allows for rows to be prefetched, as they are isolated in banks.
\end{enumerate}

\subsection{MAP-CSR Limitations}
% In spite of the major benefits afforded by \acrshort{MAPCSR}, it comes with some downsides that we address here.
While MAP-CSR offers major benefits, it also has certain shortcomings, which we address here.
While replication of rows provides faster memory read transactions (data can now be fetched from memory with relatively lower latency), this affects the writing mechanisms. Writing to a replicated row requires writing to the original copy of the row (as pointed to by the row pointer) and requires the replicated copies to be invalidated.

Conversion of the Conventional \acrshort{CSR} to \acrshort{MAPCSR} is associated with both compute and memory overheads.
Memory overhead, as discussed before (refer to Equation~\ref{eq:replication_ratio}), poses a $\Re$ times increase in the memory footprint.
The computational overhead of converting to \acrshort{MAPCSR} is associated with the replication of rows. Replication of rows requires re-computing the indices of the row pointer in \acrshort{MAPCSR} format, which is a compute-intensive process. Despite the necessary overhead associated with \acrshort{MAPCSR}, we were able to obtain an average speedup of $1.58\times$ over conventional \acrshort{CSR}, achieving an average replication ratio of $3.17$ for the SpGEMM workload.

%  ____  __  __    _    ____  _   _ 
% / ___||  \/  |  / \  / ___|| | | |
% \___ \| |\/| | / _ \ \___ \| |_| |
%  ___) | |  | |/ ___ \ ___) |  _  |
% |____/|_|  |_/_/   \_\____/|_| |_|

\section{SMASH Kernel}
\label{sec:smash}

One of the key design choices for our \acrshort{SPGEMM} kernel implementation was to select one of the four general matrix-multiplication approaches (shown in Figure~\ref{fig:matmul_types}). The inner product approach faced issues due to the cost of index-matching and low temporal reuse~\cite{pal2018outerspace}. The outer-product approach generated a large number of intermediate partial products, demanding high on-chip memory requirements. Neither of these choices provides any benefit when multiplying extremely sparse matrices.

% \begin{figure}[htbp]
% 	\centering
% 	\includegraphics[width=0.98 \textwidth]{figures/chp_05/smash/smash/ins_plot.pdf}
% 	\caption[SMASH Instructions Histogram]{\label{fig:ins_histogram} Characterization of SMASH instruction usage.}
% \end{figure}

Our novel implementation of the SpGEMM kernel is based on a row-wise product method called \acrshort{SMASH}~\cite{shivdikar2021smash}.
Our method exploits high data reuse behavior~\cite{qin2020sigma} and minimizes the number of input matrix reads, while still maintaining low on-chip memory usage.
\acrshort{SMASH} incorporates on-chip memory to store intermediate results and leverages the atomic instructions to accumulate these partial products.

\begin{figure*}[htbp]
	\centering
	\includegraphics[width=1.0\textwidth]{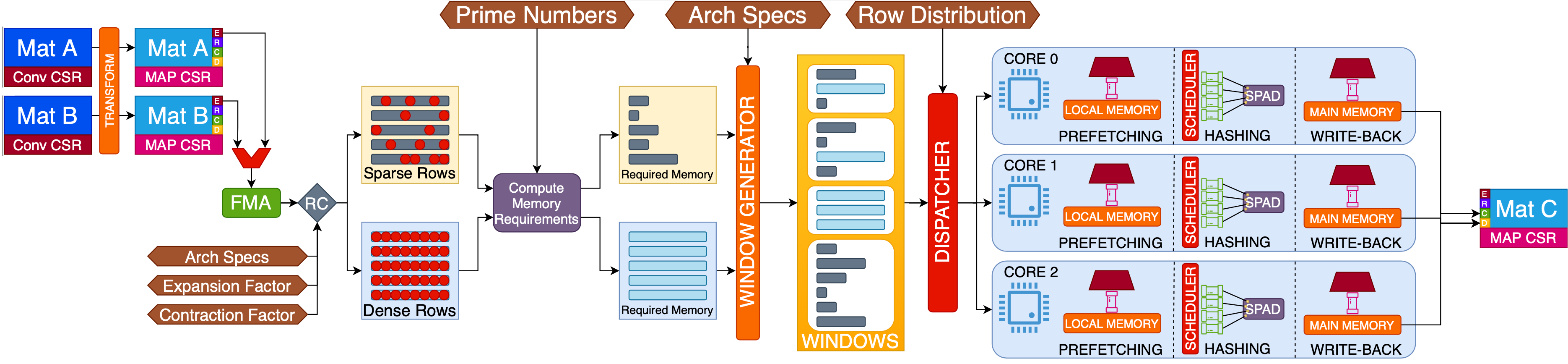}
	\caption[SMASH Architecture]{\label{fig:smash_arch} The SMASH architecture.}
\end{figure*}

In this \thesis, we present a set of successive improvements, resulting in three versions of \acrshort{SMASH}~\cite{shivdikar2021smash} (overview of SMASH architecture shown in Figure~\ref{fig:smash_arch}). In each version we identify the remaining bottlenecks, and then optimize our algorithm to mitigate them in the next version.  Each \acrshort{SMASH} implementation targets a specific performance bottleneck on the custom accelerator architecture.

The following subsections describe our implementation of \acrshort{SMASH}, along with three different optimizations.
Similar to Gustavson's two-phase matrix multiplication algorithm~\cite{gustavson1978two}, our \acrshort{SMASH} implementation is characterized by two phases:
\begin{enumerate}
    \item Memory computation phase
\begin{enumerate}
    \item Matrix Read
    \item Compute Memory Requirements
    \item Window Generation
\end{enumerate}
    \item Product computation phase
\begin{enumerate}
    \item Prefetching
    \item Hashing
    \item Write-back
\end{enumerate}
\end{enumerate}

\subsection{Memory computation phase}
Analogous to Gustavson's two-phase matrix multiplication approach~\cite{gustavson1978two}, the first phase of \acrshort{SMASH} determines the memory required for the output matrix $C$, as well as the on-chip memory requirements of the intermediate products.
For evaluation purposes, we do not include the time consumed in the memory computation phase while considering speedup over other architectures. Evaluation methodologies are further discussed in Section~\ref{sec:eval}.
This phase can be further decomposed into three tasks.
%It starts by running a single hardware thread (an \acrshort{STP}) of the first node (referred to in this paper as the head node). The task generation phase is divided into three subphases.
\begin{enumerate}
\item{\bf Matrix Read} - 
Our SMASH SpGEMM implementation starts off with reading input matrices $A$ and $B$, both of which are presumed to be in a conventional \acrshort{CSR} format.
\item{\bf Compute Memory Requirements} -
After reading the input matrix arrays in \acrshort{CSR} format, we compute the required size of memory required to store the output matrix by counting the total \acrshort{FMA} operations per row.
%To accomplish this, we use Gustavson's two-step algorithm~\cite{gustavson1978two}.
We compute the maximum number of non-zeros for every row of output matrix $C$ using Gustavson's two-step algorithm~\cite{gustavson1978two}. In this \thesis, we refer to this term as \acrfull{FMA}.
The computation of \acrshort{FMA}s per row has a computational complexity of $\mathcal{O}(n)$, where $n$ is the size of the input matrix.

\item{\bf Window Generation} -
Once the memory requirements of each row are computed, this phase then classifies each row of the output matrix $C$ as either dense or sparse. We then group multiple rows together into a single window that can be dispatched as a task to computing core.
This process of classifying and grouping rows into windows is characterized by two parameters:
\begin{enumerate}
    \item Contraction Factor ($CF$): Decides if a row of output matrix $C$ can be classified as dense or sparse. If $\frac{FMA}{CF} > threshold$, then the row is classified as dense, else it is classified as sparse. The $threshold$ value is a function of $scratchpad$ size and matrix density.
    \item Expansion Factor ($EF$): This is used to determine the memory requirements of sparse rows, where the memory requirements are equal to the higher prime number closest to the value $FMA \times EF$.
\end{enumerate}
A dense and a sparse row is evaluated differently in SMASH during the hashing phase. A sparse row will be allocated less memory than the max size of the row, as a dense row will follow a $1:1$ mapping and will be allocated memory equal to the max size of the row.
Once the classification of rows is complete, this phase groups multiple rows together into a single window, such that the intermediate partial products can fit on the on-chip memory (i.e., scratchpad).
At this phase of window generation, the input matrices are converted from conventional \acrshort{CSR} format to the new \acrshort{MAPCSR} format.
The \acrshort{MAPCSR} storage format allows for each window to consist of rows in a non-sequential order, permitting greater flexibility for this phase to generate windows. An evenly spread mix of sparse and dense rows are packed together and shipped to the next phase for computation.
Every individual compute core processes its own window independently, regardless of the status of other windows. This allows us to assign windows to compute cores in a random order and oversubscribe windows.
\end{enumerate}

\subsection{Product Computation Phase}
% Starts only after first phase.
\subsubsection{Prefetching}
Each window generated in the previous phase is scheduled on a computing core for generating intermediate products. The prefetching phase preemptively copies the input matrix rows that are required by each compute core to their respective local memory bank.   This phase of Prefetching is only possible due to the ``replicator'' property of \acrshort{MAPCSR}, which allows rows to be duplicated multiple times across each computing core.

\subsubsection{Hashing}
This phase involves the multiplication of input matrix elements, required to compute the intermediate partial product. After the partial product is computed, it needs to be stored and merged.
%The accumulation phase of matrix multiplication involves the storage and merging of partial products generated in the previous multiplication phase.
Merging partial products is a memory-intensive process requiring scanning through arrays to match indices.
In addition, on multi-threaded architectures, this class of operations needs to be synchronized to avoid data races and ensure atomicity.
Among the multiple solutions available to store and merge data, we opt for hashtables. SMASH utilizes row-partitioned hashtables to store and merge partial products.
The use of hashtables avoids the use of expensive index matching, while allowing us to merge partial products on the fly.
% Storage of Partial products

We utilize hashtables to store intermediate partial products (on the on-chip memory).  In the hashing phase, a global hashtable is created in the \acrfull{SPAD} (the on-chip memory). A single row is allocated to one thread of each compute core in a round-robin fashion. Each element of the row from the first matrix is multiplied with an entire corresponding row of the second matrix (Equation~\ref{eq:row_wise_product} and~\ref{eq_outerproduct}, where $C$ is the output matrix, $A$ and $B$ are input matrices, and $N$ is the size of the matrix). This leads to the creation of partial products. These partial products are hashed into the \acrshort{SPAD} using prime-modulo hashing.
\begin{equation}
\label{eq:row_wise_product}
    C[i,:] = \sum_{k=0}^{N}A[i,k] * B[k,:]
\end{equation}
\begin{equation}
\label{eq_outerproduct}
u \otimes v = \begin{bmatrix}
u_1 \\
u_2 \\
u_3 \\
u_4
\end{bmatrix}
\begin{bmatrix}
v_1 & v_2 & v_3
\end{bmatrix}
=
\begin{bmatrix}
u_1v_1 & u_1v_2 & u_1v_3 \\
u_2v_1 & u_2v_2 & u_2v_3 \\
u_3v_1 & u_3v_2 & u_3v_3 \\
u_4v_1 & u_4v_2 & u_4v_3
\end{bmatrix}
\end{equation}
In prime-modulo hashing, we hash the intermediate products on the \acrshort{SPAD} by indexing each to the closest highest prime number, as computed in the Window Generation phase. The use of prime numbers allows us to reduce the number of hash collisions.
A prime-modulo hash of intermediate products can result in three outcomes:
\begin{itemize}
    \item \textbf{Hash Insert}: This routine is executed when a hashed index finds an empty location on \acrshort{SPAD}. The column index value and data value of the intermediate product are stored on the \acrshort{SPAD} at the hashed index location.
    \item \textbf{Hash Update}: This routine is executed when the hashed index does not find an empty location on \acrshort{SPAD}, but the column index value of the current intermediate product matches with the one present on \acrshort{SPAD}. In this case, the \acrshort{SPAD} data value is updated with the sum of new data value and existing data value. This routine is also called the merging of partial products.
    \item \textbf{Hash Collide}: This routine indicates that the hashed index did not find an empty location on \acrshort{SPAD}, nor did the column index match the existing column index value on \acrshort{SPAD}. In this case, the algorithm probes for a new location using quadratic probing to find the next available empty location on \acrshort{SPAD} for hash insertion.
\end{itemize}
The pseudo-code for the entire hashing phase is shown in Algorithm~\ref{alg:hashing_full}.

% Pseudocode
% ____  ____  _____ _   _ ____   ___   ____ ___  ____  _____
%|  _ \/ ___|| ____| | | |  _ \ / _ \ / ___/ _ \|  _ \| ____|
%| |_) \___ \|  _| | | | | | | | | | | |  | | | | | | |  _|
%|  __/ ___) | |___| |_| | |_| | |_| | |__| |_| | |_| | |___
%|_|   |____/|_____|\___/|____/ \___/ \____\___/|____/|_____|
\begin{algorithm}[t]
\linespread{0.4}\selectfont
\SetAlgoLined
\caption{SMASH HASHING}
\DontPrintSemicolon
\label{alg:hashing_full}
%\For{$w \leftarrow Total\ Windows$}{
\tcp*[h]{READ PHASE}\;
\While{Till you reach end of window} {
\tcp*[h]{Atomically distribute work to each thread}\;
$token \leftarrow$ Each thread will receive one unique token\;
\eIf{$token\_id\ \%\ 2 = 0$}{
$row\_begin \leftarrow A\_col\_ptr\_copy\_1[\frac{token\_id}{2}]$\;
}{
$row\_begin \leftarrow A\_col\_ptr\_copy\_2[\frac{token\_id}{2}]$\;
}
$row\_end \leftarrow A\_col\_ptr[\frac{token\_id}{2} + 1]$\;
\For{$i \leftarrow$ Iterate from $row\_begin$ to $row\_end$}{
\If{Check if we are within our assigned window}{
$col\_begin \leftarrow B\_row\_ptr[\frac{token\_id}{2}]$\;
$col\_end \leftarrow B\_row\_ptr[\frac{token\_id}{2} + 1]$\;
\eIf{$token\_id\ \%\ 2 = 0$}{
\tcp*[h]{Hash EVEN Section}\;
}{
\tcp*[h]{Hash ODD Section}\;
}
}
}
}	
%}
$A\_col\_ptr\_copy\_1$ and $A\_col\_ptr\_copy\_2$ will now reflect new positions\;
\end{algorithm}

\begin{algorithm}[t]
\linespread{0.4}\selectfont
\SetAlgoLined
\caption{SMASH HASHING Even and Odd Section}
\DontPrintSemicolon
\label{alg:hashing_even_odd}
\For{$k \leftarrow$ Iterate from \textcolor{red}{ $col\_begin$ } to \textcolor{red}{ $col\_end$ } }{
\tcp*[h]{Multiply element from $mat\_A$ with that from $mat\_B$ and store its tag and value}\;
$tag \leftarrow$ $X$ coordinate from $mat\_A$ element and $Y$ coordinate from $mat\_B$ element\;
\tcp*[h]{Hash the Tag}\;
$tag \leftarrow tag\ \%\ prime\_modulo$\;
\eIf{$SPAD\_tag[tag] = EMPTY$} {
$SPAD\_tag[tag] \leftarrow tag$  \tcp*[h]{Store Tag on scratchpad}\;
$SPAD\_val[tag] \leftarrow value$  \tcp*[h]{Store Value on scratchpad}\;
}
{
\eIf{$SPAD\_tag[tag] = tag$}
{
$SPAD\_val[tag] += value$  \tcp*[h]{Accumulate Value}\;
}
{
\tcp*[h]{Probe for empty space on Scratchpad}\;
}
}
}
\end{algorithm}

\subsubsection{Write-back}
The write-back phase moves the partial products from the hashtable to their final output matrix, stored in \acrshort{DRAM} in the \acrshort{MAPCSR} format.
The use of the \acrshort{MAPCSR} storage format allows us to asynchronously move rows of the output matrix $C$ in non-sequential order from the \acrshort{SPAD} to their final \acrshort{DRAM} memory location.

The \acrshort{SMASH} implementation discussed so far is considered as our ``base'' implementation. We iteratively add three more optimizations on top of this base implementation, addressing key performance bottlenecks observed during each implementation.

\subsection{SMASH Version 1: Atomic Hashing}

A row-wise product method multiples each element of the first input matrix with an entire row of the second input matrix, generating a row of partial products of the output matrix. These partial products are then merged to form the output matrix elements using a hashtable.
% Each row of the first input matrix, when multiplied by their corresponding rows from the second input matrix, will generate a series of partial product matrices.
This is one of the disadvantages of using a row-wise product method. The intermediate results  (i.e., partial products) need to be stored and merged into the output matrix atomically.
The base version of \acrshort{SMASH} only allowed a single compute core to work on each row of output matrix $C$,  avoiding data races.
This leads to a lower degree of parallelism in each window, as the maximum number of compute cores concurrently working on any window depends on the number of rows in that window.
We overcome this obstacle with our first version $V1$ of the \acrshort{SMASH} kernel by using atomic hashing.
We make use of atomic compare and exchange instructions and atomic fetch and add instructions, enabling us to use multiple cores simultaneously to produce a single output row of matrix $C$.
Optimizing with atomic instructions leads to a $2.48\times$ speedup over the base \acrshort{SMASH} implementation for synthetic datasets.

\subsection{SMASH Version 2: Tokenization}
\label{chap:smash:section:tokenization}

\begin{figure*}[htbp]
	\centering
	\includegraphics[width=1.0\textwidth]{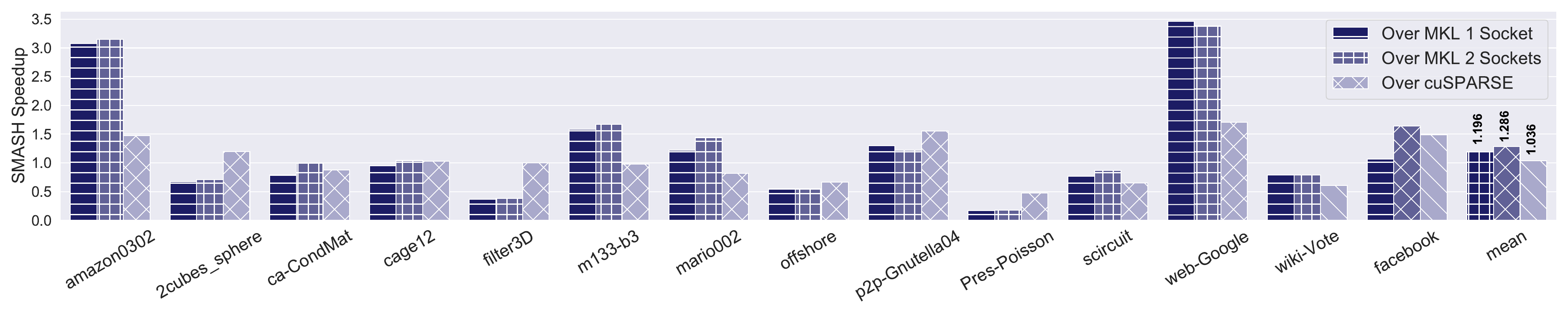}
	\caption[SMASH Speedup]{\label{fig:real_world_speedup} Speedup of SMASH over MKL using 1 CPU, 2 CPUs, and over cuSPARSE using an A100 GPU}
\end{figure*}

SpGEMM workloads, when working with extremely sparse matrices that possess a highly irregular non-zero distribution, experience load imbalance on multi-core architectures.
Although our implementation is not immune to the effects of such irregular sparsity patterns, we aim to reduce the performance impacts of load imbalance with an on-the-fly row scheduler that is based on the classic producer-consumer model.

%The previous version allows one row of the input matrix to be assigned to one thread on the \acrshort{PIUMA} block. This leads a high degree of imbalance in terms of work across all threads in a block. The \acrshort{SpGEMM} datasets are notorious for producing load imbalance during kernel execution.

We tackle this issue by adding a dynamic work scheduler layer into our hashing phase. Instead of statically assigning rows to threads in a round-robin fashion, we adopt a Producer-Consumer for model row allocation.  The dynamic row allocation works as follows:
\begin{enumerate}
	\item Generate two tokens for every row present in the window.
	\item Each compute core polls for a single token. Thus, every row is allocated 2 compute cores.
	\item The 2 compute cores start hashing the row. The first core starts from the beginning of the row and hashes the first half of the row (i.e., the even section). The second thread applies the same steps over the second half of the row (i.e., the odd section).
	\item Partial products from both threads are hashed into a common hashtable, stored in the \acrshort{SPAD} memory.
	\item When all of the tokens have been polled, the window execution is completed. 
\end{enumerate}
%The split of workload between even and odd sections can be seen in  Algorithms~\ref{alg:hashing},~\ref{alg:hashing:even}, and~\ref{alg:hashing:odd}.
Despite the overhead of polling tokens, tokenization produces a $1.5\times$ speedup over static allocation, as it achieves a near-perfect distribution of workload across threads. More details of the performance benefits are presented in Section~\ref{sec:eval}. 

% Another optimization carried out in \acrshort{SMASH} Version 2 is to change the selected bits used for hashing.
% In \acrshort{SMASH} Version 1, we used the high-order bits for hashing in the hashtable. The downside of using high-order bits is that if two elements with their $tag$ values close to each other need to be hashed, they will end up hashing to the same position, thus invoking the collision handler. Hashing on high-order bits groups clusters of adjacent elements together. Instead, in this version, we chose to hash on low-order bits by setting the top $n$ high-order bits to $zero$. 
% Using low-order bits evenly distributes a cluster of elements over the entire hashtable, thus sharply reduces the number of collisions. This can be seen in Figure~\ref{fig:hash_lower_bits}

% The disadvantage of using low-order bits is that the order of hashing is no longer preserved. The hashtable is no longer partially sorted, as in the case of the previous version.
% We overcome this problem by merging all partial products of the same $tag$ before writing them to the hashtable.
% Even though the order is not preserved and the output matrix in the \acrshort{CSR} format is not sorted, the correctness of the solution is maintained, as all partial products are properly merged.

% \begin{figure}[htbp]
% 	\centering
% 	\includegraphics[width=0.8\textwidth]{fig/SMASH_hash_no_collision.pdf}
% 	\caption[Hashing on low-order bits.]{\label{fig:hash_lower_bits} Hashing on low-order bits.}
% \end{figure}

\subsection{SMASH Version 3: Pipelining}

\begin{figure}[htbp]
	\centering
	\includegraphics[width=0.93 \textwidth]{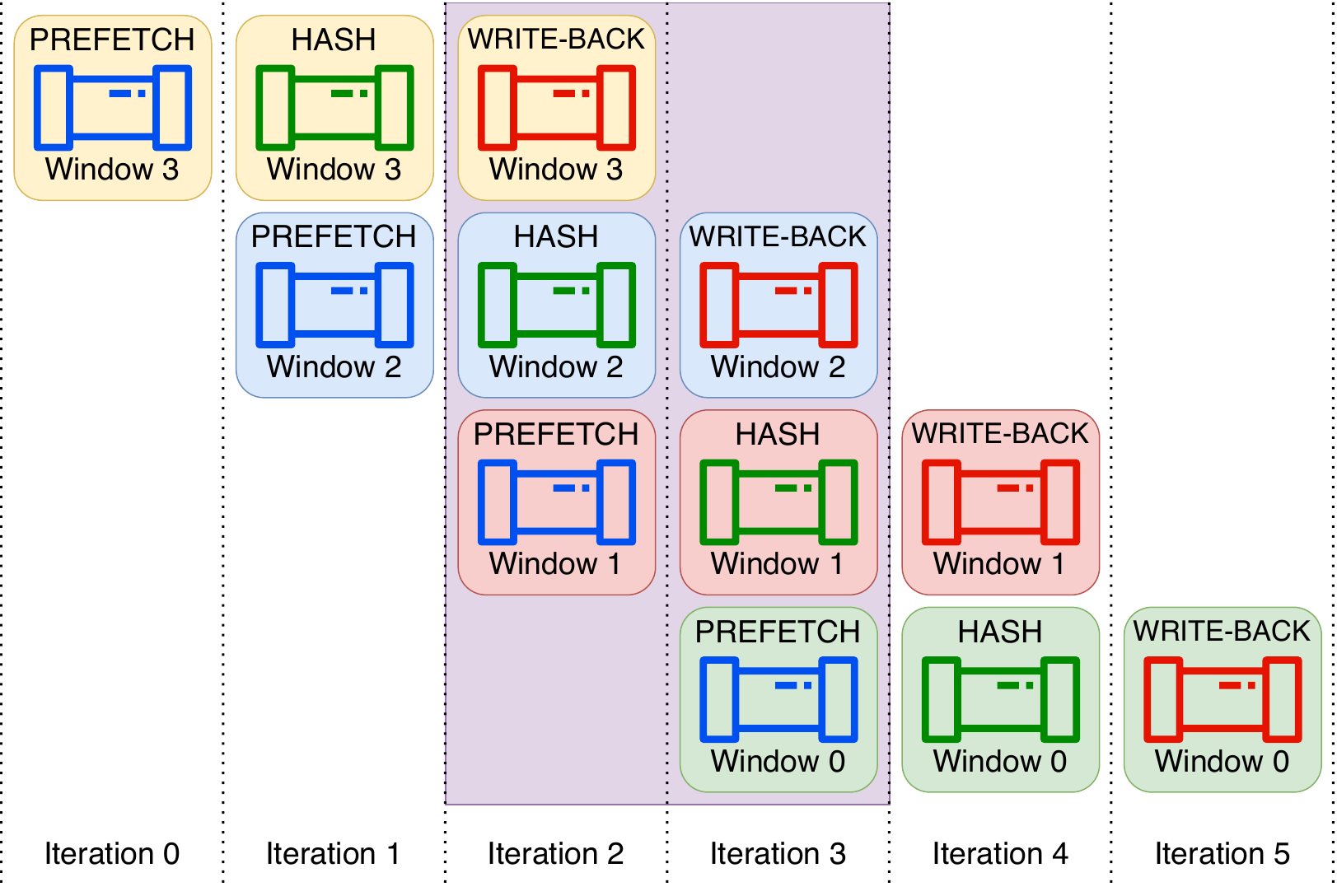}
%\vspace{-0.5em}
	\caption[SMASH Pipeline Stages]{\label{fig:smash_pipeline} SMASH Pipeline Stages}
\end{figure}

Previous versions describe how \textit{atomic hashing} exposes more parallelism and how \textit{tokenization} balances workload across compute core.
Next, we describe an optimization to increase resource utilization, where we adopt pipelining at the phase level. We divide the local memory and \acrshort{SPAD} into $2$ equal parts.
In the first iteration, the first window is processed by the prefetching phase. In the next iteration, that hashing phase processes the previously prefetched window and the prefetching phase works on the next upcoming window.
Once the hashing phase is completed, the write-back phase starts moving data out of the \acrshort{SPAD}. The hashing phase starts working on the next window, preemptively loaded by the prefetching phase (Figure~\ref{fig:smash_pipeline}).
% \vspace{-0.70em}
% \vspace{-2.2em}
As compared to our previous \acrshort{SMASH} versions, where at each stage only one phase is active, this phase enables all three phases to be active simultaneously.
This comes at the cost of increased resources required in terms of local memory and \acrshort{SPAD} memory.
Despite the increased resource utilization, with \acrshort{SMASH} $V3$ pipelining, we were able to obtain a $1.3\times$ speedup as compared to \acrshort{SMASH} $V2$, for synthetic datasets.

\section{SMASH Evaluation}
\label{sec:eval}

We designed \acrshort{SMASH}, a SpGEMM kernel implementation, to expose the performance improvements provided by the custom accelerator architecture. We compare its performance to Intel's MKL implementation on a dual socket server (with Intel Xeon E5-2630), as well as against NVIDIA's A100 GPU.

We evaluate the performance of our \acrshort{SMASH} \acrshort{SPGEMM} kernel implementation on synthetic, as well as real-world, datasets.
For synthetic datasets, we chose the \acrshort{RMAT}, as it produces datasets possessing a non-zero power-law  distribution~\cite{chakrabarti2004r, srinivasan2016dynamic}, making it harder to find patterns in the non-zero values.

\begin{figure}[htbp]
	\centering
	\includegraphics[width=0.98 \textwidth]{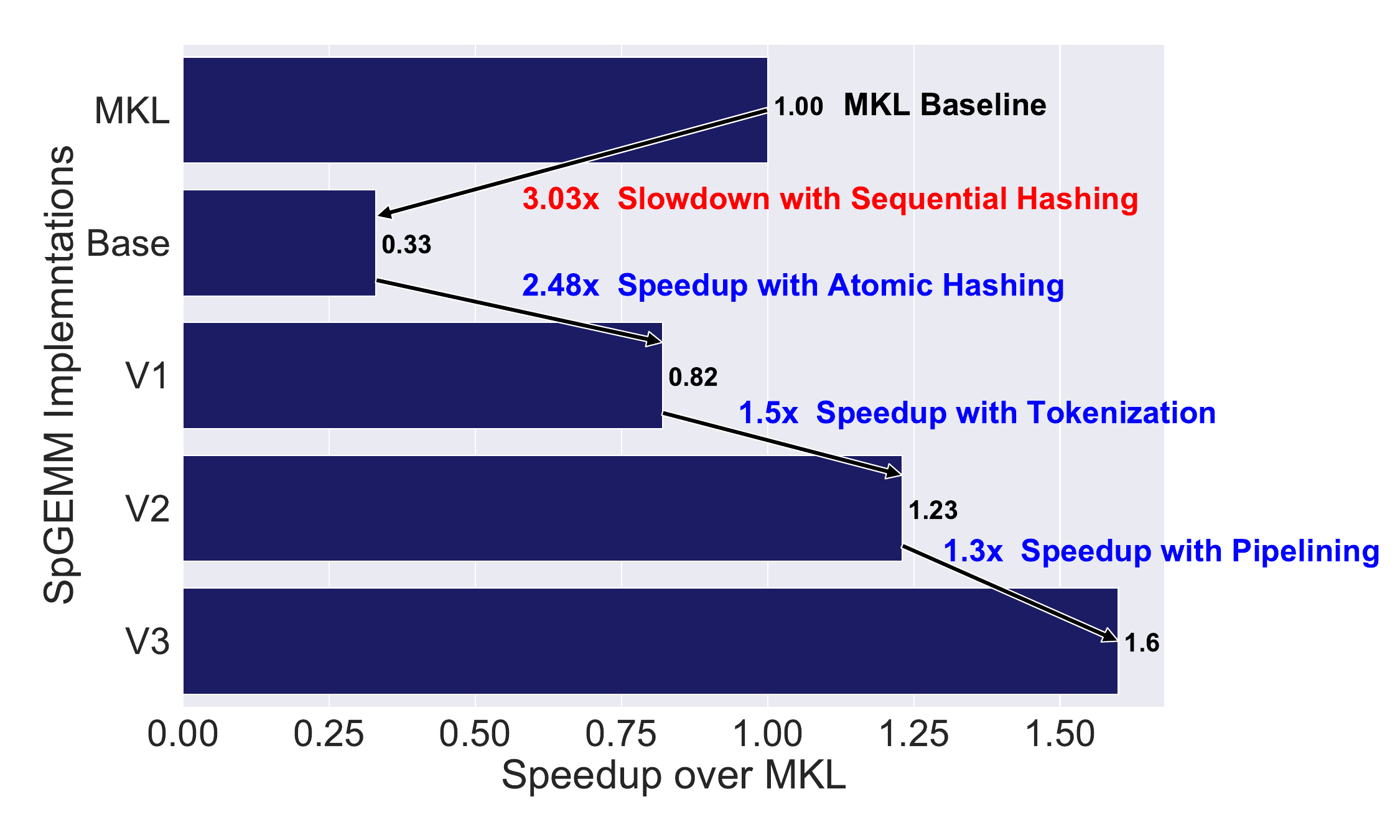}
% \vspace{-2em}
	\caption[Speedup Matrix]{\label{fig:version_speedup} Speedup over MKL, as compared to different versions of SMASH exploiting various architectural features.}
\end{figure}
% \vspace{-0.5em}

For real-world datasets, we experimented with datasets from the Stanford Network Analysis Platform (SNAP).
For all our experiments, we compare the performance of a single CPU core to a single compute core, a single Intel CPU with $8$ cores, to $8$ custom accelerator cores, a dual-socket server with $16$ compute cores, and an A100 GPU to a system with $64$ compute cores.
For our first experiment, we compare the speedup obtained for our \acrshort{SMASH} implementations on $16$ compute cores, and compare against Intel MKL's performance on a dual-socket server, as seen in~Figure~\ref{fig:version_speedup}.
Our base implementation of \acrshort{SMASH} ends up with $3.03\times$ slowdown, but after iterative optimizations, exploiting various architectural features, we end up with an average speedup of $1.6\times$ over MKL for synthetic datasets.

\begin{figure}[htbp]
	\centering
	\includegraphics[width=0.80 \textwidth]{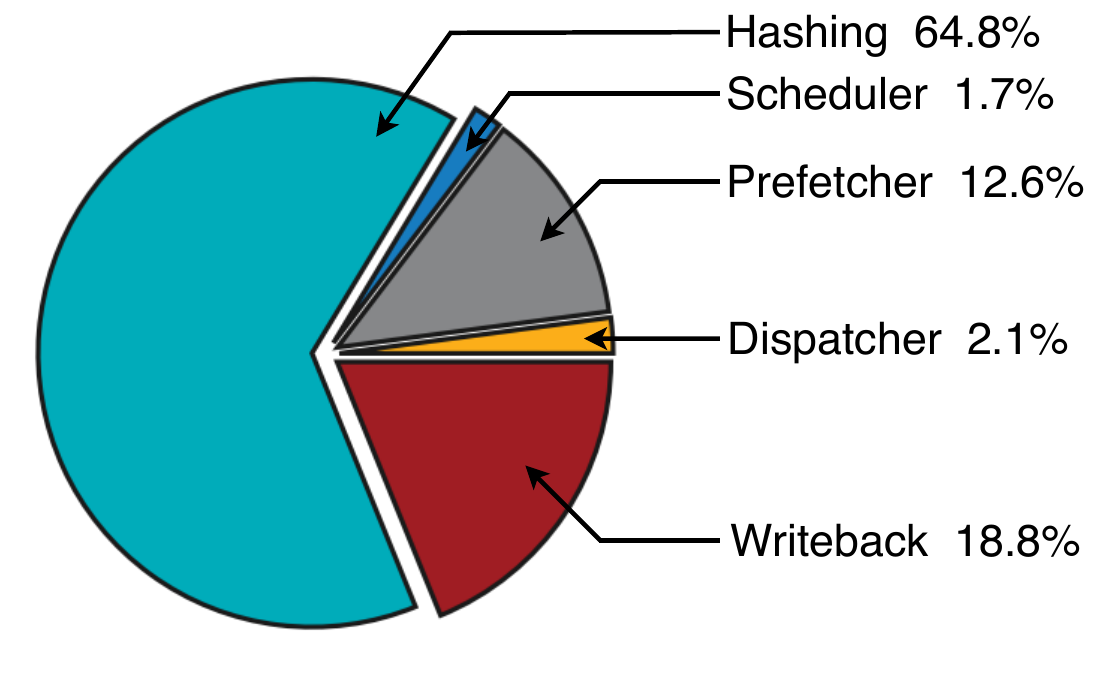}
	\caption[Cycle Distribution]{\label{fig:cycle_dist} Cycle consumption breakdown of SMASH phases}
\end{figure}

\begin{figure}[htbp]
	\centering
	\includegraphics[width=0.65 \textwidth]{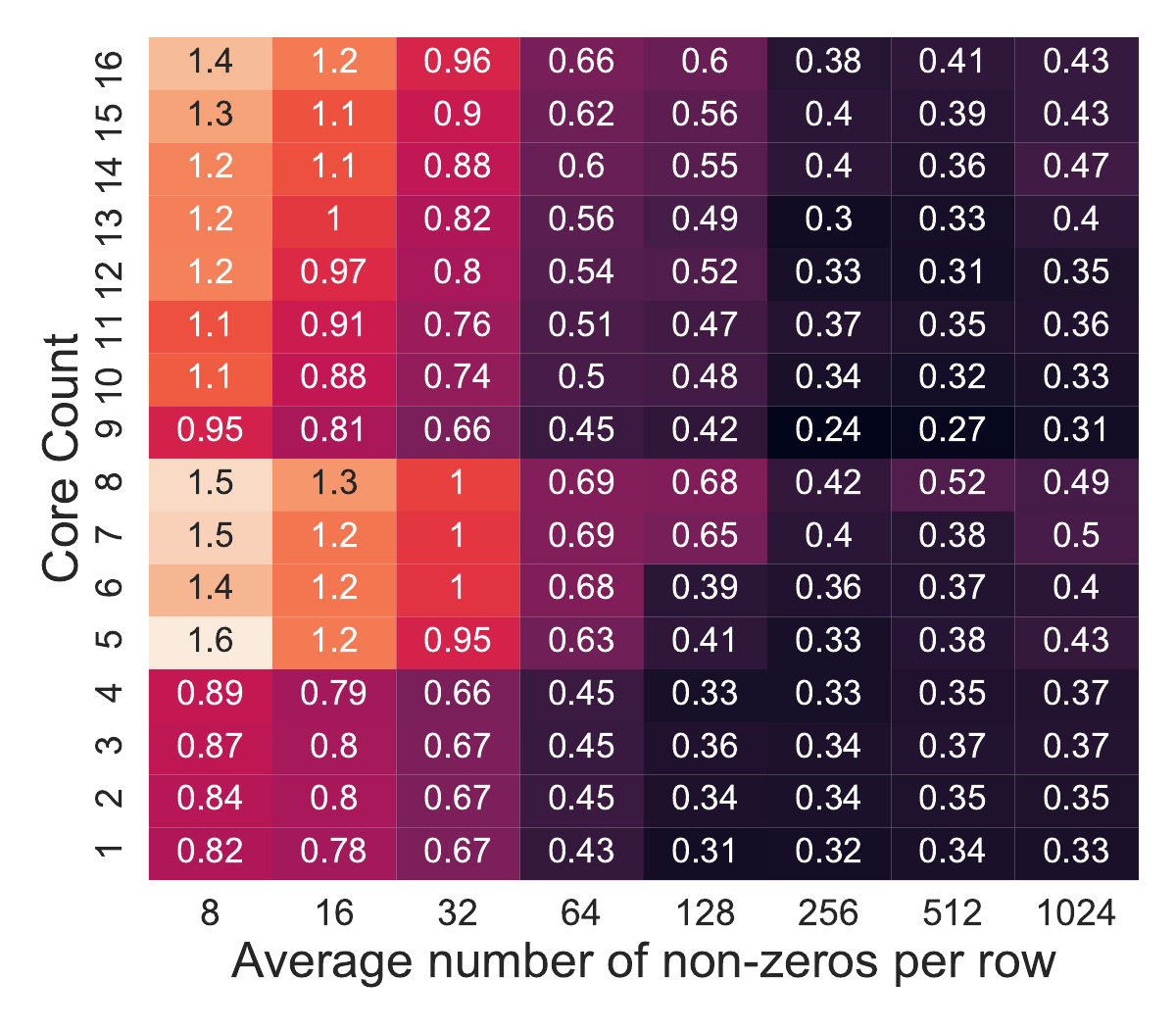}
	\vspace{-1.5em}
	\caption[Speedup Matrix]{\label{fig:speedup_matrix} Speedup over MKL, which varies as a function of matrix density and core count.}
\end{figure}

Next, we focus on the ``Tokenization'' optimization. Tokenization of the hashing phase led to better workload balance between threads. With tokenization, almost all threads have near-perfect utilization, leading to an average  $1.5\times$ speedup over the non-optimized version.
% \begin{figure}[htbp]
% 	\centering
% 	\includegraphics[width=0.98 \textwidth]{figures/chp_05/smash/results/individual_thread_utilization_unbalanced.pdf}
% 	% \vspace{-2em}
% 	\caption[Unbalanced Distribution]{\label{fig:threads_unbalanced} Thread utilization plots for unbalanced workload}
% \end{figure}
% % \vspace{-2.5em}
% \begin{figure}[htbp]
% 	\centering
% 	\includegraphics[width=0.98 \textwidth]{figures/chp_05/smash/results/individual_thread_utilization_balanced.pdf}
% 	% \vspace{-1.5em}
% 	\caption[Balanced Distribution]{\label{fig:threads_balanced} Thread utilization plots for balanced workload}
% \end{figure}
% \vspace{-1.5em}
We next analyzed the performance improvements provided by pipelining. Ideally, if a workload is divided into $3$ stages of a pipeline, the highest speedup achievable is $3$. For this case to hold true, the work across each stage of the pipeline would need to be completely balanced.
In our case, the prefetcher consumes $12.1\%$ of cycles, the hashing phase consumes $64.8\%$ of cycles, while the writeback phase consumes $18.8\%$ of overall cycles, as shown in Figure~\ref{fig:cycle_dist}.
Despite this imbalance of cycles taken by each of the phases, we obtained a $1.3\times$ speedup over the non-pipelined version of \acrshort{SMASH} for the synthetic datasets.
%\vspace{-0.5em}
We also performed scaling experiments by measuring the performance improvements as a function of both the number of cores, as well as the matrix density (this experiment utilizes the synthetic dataset).
Figure~\ref{fig:speedup_matrix} plots the density of the input matrix on the X-axis and the number of cores utilized on the Y-axis. Each number in this heat map is representative of the speedup acquired by \acrshort{SMASH} V3 over MKL with the same number of cores.
This plot indicates that \acrshort{SMASH} outperforms the MKL implementation for sparse matrices at higher core counts.
% \vspace{-1em}
Finally, we compare the performance of real-world datasets from the SNAP library against an MKL single socket, an MKL dual-socket, and a cuSPARSE single GPU A100.
We obtain an average speedup of $1.20\times$ over a single socket MKL kernel, $1.29\times$ speedup over a dual-socket MKL kernel, and $1.04\times$ speedup over the cuSPARSE kernel.

\section{SMASH Summary}
\label{sec:conclusion}
SpGEMM workloads are memory-intensive workloads that possess highly irregular memory access patterns. In this work, we presented \acrshort{SMASH}, a scalable \acrshort{SPGEMM} kernel implementation targeting a custom graph accelerator. Our $3$ iterative optimizations exploit the architectural features of the graph accelerator and provide $2.48\times$, $1.5\times$ and $1.3\times$ speedups, respectively. Our atomic hashing optimization, tokenization, and pipelining of \acrshort{SMASH} kernels provided us with an average of $1.20\times$, $1.29\times$, and $1.04\times$ speedup over MKL single socket, MKL dual-socket, and GPU A100 hardware, respectively.

\chapter{Hardware Acceleration of GNNs}
\label{chp6}
% \section{NeuraChip}

Graph Neural Networks (GNNs) are emerging as a formidable tool for processing non-euclidean data across various domains, including bioinformatics, financial networks, energy networks, telecommunication and social network analysis. 
Despite their effectiveness, their adoption has not been pervasive because of scalability challenges associated with large-scale graph datasets, particularly when leveraging message passing, posing significant computational bottlenecks.
This class of large-scale workloads exhibits irregular sparsity patterns, resulting in unbalanced utilization of computational resources.

% GitHub\footnote{\url{https://github.com/NeuraChip/neurachip}}.

%  ___       _             
% |_ _|_ __ | |_ _ __ ___  
%  | || '_ \| __| '__/ _ \ 
%  | || | | | |_| | | (_) |
% |___|_| |_|\__|_|  \___/ 

\section{Key Bottlenecks in Accelerating GNNs}

% \review{Modern trends in Big Data have witnessed an increase in data sparsity, along with an increase in data set size~\cite{rong2019feature}. In 2022, Facebook reported $2.89$ billion monthly active users, with studies suggesting an average of 338 friends per user~\cite{smith_2020}. 
% The resulting adjacency matrix of the Facebook user graph would approach $99.999\%$ sparsity.}
% Graph analytics of such highly sparse datasets push the limits of the current computing infrastructure and expose inherent fallacies exhibited by traditional architectures.  Sparse graph workloads are dominated by highly irregular and uncoalesced memory access patterns.
Deep Neural Networks have proven to be powerful models for solving problems that rely on data with an underlying Euclidean or grid-like structure~\cite{yosinski2014transferable}, such as computer vision, natural language processing, and audio vision.
In contrast, Graph Neural Networks (GNNs) have emerged as powerful frameworks for handling non-Euclidean data (e.g., social networks on the scale of billions~\cite{smith_2020}), achieving impressive performance across various domains such as social science, chemistry, and bioinformatics~\cite{xu2018powerful}.
However, the computational complexity of GNNs, especially when working with ultra-sparse, large-scale graph datasets, poses challenges due to architectural limitations of traditional hardware (i.e., CPUs / GPUs)~\cite{peng2023maxk}.
Moreover, GNNs predominantly adopt a recursive neighborhood aggregation methodology, in which each node aggregates the feature vectors of its neighboring nodes to derive its own updated feature vector.
The scalability of message passing in GNNs, when applied to large graph structures, poses a significant bottleneck, especially as the size of the graphs surpasses the capacity of on-chip memory hierarchies in today's CPUs and GPUs~\cite{shivdikar2023gme}. 
This leads to redundant and time-consuming memory transactions to fetch data from the main memory~\cite{baruah2021gnnmark, peng2024maxk}.

We identify the following three key bottlenecks for GNN workloads:
\begin{enumerate}
\item Diverse Data Dependence Patterns: GNN workloads feature multiplication and accumulation operations, each demonstrating unique data dependency patterns.
\item Poor Hardware Resource Utilization: Compute units suffer from data starvation and load imbalance due to irregular sparsity patterns exhibited by large input graphs.
\item Memory bloat: The matrix multiplication methods generate a large number of intermediate partial products that require efficient merging to avoid redundant accesses to higher level memory.
We further elaborate on each of these three critical bottlenecks identified for GNN workloads.
\end{enumerate}

\begin{figure}[htbp]
	\centering
	\includegraphics[width=0.98\textwidth]{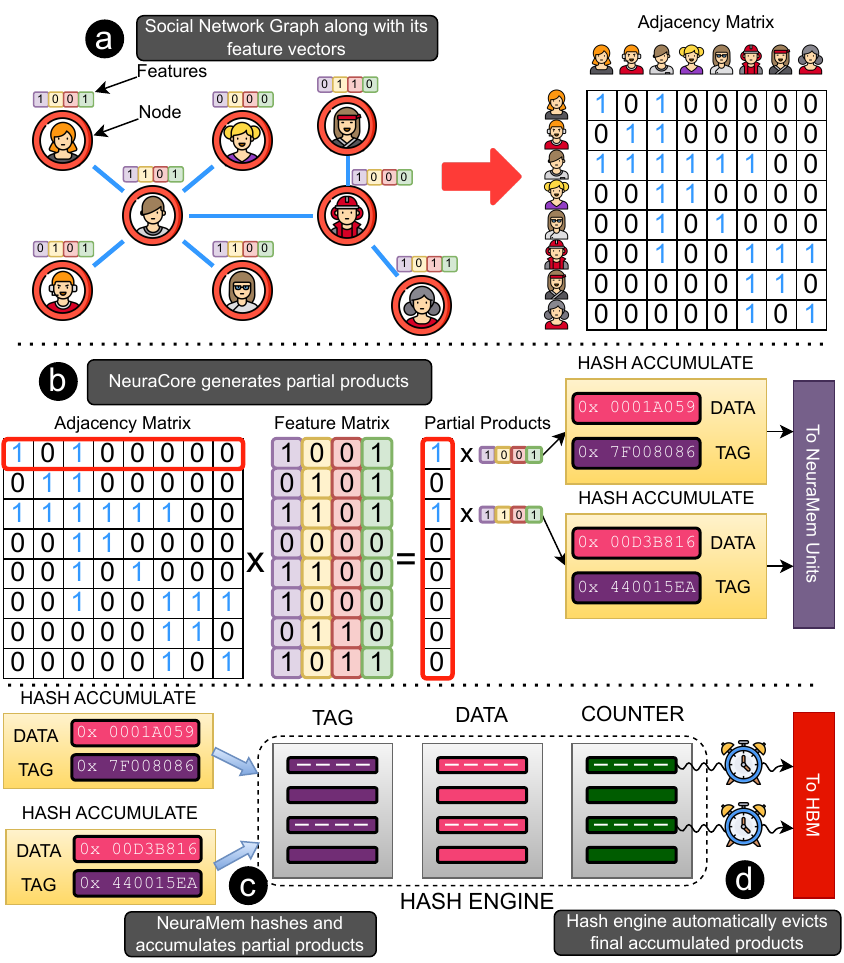}
	% \vspace{-1.0em}
	\caption{NeuraChip overview, illustrating aggregation phase of graph convolution on a social network graph (a). NeuraCore generates partial products based on input graph (b). NeuraMem accumulates partial products (c) and writes back to HBM (d).}
	\label{fig:teaser}
% \vspace{-1.8em}
\end{figure}

\begin{figure*}[htbp]
	\centering
	\includegraphics[width=0.99\textwidth]{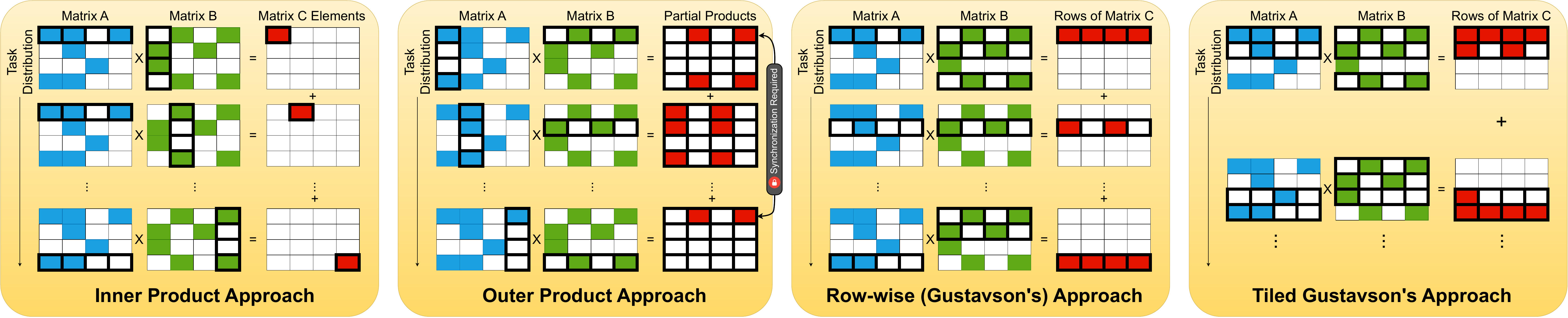}
	% \vspace{-2.2em}
	\caption{Various approaches to matrix multiplication, each showcasing different degrees of data reuse for input and output matrices.}
	\label{fig:matmul_types2}
% \vspace{-1.7em}
\end{figure*}

\textbf{Diverse Data Dependence Patterns:} The process of neighborhood aggregation in graphs can be split into a multiplication stage, followed by a merge/reduction stage.
The multiplication stage creates partial products by multiplying the adjacency matrix of the input graph with the feature matrix.
The reduction stage accumulates (i.e., merges) the partial products to update the node feature vectors.
The multiplication stage's operands depend on data stored in the high-bandwidth memory (HBM)~\cite{pal2018outerspace}. 
In contrast, the reduction stage's operands depend on data located within the on-chip memory.
Utilizing a singular computational resource for both multiplication and accumulation operations proves suboptimal, as mapping multiplication operations on computing resources tends to compromise the efficiency of mapping the accumulation operations (due to varying data dependency patterns).
%$ To effectively address these distinct data dependencies during these stages, we present two dedicated components in our custom accelerator, tailored to specific data dependencies: NeuraCore, which executes multiplication tasks, and NeuraMem, which accumulates the intermediate partial products generated by NeuraCore (\figref{fig:teaser}).

% \textbf{Asymmetric Hardware Resource Utilization:} The NeuraCore frequently encounters delays arising from input data availability, while NeuraMem experiences slowdowns when nearing the full capacity of its on-chip memory hashtable. To enhance efficiency, our architecture strategically allocates multiplication tasks to NeuraCore, capitalizing on the data locality of input graph data. Simultaneously, we employ a \textit{dynamically reseeding hashing function} in NeuraMem. This approach evenly distributes the workload of accumulating partial products across all components of NeuraMem, effectively addressing the imbalance in partial product accumulation.

\textbf{Poor Hardware Resource Utilization:} The multiplication and accumulation stages are characterized by distinct architectural implications.
The multiplication stage typically stalls due to data starvation (stalls accessing the input graph and feature matrix elements), whereas the accumulation stage suffers from uneven partial product distribution due to the sparsity patterns. 
Prior accelerators~\cite{zhang2021gamma, sarkar2023flowgnn} have adopted look-ahead buffers for prefetching data, aiming to prevent compute stalls caused by data starvation. While these solutions reduce compute stalls, aggressive prefetching leads to cache pollution as redundant data resides in the cache~\cite{seshadri2012evicted}.
These issues can be addressed using two strategies, each catering to their respective problems.
(a) \textit{Multiplication mapping:} Implementing a \textit{tiled row-wise product approach} to partition the computation into distinct tasks, which are then dynamically allocated to NeuraCore (computing elements), depending on its utilization.
The row-wise product method, also known as Gustavson's algorithm, is a popular choice among recent accelerators such as Gamma~\cite{zhang2021gamma}, MatRaptor~\cite{srivastava2020matraptor}, and SPADA~\cite{li2023spada} as this approach has shown high-efficiency when targeting sparse matrix computations in the aggregation stages of GNNs.
Developing dedicated components for multiplication enables mapping these operations to NeuraCore, independent of the accumulation stage, thus leveraging the locality of the input data. 
(b) \textit{Accumulation Mapping:} Using a dynamic reseed hash-based mapping agnostic to sparsity patterns.
This allows even distribution of the partial products among the NeuraMem (on-chip memory) accumulation units.

\textbf{Memory Bloat:} Incorporating the row-wise product approach enhances input data locality but creates a large number of partial products. 
 Table~\ref{tab:bloat} presents memory bloat for SpGEMM workload across various sparse graph datasets. We define bloat percent as shown in Equation~\ref{eq:bloat}
\begin{equation}
\label{eq:bloat}
\mathrm{Bloat\ Percent} = \frac{pp_{\mathrm{interim}} - nnz_{\mathrm{output}}}{nnz_\mathrm{output}} * 100
\end{equation}
% where $pp_{\mathrm{interim}}$ represents intermediate partial products and $\mathrm{output}$ represents total non-zeros in the output product matrix.
wherein $pp_{\mathrm{interim}}$ denotes the count of intermediate partial products and $nnz_{\mathrm{output}}$ signifies the count of non-zero elements in the resultant product matrix.
Although tiling the computation partially addresses this issue, it does not fully resolve it.
Prior solutions such as Gamma~\cite{zhang2021gamma} have relied on large explicitly managed cache systems similar to FiberCache~\cite{zhang2021gamma}, which consumes up to $72\%$ of the total chip's area. The memory bloat issue can be addressed using a rolling eviction strategy, which automatically evicts a partial product from the on-chip memory once all contributing partial products have been fully accumulated. 
We enable a strategy using an eviction counter integrated with the on-chip memory hashtables.

\begin{table}
\centering
\caption{SpGEMM bloat analysis across hyper-sparse graph datasets}
\label{tab:bloat}
\begin{tabular}{l | x{0.7in} x{0.695in} x{0.695in} x{0.695in}}
\arrayrulecolor{black}\toprule
   \textbf{Dataset} & \textbf{Node Count} & \textbf{Edge Count} & \textbf{Sparsity (\%)} & \textbf{Bloat Percent} \\ [0.5ex] 
\arrayrulecolor{black}\toprule
\arrayrulecolor{black}\toprule

2cubes\_sphere & $101492$ & $1647264$ & $99.9840$ & $205.87$ \\
\arrayrulecolor{black!20}\midrule
ca-CondMat & $23133$ & $186936$ & $99.9651$ & $75.23$ \\
\arrayrulecolor{black!20}\midrule
cit-Patents & $3774768$ & $16518948$ & $99.9999$ & $19.32$ \\
\arrayrulecolor{black!20}\midrule
email-Enron & $36692$ & $367662$ & $99.9727$ & $68.90$ \\
\arrayrulecolor{black!20}\midrule
filter3D & $106437$ & $2707179$ & $99.9761$ & $326.34$ \\
\arrayrulecolor{black!20}\midrule
mario002 & $389874$ & $2101242$ & $99.9986$ & $99.43$ \\
\arrayrulecolor{black!20}\midrule
p2p-Gnutella31 & $62586$ & $147892$ & $99.9962$ & $10.21$ \\
\arrayrulecolor{black!20}\midrule
poisson3Da & $13514$ & $352762$ & $99.8068$ & $297.92$ \\
\arrayrulecolor{black!20}\midrule
scircuit & $170998$ & $958936$ & $99.9967$ & $66.13$ \\
\arrayrulecolor{black!20}\midrule
web-Google & $916428$ & $5105039$ & $99.9994$ & $104.27$ \\
\arrayrulecolor{black!20}\midrule
amazon0312 & $400727$ & $3200440$ & $99.9980$ & $97.21$ \\
\arrayrulecolor{black!20}\midrule
cage12 & $130228$ & $2032536$ & $99.9880$ & $127.23$ \\
\arrayrulecolor{black!20}\midrule
cop20k\_A & $121192$ & $2624331$ & $99.9821$ & $327.07$ \\
\arrayrulecolor{black!20}\midrule
facebook & $4039$ & $60050$ & $99.1519$ & $2872.80$ \\
\arrayrulecolor{black!20}\midrule
m133-b3 & $200200$ & $800800$ & $99.9980$ & $26.93$ \\
\arrayrulecolor{black!20}\midrule
offshore & $259789$ & $4242673$ & $99.9937$ & $205.45$ \\
\arrayrulecolor{black!20}\midrule
patents\_main & $240547$ & $560943$ & $99.9990$ & $14.18$ \\
\arrayrulecolor{black!20}\midrule
roadNet-CA & $1971281$ & $5533214$ & $99.9999$ & $35.75$ \\
\arrayrulecolor{black!20}\midrule
webbase-1M & $1000005$ & $3105536$ & $99.9997$ & $36.02$ \\
\arrayrulecolor{black!20}\midrule
wiki-Vote & $8297$ & $103689$ & $99.8494$ & $148.09$ \\
     \arrayrulecolor{black}\bottomrule
% \vspace{-2.7em}
\end{tabular}
\vspace{0.2mm}

\scriptsize

\centering
% $^{*}$Values represent the count of components/elements across the entire NeuraChip accelerators for three different tile configurations.

\end{table}

 \noindent
The work here develops NeuraChip, an innovative GNN spatial accelerator featuring a decoupled computation pipeline. Decoupling multiplication and accumulation operations into dedicated components, we optimize data reuse through strategic mapping.  We explore an adaptive hash-based compute mapping. Our approach introduces a flexible, dynamic reseeding hash-based compute mapping (DRHM) tailored for GNN workloads. DRHM benefits from the constant lookup times characteristic of hash functions, while also mapping tasks evenly across all computing resources by generating a new seed at predetermined intervals of computation.
We also explore how to use rolling evictions in order to address memory bloat. We explore on-chip hash tables to manage partial products, effectively reducing memory congestion caused by their generation.

\section{Fundamentals of GNN Workloads}

Graph neural networks (GNNs) are capable of extracting important features, such as structural motifs (i.e., arrangements of nodes, edges, and metadata)~\cite{jin2020gralsp}, learning not only the individual characteristics of each element (i.e., a node in the graph), but also the interconnections (i.e., the interrelationships between nodes) between elements~\cite{zhou2020graph}. GNNs use convolution operations to extract various features from the graph~\cite{kipf2016semi}. 
The methodology employed is called {\em neighborhood aggregation}, where the final feature vector for each vertex is computed by iteratively aggregating and transforming the input feature vectors of adjacent vertices~\cite{zhou2020graph}. 
This process includes two steps, which are called the {\em aggregation and combination stages}. 
This process is carried out iteratively, and after $k$ iterations through these stages, the resultant feature vector of the target vertex signifies the distinct structural data of the vertex's $k$-hop vicinity~\cite{wu2020comprehensive}.

For instance, a Graph Convolutional Network (GCN) is one such GNN model. 
Equation~\ref{eq:gcn} below computes the forward propagation for a single layer in a GCN.
\begin{equation}
\label{eq:gcn}
    X^{(l+1)} = \sigma(AX^{(l)}W^{(l)})
\end{equation}
where $A$ represents the adjacency matrix of the graph, where each row lists the interconnections of a vertex to all other vertices in the graph. $X^{(l)}$ refers to the input feature vectors of every vertex in the $l^{th}$ layer of matrix $X$. 
% In other words, each column of $X$ signifies a feature, and each row embodies the feature vector of a specific vertex.
$W$ contains the GNN's model parameters, which are obtained through model training. $\sigma()$ represents the non-linear activation function, for instance, ReLU (Rectified Linear Unit).

% A variety of GNN models have been designed to serve specific purposes, including node or edge prediction, graph clustering, and recommendation modeling. A few examples of these application-oriented GNN models include the Graph Isomorphism Network (GIN), Graph Attention Network (GAT), GraphSAGE (SAGE), Principal Neighbourhood Aggregation (PNA), and Deep Graph Network (DGN).
% Table~\ref{tab:gnn_equations} lists GNN models and their corresponding message-passing equations.

% \input{tables/gnn_equations}

\section{Architectural Implications of GNN Workloads}

\textbf{Aggregation Stage}: 
The aggregation stage in GNN workloads is critical for capturing the structural information of graphs. It involves gathering and summarizing information from a node's neighbors, which can be a challenging task given the irregular data structures common in graph-based data.
This is typically computed with sparse matrix multiplication kernels.
% With input graphs exhibiting over $99\%$ sparsity, this stage often exhibits random memory access patterns and may lead to non-coalesced memory accesses, posing challenges for traditional architectures optimized for sequential data access.
% Furthermore, the skewed sparsity patterns cause uneven workload distribution across compute resources, leading to further performance degradation.
Given the high level of sparsity in input graphs, typically above $99\%$, this stage is characterized by random access patterns in memory, which presents a challenge for traditional architectures that are more suited for linear data access. Additionally, the skewed sparsity patterns often lead to workload imbalance on computing resources, which can impact performance efficiency.

\textbf{Combination Stage}: 
The combination stage in GNNs involves the integration of node features with neighborhood information. This process is computationally intensive and typically comprises dense matrix multiplications, nonlinear activations, and dimensionality reduction operations. Architecturally, this stage demands high memory bandwidth and efficient data reuse mechanisms to handle large matrices. It also necessitates a balance between compute utilization and memory access, as the combination of features from large graphs can lead to memory bottlenecks.
While prior accelerators~\cite{zhang2021gamma, zhang2020sparch, li2023spada} often focus on sparse matrix multiplication tasks, they do not adequately address dense workload demands. Our NeuraChip accelerator model provides a more generalized solution, addressing the needs of both sparse graph computations and dense workloads. This approach positions NeuraChip as a versatile GNN accelerator, adept at handling both the aggregation and combination stages.

\section{Sparse Matrix Multiplication: Algorithmic Overview}
\begin{figure}[htbp]
	\centering
	\includegraphics[width=0.98\textwidth]{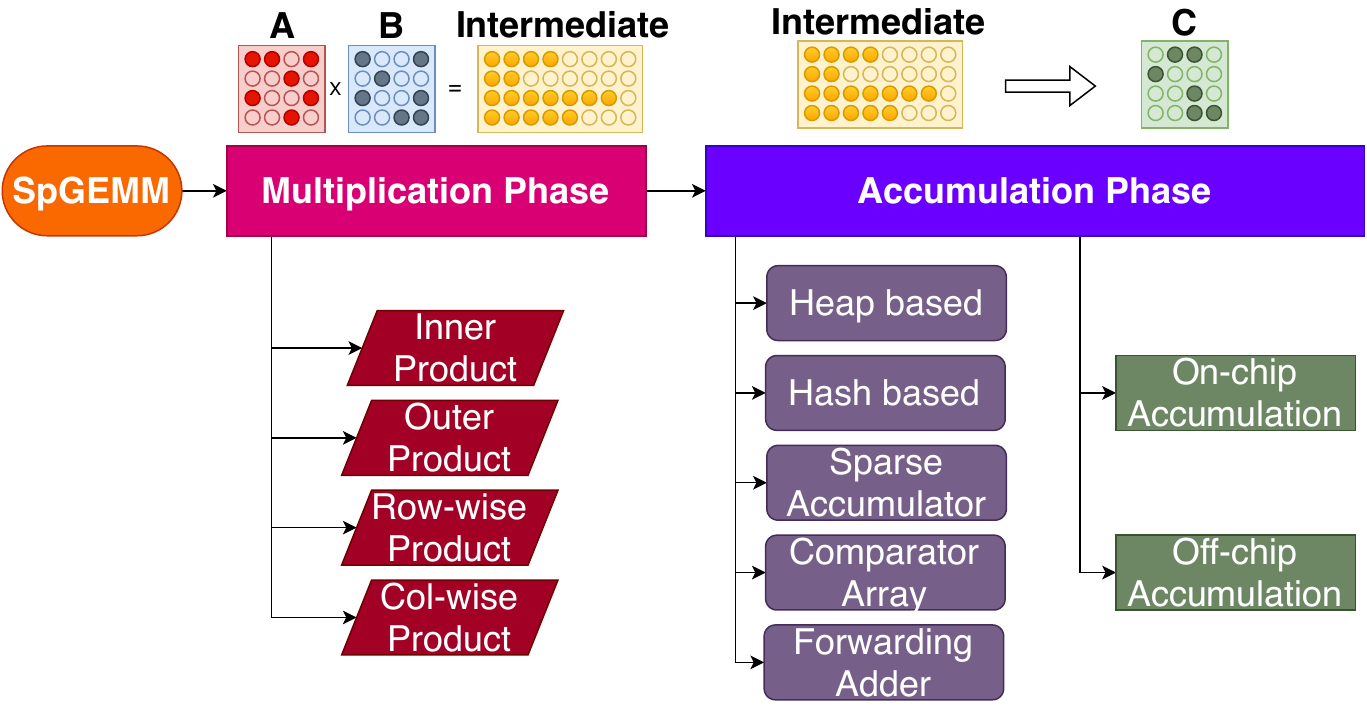}
	% \vspace{-1.0em}
	\caption{Various methods employed in multiplication and accumulation stages.}
	\label{fig:spgemm_stages}
% \vspace{-2.0em}
\end{figure}

The Sparse General Matrix-Matrix Multiplication (SpGEMM) kernel execution is characterized by two main stages: the multiplication stage and the accumulation stage as visualized in \figref{fig:spgemm_stages}. The implementation variations in these stages lead to distinct SpGEMM algorithms. We describe the four approaches to execute the initial multiplication stage, as illustrated in Figure~\ref{fig:matmul_types2}. These approaches vary in their memory access patterns and the level of parallelism they expose.

The inner product approach, incorporated in InnerSP~\cite{baek2021innersp} computes elements of the output matrix directly but is hindered by inefficient input reuse. Conversely, the outer product approach, utilized in OuterSPACE accelerator~\cite{pal2018outerspace} is hampered by suboptimal output locality due to the creation of numerous batches of intermediate partial product matrices~\cite{zhang2020sparch}. Our research adopts the row-wise multiplication approach (i.e., Gustavson's algorithm), selected for the extensive parallelism it provides. Notably, this approach efficiently avoids the memory bloat issue associated with handling numerous intermediate partial products~\cite{zhang2020sparch}.

The subsequent stage, known as the accumulation stage, merges the generated intermediate partial products. Various accumulation methods include heap-based~\cite{azad2016exploiting}, hash-based~\cite{nagasaka2017high}, sparse accumulator (SPA) based~\cite{gilbert1992sparse}, comparator array based~\cite{zhang2020sparch}, and Forwarding Adder Network (FAN) based~\cite{qin2020sigma,stift-jetc23}, among others (illustrated in~\figref{fig:spgemm_stages}). This stage can also be subdivided into on-chip and off-chip accumulation, based on the utilized memory hierarchy.
NeuraChip merges partial products using on-chip accumulation to reduce redundant main memory data fetches. 
For sparse matrices with skewed non-zero distributions, the on-chip accumulation stage can result in uneven workload distribution, a factor that significantly impacts the overall performance and efficiency of SpGEMM operations.
% This uneven distribution of workloads across the computational resources poses challenges in achieving optimal parallelism and efficiency. To address this issue and streamline the process, the subsequent subsection will delve into the implementation of hash-based mapping in NeuraChip, showcasing how it effectively balances workloads and enhances performance in sparse matrix multiplication.

% \begin{figure}[tbp]
% 	\centering
% 	\includegraphics[width=0.38\textwidth]{figures/chp_05/neurachip/introduction/matmul_types.pdf}
% 	\caption{Four approaches to matrix multiplication, each showcasing different degrees of data reuse for input and output matrices.}
% 	\label{fig:matmul_types}
% \end{figure}

\section{Mapping Algorithms: Design and Requirements}

Mapping algorithms play a crucial role in efficiently handling computational tasks, particularly in scenarios involving sparse data structures such as those found in Graph Neural Networks (GNNs). These algorithms are tasked with assigning tasks or data elements to computational nodes or memory locations. The key requirements for effective mapping algorithms include:

\textbf{Consistency}: The algorithm must consistently map the same index to the same node. This ensures correctness in data processing.

\textbf{Low Computational Overhead}: The lookup process should be relatively fast, with minimal computational and memory overheads.
This efficiency facilitates cost-effective index matching, streamlining partial product reduction.
% This is essential to maintain overall system efficiency.

\textbf{Sparsity Agnostic}: Regardless of skewed sparsity patterns, the mapping algorithm should remain impartial to these variations. This ensures uniform performance across different data sets~\cite{camacho2008strong}.

Given these requirements, hash-based mapping emerges as a viable solution~\cite{chi2017hashing}. However, traditional hash-based methods such as Round Robin Hashing (or Ring Hashing)~\cite{takenaka2004adaptive} and Prime Number Based Modular Hashing~\cite{bhullar2016novel} have limitations~\cite{cao2000performance}. Neither is fully insensitive to sparsity patterns; a specific set of indices might consistently map to the same node, leading to potential workload imbalance.

An alternative approach is random mapping, which ideally achieves sparsity-agnostic mapping by randomly distributing indices. However, to ensure consistency, this method requires maintaining a large lookup table, which is not practical due to memory constraints.

To address these challenges, we propose a novel approach: Dynamic Reseed Hash-Based Mapping (DRHM)~\cite{malith2024neurachip}. This method is similar to prime modular hashing, but with a significant enhancement. After processing a predetermined set of computations, we reseed the hash function. The updated seed values are then stored in a compact lookup table. This dynamic reseeding ensures that the distribution of indices does not become predictable or skewed, effectively mimicking the sparsity-agnostic property of random hashing.

Dynamic Reseed Hash-Based Mapping strikes a balance between the ideal characteristics of random mapping and the practical limitations of traditional hash-based methods. By only storing seed values rather than the entire mapping of indices, it maintains a small memory footprint. Concurrently, it offers the sparsity-agnostic mapping necessary for handling diverse and skewed data sets efficiently. This method significantly enhances the performance of computational tasks, particularly in environments where data sparsity and distribution can vary widely.

% Consistency
% Lookup time
% Workload Balance

% \subsection{Memory Bloat Issue}..
% Our study employs Gustavson's matrix multiplication method~\cite{gust1978matmul}, also known as the row-wise multiplication technique which has been previously used in \cite{zhang2021gamma,srivastava2020matraptor,li2023spada}. One primary issue associated with sparse matrix multiplication, especially when using the outer-product or row-wise multiplication, is the problem of memory expansion (memory bloating). This problem arises from the generation of intermediate partial products during matrix multiplication, which need to be accumulated for calculating the final results of the matrix. Memory bloating refers to the additional off-chip memory consumption required to store these intermediate products created during the matrix multiplication process. This issue is not addressed by some  of the previously proposed GNN accelerators~\cite{srivastava2020matraptor, zhang2020sparch}, leading to unnecessary memory reads and writes to the main memory. NeuraChip tackles this problem using a technique called on-the-fly hash-accumulation, which allows for the accumulation of intermediate products as they are generated. A full description of our hash-accumulation can be found in Section~\ref{ssec:neuramem}.

% \todo{GCN layers, aggregation phase, combination phase}

% \todo{Explain kernels and sparsity levels}

% \todo{Perhaps including a kernel/runtime breakdown of a common GCN model as motivation/baseline}

% \todo{Focus on the general compilation view, and modified compilation steps to map on a neurachip}

\section{NeuraChip Architecture}
\label{sec:neurachip}

\begin{figure*}[htbp]
	\centering
	\includegraphics[width=0.98\textwidth]{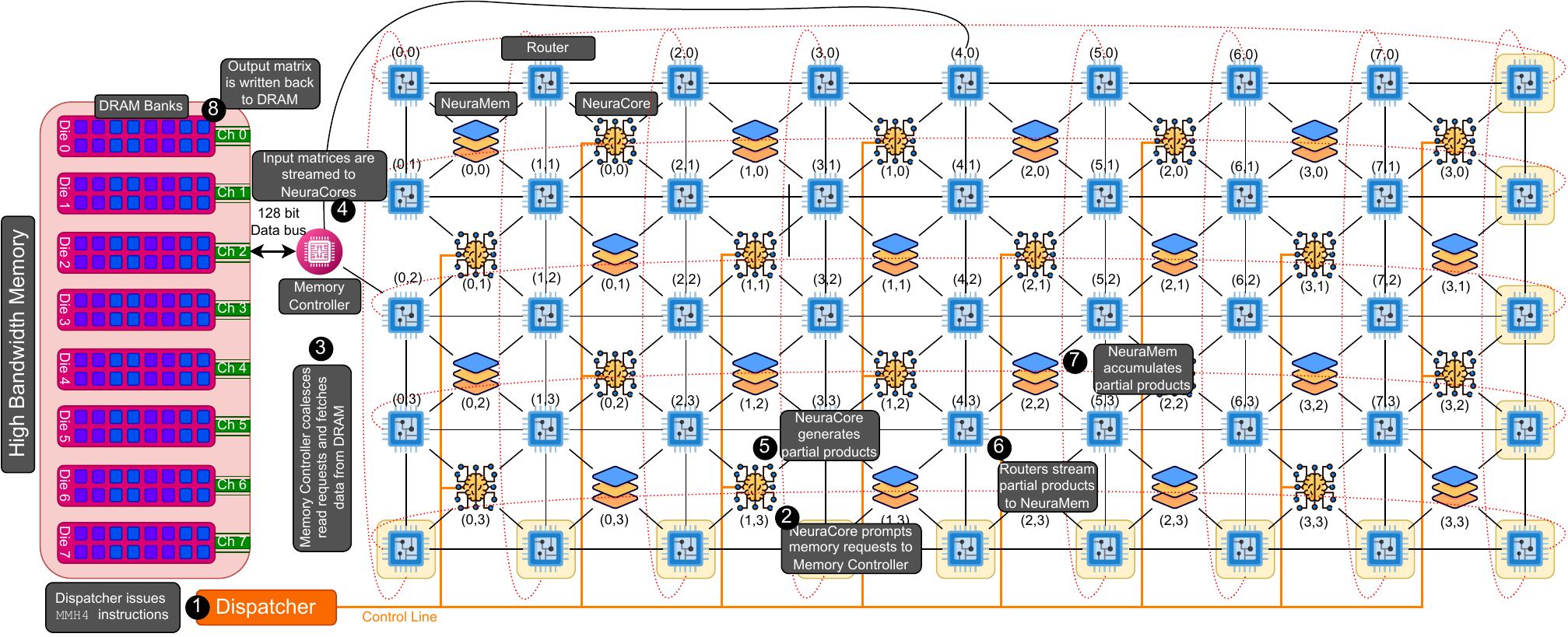}
	% \vspace{-2.2em}
	\caption{Overview of NeuraChip architecture with Tile 16 configuration (16 NeuraCores and NeuraMems per tile with a total of 8 tiles).}
	\label{fig:tile_16}
% \vspace{-1.7em}
\end{figure*}

NeuraChip~\cite{shivdikar2024neurachip} is a decoupled spatial accelerator.
Its two primary components include: i) the NeuraCore and ii) the NeuraMem. 
The NeuraCore is specifically tailored for multiplication tasks, whereas the NeuraMem focuses on accumulating data on-chip~\cite{shivdikar2024neurachip2}. 
They are arranged in an interleaved pattern and connected through a 2D torus network fabric, as shown in Figure~\ref{fig:tile_16}. 
To facilitate efficient communication among these components, on-chip routers have been incorporated. 
NeuraCores and NeuraMems are organized into clusters known as tiles~\cite{shivdikar2024neurachip}. 
The accelerator includes a total of eight tiles, each linked to a single Double Data Rate (DDR) channel. 
Each tile features a memory controller responsible for interfacing with the DRAM banks.

Buffers play a critical role in the functionality of the four major components of our accelerator~\cite{shivdikar2023gme}. 
Both the NeuraCore and the NeuraMem are equipped with instruction buffers~\cite{shivdikar2024neurachip2}. 
Additionally, the on-chip routers incorporate packet buffers, and the memory controllers are fitted with buffers for managing both reading and writing operations.

The incorporation of these on-chip buffers enhances the accelerator's flexibility, allowing it to adapt to diverse sparsity patterns. 
In scenarios where irregular graph structures could lead to network congestion, these on-chip buffers prove beneficial. 
They ensure that the components consistently have instructions to execute, thus avoiding potential delays or bottlenecks in processing.

\subsection{Tiled Gustavson's Multiplication Algorithm}
% \subsection{Tiled Row-wise Multiplication}

\begin{figure}[htbp]
	\centering
	\includegraphics[width=0.98\textwidth]{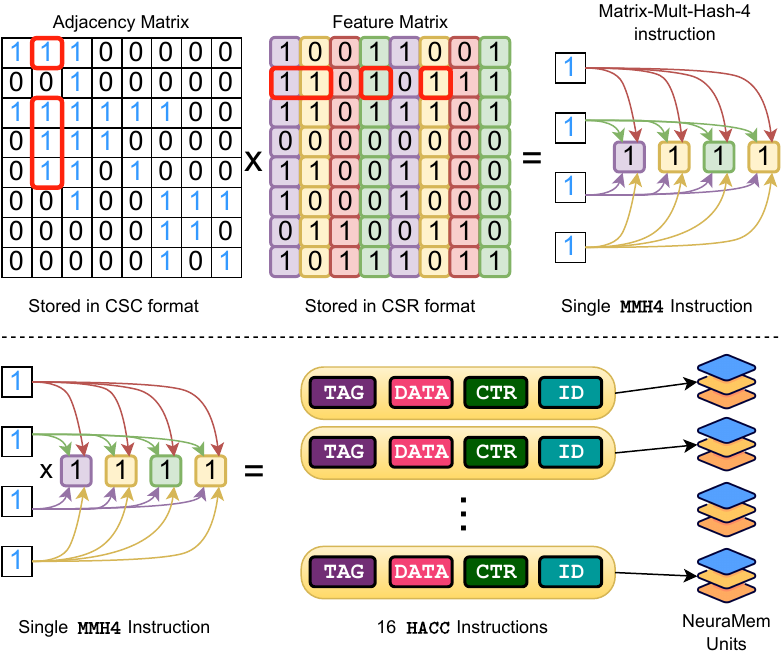}
	% \vspace{-1.2em}
	\caption{Implementation of tiled Gustavson's algorithm using NeuraCore for multiplication and NeuraMem for accumulation.}
	\label{fig:tiled_algorithm}
% \vspace{-1.7em}
\end{figure}

GNNs typically employ two primary layers (phases) in their architecture: the neighborhood aggregation phase, which gathers information from a node's neighbors in the graph, and the combination phase, where a node's representation is updated by integrating its own features with those aggregated from its neighbors~\cite{gama2018convolutional}. 
This discussion focuses on the aggregation phase, which predominantly involves sparse matrix multiplications.

In this \thesis, we implement a modified version of Gustavson's matrix multiplication algorithm~\cite{shivdikar2024neurachip2}. 
Gustavson's algorithm operates on a row-stationary approach, processing the output matrix one row at a time. 
Specifically, it traverses the adjacency matrix row by row, performing a linear combination of these rows as illustrated in~\figref{fig:tiled_algorithm}.

Gustavson's approach multiplies each element in a row of the adjacency matrix with all elements in the corresponding row in the feature matrix that has the same row index as the element's column index. 
Our adaptation enhances Gustavson's method by simultaneously processing multiple rows. 
We execute the multiplication of four rows at a time, aligning four elements from a column of the adjacency matrix with four elements from a row of the feature matrix. 
This is achieved using a specialized instruction, denoted as the \(\texttt{MMH4}\) instruction.

Our technique represents a fusion of Gustavson's algorithm and the outer-product method. 
Unlike the outer-product approach which finalizes the multiplication of an entire column with a row before moving to the next, our strategy concurrently processes four rows by employing the Gustavson method. 
The selection of the number `four' for simultaneous row processing results from design space exploration specific to the NeuraChip accelerator.

To implement this modified Gustavson's approach, the adjacency matrix is stored in a compressed sparse column (CSC) format, and the feature matrix is stored in a compressed sparse row (CSR) format.
However, this approach presents two primary challenges:

\textbf{Unavoidable Index Matching}: Employing Gustavson's algorithm and compressed matrix storage formats such as CSR and CSC inherently leads to the necessity of index matching~\cite{srivastava2020matraptor}. 
We address the index-matching overhead with a constant lookup hash function, facilitating the on-chip accumulation of partial products with a constant lookup time. 
The low overhead provided by our hash function is further optimized by adding a dedicated hash engine, as described in \secref{ssec:neuramem}.

\textbf{Memory Bloat Issue}: The tiled Gustavson method can result in memory bloat, characterized by the generation of a large number of partial products~\cite{baek2021innersp}.
To tackle this issue, we have implemented a rolling eviction mechanism. 
This system accumulates partial products as they are generated and promptly evicts them once the reduction is complete, with further details provided in Section~\ref{ssec:neuramem}.

\begin{figure}[tbp]
	\centering
	\includegraphics[width=0.55\textwidth]{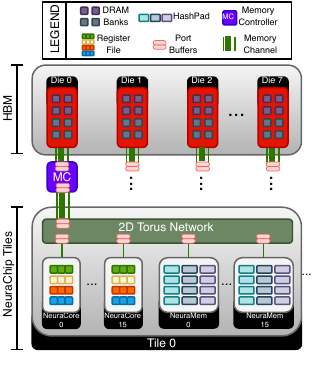}
        \vspace{-0.8em}
	\caption{NeuraChip memory hierarchy}
	\label{fig:mem_hierarchy}
\end{figure}

\subsection{On-chip Dataflow}

To illustrate the data flow within NeuraChip, we walk through an example of an SpGEMM kernel executed on the NeuraChip accelerator (see Figure~\ref{fig:tile_16}~and~\ref{fig:mem_hierarchy}).
Step \circled{1} The process begins with the \textit{Dispatcher} issuing \textbf{matrix\_mult\_hash\_4} (\texttt{MMH4}) instructions to every \textit{NeuraCore}.
Step \circled{2} The \textit{NeuraCores} trigger memory read requests that are routed to the memory controller.
Step \circled{3} The \textit{Memory Controller} coalesces requests for contiguous memory locations into a singular transaction and reorganizes memory transactions to enhance spatial locality.
Step \circled{4} Input matrix data, fetched from DRAM, is streamed onto respective \textit{NeuraCore} components.
Step \circled{5} The \textit{NeuraCores} compute the partial products, along with their corresponding rolling counters (further details in Section~\ref{ssec:neuracore}), subsequently generating the \textbf{hash\_accumulate} (\texttt{HACC}) instructions.
Step \circled{6} \texttt{HACC} instructions are streamed over on-chip routers into NeuraMem components, based on a hash-based mapping.
Step \circled{7} The \textit{NeuraMem} component employs another hash function to hash and accumulate these partial products onto their on-chip memory. 
Consecutive hashes of partial products with the same TAG are merged within NeuraMem, with each hash insertion decrementing the counter by $1$.
Step \circled{8} When the counter reaches zero; this triggers the eviction of the hashline, and the resultant data is written back to the High Bandwidth Memory (HBM).

\subsection{NeuraCore}
\label{ssec:neuracore}

\begin{figure}[htbp]
	\centering
	\includegraphics[width=0.98\textwidth]{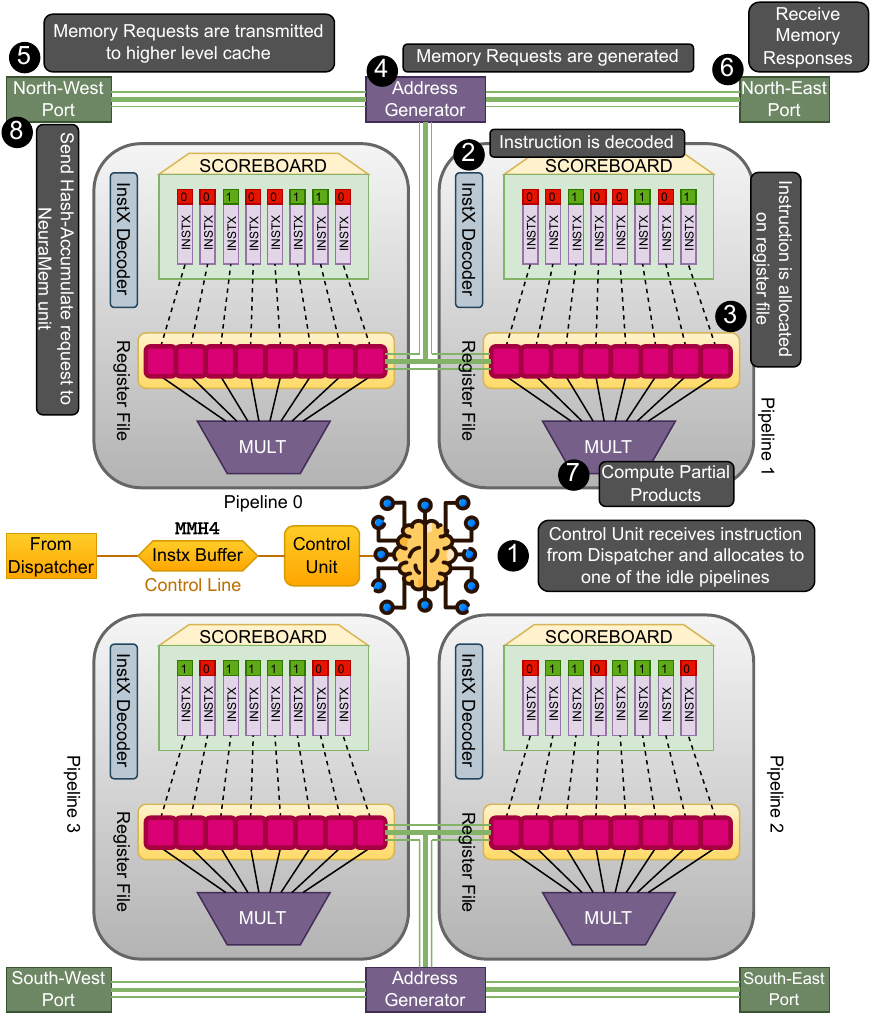}
	% \vspace{-2.2em}
	\caption{Block diagram showing NeuraCore's quad-pipeline layout.}
	\label{fig:neuracore}
% \vspace{-1.7em}
\end{figure}

The NeuraCore is the primary compute engine in our accelerator. It computes the multiplication operation and generates the partial products of matrix multiplication operations. It is a simple in-order core with support for matrix instructions.
% The supported instructions by NeuraCore are shown in Table~\ref{XX}.
NeuraCore supports a special matrix instruction called \texttt{matrix\_mult\_hash\_4} or simply \texttt{MMH4}.

\begin{figure}[hbp]
	\centering
        % \vspace{-1.0em}
	\includegraphics[width=0.98\textwidth]{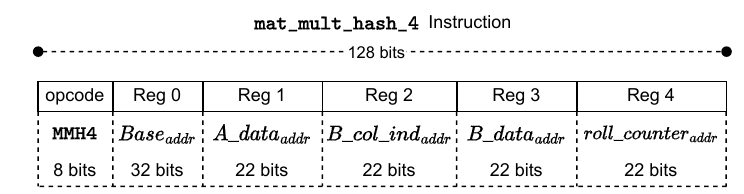}
	% \vspace{-0.6em}
	\caption{\texttt{MMH4} instruction bit layout.}
	\label{fig:mmh4}
% \vspace{-1.7em}
\end{figure}

where
$opcode$ represents the operation code, which specifies the MMH4 instruction to be executed by NeuraCore.
$Base_{addr}$ denotes the base address used to offset the address of all other addresses involved in this instruction.
$A\_data_{addr}$ refers to the memory address where the data of matrix A is located (matrix A is stored in CSC storage format).
$B\_col\_ind_{addr}$ points to the memory address containing the column indices of matrix B (matrix B is stored in CSR storage format).
$B\_data_{addr}$ indicates the memory address where data from matrix B is stored.
$roll\_counter_{addr}$ denotes the memory address where the rolling eviction counter is located.
The pseudocode for executing the \texttt{MMH4} instruction is shown in Algorithm~\ref{algo:mmh4}. Each \texttt{MMH4} instruction has the capability to dispatch up to $16$ \texttt{HACC} instructions (further elaborated in the NeuraMem section).

\begin{algorithm}
\caption{\texttt{MMH4} instruction execution}
\label{algo:mmh4}
\begin{algorithmic}[1]
\FOR{$i = 0$ \TO $3$}
    \FOR{$j = 0$ \TO $3$}
        \STATE $TAG \gets \text{Mem}[(Base_{addr} + B_{col\_ind\_addr} + j)]$
        \STATE $DATA \gets \text{Mem}[(Base_{addr} + A_{data\_addr} + i)]$
        \STATE \hspace{\algorithmicindent} $\times \text{Mem}[(Base_{addr} + B_{data\_addr} + j)]$
        \STATE $CTR \gets \text{Mem}[(Base_{addr} + roll\_counter + i*4 + j)]$%
        \STATE $\text{Dispatch} \ \texttt{HACC}(TAG, DATA, COUNTER)$
    \ENDFOR
\ENDFOR
\end{algorithmic}
\end{algorithm}

The operational sequence within NeuraCore is shown in Figure~\ref{fig:neuracore}, and can be broken down into the following steps:
Step \circled{1}: The operation starts with the dispatcher transmitting a \texttt{MMH4} instruction to NeuraCore, allocating the instruction to one of the available pipelines. Pipelines are allocated using a round-robin scheme.
Step \circled{2}: The \texttt{MMH4} instruction is decoded by the on-chip decoder.
Step \circled{3}: Following decoding, NeuraCore maps instruction variables to the register file, utilizing dynamic register allocation.
Step \circled{4}: Post register allocation, the NeuraCore's internal address generator constructs memory requests to fetch elements from the input matrices.
Step \circled{5}: An adaptive routing algorithm~\cite{ascia2008implementation} selects the best port to dispatch the memory request, which is then forwarded to a higher-level cache.
Step \circled{6}: Upon completing the memory request, a response is received at one of the NeuraCore's four ports. This response is then routed toward its respective pipeline.
Step \circled{7}: As soon as all memory responses corresponding to a particular instruction are received, the instruction is deemed ready for execution by the scoreboard. Subsequently, the multiplication pipeline calculates the partial product and generates up to $16$ \texttt{HACC} instructions.
Step \circled{8}: Lastly, the \texttt{HACC} instructions are relayed to NeuraMem units using the most suitable port, as determined by the on-chip hash-based mapping function.

\subsection{NeuraMem}
\label{ssec:neuramem}

\begin{figure}[htbp]
	\centering
	\includegraphics[width=0.98\textwidth]{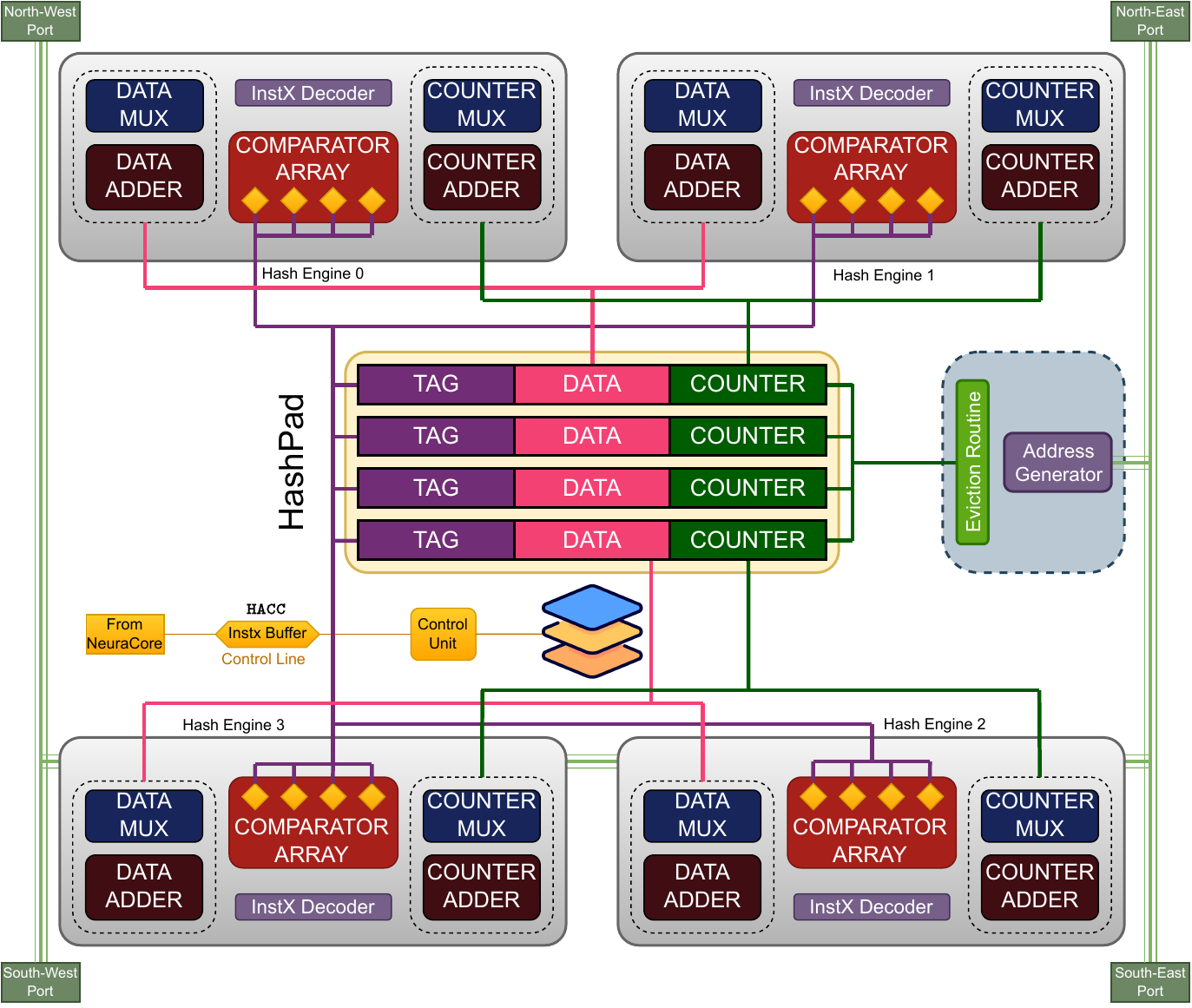}
	% \vspace{-2.2em}
	\caption{Block diagram showing NeuraMem's quad-hash-engine layout.}
	\label{fig:neuramem}
% \vspace{-1.7em}
\end{figure}

\begin{figure}[hbp]
	\centering
 % \vspace{-1.7em}
	\includegraphics[width=0.98\textwidth]{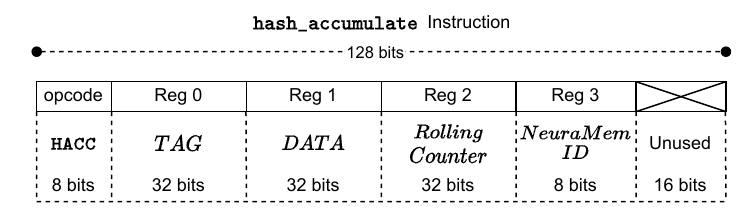}
        % \vspace{-0.6em}
	\caption{\texttt{HACC} instruction bit layout.}
	\label{fig:hacc}
% \vspace{-1.7em}
\end{figure}

NeuraMem is a crucial component of the NeuraChip accelerator. While NeuraCore units generate partial products, NeuraMem units handle the on-chip accumulation of these partial products. The central component of NeuraMem units is the Hash-Engine. The layout of various components within NeuraMem is as shown in Figure~\ref{fig:neuramem}.

\textbf{HashPad}: The Hash-Engine operates on what we refer to as ``hash-lines''~\figref{fig:neuramem}. A hash-line comprises a single TAG, DATA, and COUNTER entry. The collective TAG array, DATA array, and COUNTER array, essentially the whole set of hash-lines, form what is known as the HashPad, as shown in Figure~\ref{fig:neuramem}.
% This figure also shows a typical sequence of events during the execution of a \texttt{hash\_accumulate} instruction by the Hash-Engine.

\textbf{\texttt{HACC} instruction}: NeuraMem supports a special instruction for partial product accumulation called \texttt{hash\_accumulate}, or simply \texttt{HACC} instruction. The bit layout of \texttt{HACC} instruction is illustrated in \figref{fig:hacc}. \algoref{algo:hacc} presents a pseudocode of the \hacc  instruction, providing clearer insight into its functionality.

%% \texttt{HACC} instruction execution algorithm
%% index = Hash(TAG)
%% if tag_array[index] == EMPTY
%%     data_array[index] = DATA
%%     counter_array[index] = COUNTER
%% else if tag_array[index] == TAG:
%%     data_array[index] += DATA
%%     counter_array[index] -= 1
%%     if counter_array[index] == 0
%%         Hash Line Eviction Routine
%% else:
%%     Hash Collision Routine

\begin{figure}[htbp]
	\centering
	\includegraphics[width=0.98\textwidth]{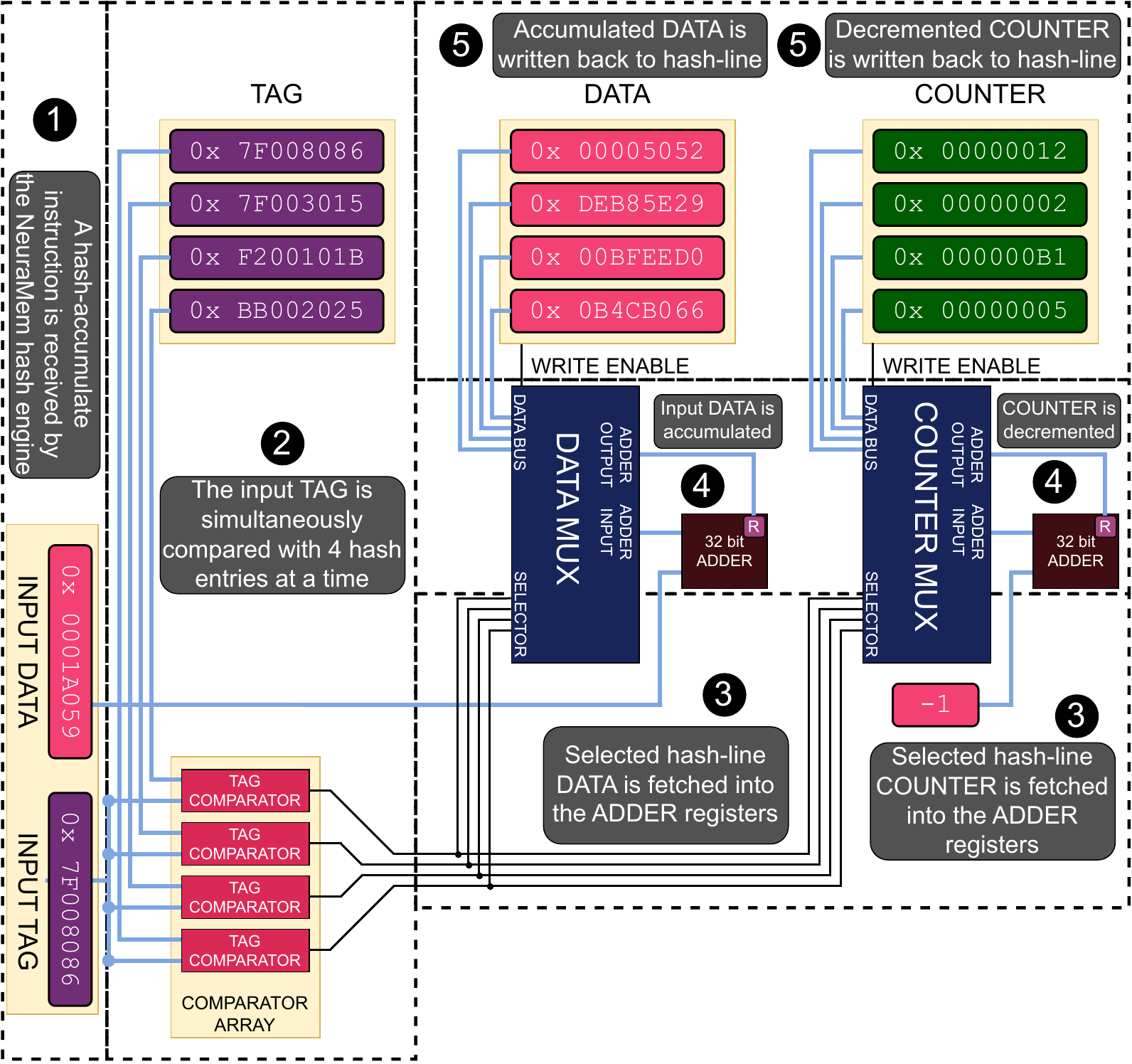}
	% \vspace{-2.2em}
	\caption{The NeuraMem Hash-Engine accumulates a single partial product using the \hacc instruction.}
	\label{fig:hash_engine}
% \vspace{-1.7em}
\end{figure}

\textbf{Hash-Engine workflow}: \figref{fig:hash_engine} shows a typical sequence of events during the execution of a \texttt{HACC} instruction by the Hash-Engine (illustrated using pseudocode in \algoref{algo:hacc}). The process starts in step \circled{1}, where the Hash-Engine receives a \hacc instruction from the NeuraCore units. This instruction's \mtag is simultaneously compared with all the TAGs currently present on the HashPad (step \circled{2}). The multiplexers select the hash-line with the matching TAG in step \circled{3}. The corresponding hash-line's DATA gets accumulated with the \hacc instruction's data. Simultaneously, the counter for that hash-line is decremented by one (step \circled{4}). The accumulated data and the updated counter are then written back to the HashPad in step \circled{5}.
If the TAG from the instruction does not match any of the TAGs in the HashPad in step \circled{2}, the Hash-Engine creates a new entry for the hash instruction and stores its content in a new hash-line.

\begin{algorithm}
\caption{\texttt{HACC} Instruction Execution}
\label{algo:hacc}
\begin{algorithmic}[1]
\STATE $index \gets \text{Hash}(TAG)$
\IF{$tag\_array[index] == \text{EMPTY}$} 
    \STATE $data\_array[index] \gets DATA$
    \STATE $counter\_array[index] \gets COUNTER$
\ELSIF{$tag\_array[index] == TAG$}
    \STATE $data\_array[index] \mathrel{+}= DATA$
    \STATE $counter\_array[index] \mathrel{-}= 1$
    \IF{$counter\_array[index] == 0$}
        \STATE $\text{Hash Line Eviction Routine}$
    \ENDIF
\ELSE
    \STATE $\text{Hash Collision Routine}$
\ENDIF
\end{algorithmic}
\end{algorithm}

% \textbf{Timed Eviction} (Hot-Potato Caching): An advantage of such a system is that the Hash-Engine keeps track of when the partial product accumulation is completed (using the COUNTER, as seen in Figure~\ref{fig:hash_engine}).
% When the COUNTER hits zero (when all partial products for the particular TAG are accumulated), the hash-engine automatically evicts the hash-line where the accumulated final result is written back to the main memory (HBM).
% This means, that the hashed partial product spends the least number of cycles on the on-chip HashPad, thus saving on the precious on-chip storage.

\textbf{Rolling Evictions}: The Hash-Engine monitors the completion of partial product accumulation (via the COUNTER, as seen in \figref{fig:neuramem}. Once the COUNTER reaches zero, indicating that all partial products for a particular TAG have been accumulated, the Hash-Engine automatically evicts the corresponding hash-line, and the accumulated result is written back to the main memory (HBM). This ensures that the hashed partial product spends the minimal possible number of cycles in the HashPad, addressing the memory bloat issue.

% The intermediate results of the matrix multiplication are streamed into corresponding NeuraMem units based on a dynamic hashing algorithm.
% The hashing algorithm utilized in NeuraChip addresses two main concerns, avoiding hash collisions and load imbalance across multiple NeuraMems.
% The partial products received at each NeuraMem unit are accumulated together to generate the final product. This final product is then streamed out to the HBM.

\subsection{Dynamically Reseeding Hash-based Mapping}
\label{ssec:drhm}
%
% A significant portion of our speedup in NeuraChip hash-based accelerator can be contributed to our collision reduction algorithm called Dynamic Collision Reduction Hashing (DRHM).
% DRHM is based on the notion that if certain nodes within the input graph are comparatively densely connected, they have to be allocated more resources (both in terms of compute and memory). DRHM handles the memory aspect of this. While hashing dense rows, DRHM maps relatively more number of hash-lines to these rows, leading to a significant reduction in number of hash collisions, see Figure~\ref{fig:dynamic_hashing}~\circled{h}.
%

The performance benefits provided by the NeuraChip accelerator are primarily due to our sparsity-agnostic mapping algorithm, named Dynamically Reseeding Hash-based Mapping (DRHM). DRHM is designed to eliminate computational patterns, promoting an even distribution of workload across all computational resources. Traditional hash-based mappings often lead to concentrated areas of high activity, known as hot spots, especially when the hash function is optimized for a specific sparsity pattern but encounters a different one. An ideal solution would involve uniformly distributing computational tasks across resources. One such method is random mapping, where tasks are allocated to random resources. However, maintaining consistency in random mapping requires extensive record-keeping (a large lookup table), which is impractical.

We introduce a hybrid approach, Dynamically Reseeding Hash-based Mapping (DRHM), which combines the advantages of consistent lookup times in hashing, a distribution akin to random mapping, and minimal overhead similar to small lookup tables. This method significantly reduces the occurrence of hot spots in the allocation of computational resources.

% DRHM operates on the principle that nodes in the input graph, which are relatively more densely connected, should be assigned more resources, both computationally and memory-wise. DRHM primarily manages the memory aspect. During the hashing of dense rows, DRHM assigns a relatively larger number of hash-lines to these rows. This approach considerably decreases the number of hash collisions, as demonstrated in Figure~\ref{fig:dynamic_hashing}~\circled{h}.

% DRHM leverages a dynamic mapping that shifts based on a key parameter, denoted as $\gamma$. This parameter is designed to adjust the mapping and, therefore, the hash function dynamically to mitigate hash collisions.
% By having an uneven mapping of the number of hash-lines to each row of input graph, DRHM ensures that dense rows are allocated the most number of hash-lines where as sparse rows are  allocated far fewer number of hash-lines.
% The equation for hashing based on $\gamma$ scaling is show in Equation~\ref{eq:DRHM}
% \begin{equation}
% \label{eq:DRHM}
% H(\mathrm{TAG}, \gamma) = ((\mathrm{TAG} >> k) \mod (\gamma \times N))
% \end{equation}
% where ...

% OLD \begin{equation}
% \label{eq:DRHM_lower}
% H_{l}(\mathrm{TAG}, \gamma) = ((\mathrm{TAG} << k) >> k) \mod (\gamma \times N))
% \end{equation}
% \begin{equation}
% \label{eq:DRHM_higher}
% H_{h}(\mathrm{TAG}, \gamma) = ((\mathrm{TAG} >> k) << k) \mod (\gamma \times N))
% \end{equation}

DRHM utilizes a flexible mapping that adjusts based on a `seed' parameter, denoted as $\gamma$. This parameter is specifically designed to alter the mapping, and consequently, the hash function dynamically. After each row of the input sparse matrix is computed, $\gamma$ is initialized with a random number.
DRHM offers two implementation approaches: one using the $k$ upper bits of the \mtag, and the other utilizing the $k$ lower bits of the \mtag. The lower-bit and upper-bit hashing equations that accommodate $\gamma$ seed are presented in Equations~\ref{eq:DRHM_lower} and \ref{eq:DRHM_higher}.

% DRHM achieves an unbalanced allocation of hash-lines for each row of the input graph, ensuring that the denser rows receive a more substantial number of hash-lines, while sparser rows are granted far fewer.

\begin{equation}
\label{eq:DRHM_lower}
H_{l}(\mathrm{TAG_{32}}, \gamma) = ((\mathrm{TAG_{32}} \ll k) \gg k) \cdot \gamma \mod N
\end{equation}
\begin{equation}
\label{eq:DRHM_higher}
H_{h}(\mathrm{TAG_{32}}, \gamma) = ((\mathrm{TAG_{32}} \gg k) \ll k) \cdot \gamma \mod N
\end{equation}

where \mtag represents the unique identifier for each row of the input graph. The term $\gamma$ acts as a `seed' to introduce randomness in the mapping. $N$ signifies the total number of available output hash spaces. The operations ``$\ll k$'' and ``$\gg k$'' refer to bitwise left and right shifts by $k$ positions, respectively. The modulus operation $\mod$ ensures that the result of the hash function falls within the predefined range of the hash table.
% Consequently, this approach provides a dynamic hashing function that effectively reduces collisions and enhances the efficiency of hash-based operations.
These equations assume that the bit-shift operations conform to standard behavior where bits shifted beyond the boundary of the number's bit-width are discarded.

In our experiments, we assessed both upper $k$-bit address hashing and lower $k$-bit address hashing. We found that the lower $k$-bit address hashing method had a lower incidence of hash collisions, due to the higher variability in the lower bits of the address. Consequently, in all the work presented here, we employ the lower $k$-bit address hashing technique (Equation~\ref{eq:DRHM_lower}).
% Figure~\ref{fig:dynamic_hashing} provides a visual representation of the $\gamma$ map for several of the datasets used in this paper.
The efficiency of compute mapping using our DRHM approach is evaluated in detail in \secref{sec:dse}.

\section{Exploring the Design Space of NeuraChip}
\label{sec:dse}

The flexibility of our NeuraSim simulator, which is used to simulate our NeuraChip accelerator, enables us to evaluate multiple NeuraChip configurations. We have two primary design goals: i) optimizing resource utilization across the accelerator to enhance speedup and ii) striking a balance between performance, chip area, and power consumption to make sure the advantages outweigh the costs.

\begin{table}
\centering
\caption{Individual Component Configuration}
% \vspace*{-2mm}
\label{tab:comp_config}
\begin{tabular}{p{0.99in} p{1.5in} x{0.7in} x{0.7in} x{0.7in}}
\arrayrulecolor{black}\toprule
    \textbf{Component}&  \textbf{Elements} & \textbf{Tile-4$^*$} & \textbf{Tile-16$^*$} & \textbf{Tile-64$^*$} \\ [0.5ex] 
\arrayrulecolor{black}\toprule
% \arrayrulecolor{black}\toprule
    % \arrayrulecolor{black!20}\midrule
    \multirow{5}{*}{\textbf{NeuraCore}} & Registers per pipeline & $4$ & $8$ & $16$ \\
                                        & Pipelines & $2$ & $4$ & $8$ \\
                                        & Multipliers & $2$ & $4$ & $8$ \\
                                        & Address Generators & $1$ & $2$ & $2$ \\
                                        & Ports & $4$ & $4$ & $4$ \\                                        
    \arrayrulecolor{black!20}\midrule
    \multirow{5}{*}{\textbf{NeuraMem}}  & TAG Comparators & $1$ & $4$ & $8$ \\
                                        & Hash-Engines & $2$ & $4$ & $8$ \\    
                                        & Hashlines & $4096$ & $2048$ & $2048$ \\
                                        & Accumulators & $128$ & $256$ & $512$ \\
                                        & Ports & $4$ & $4$ & $4$ \\                                        
    % \arrayrulecolor{black!20}\midrule
    % \multirow{2}{*}{\parbox{0.6in}{\textbf{NeuraChip Accelerator}}} & Total NeuraCores & 32 & 128 & 512 \\
    %                                                                 & Total NeuraMems & 32 & 128 & 512 \\
                                                                    % & Routers & 128 & 64 & 256 \\
                                                                    % & Mem Controllers & 8 & 8 & 8 \\
                                                                    % & HBM Channels & 8 & 8 & 8 \\

     \arrayrulecolor{black}\bottomrule

\end{tabular}

% \vspace{0.2em}

\scriptsize

\centering
$^{*}$Values represent the count of elements per component across three tile configurations.

% \vspace{-5mm}
\end{table}

\begin{figure}[hbp]
	\centering
 % \vspace{-1.2em}
	\includegraphics[width=0.58\textwidth]{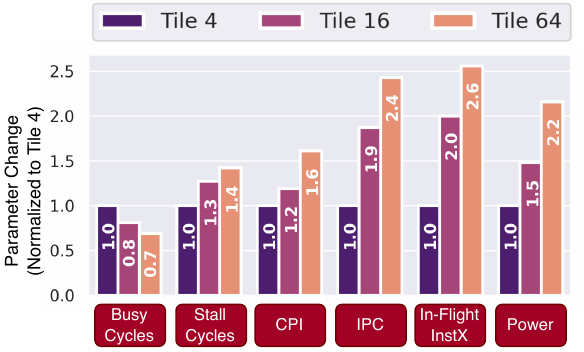}
	% \vspace{-0.5em}
	\caption{Architectural impact of GCN model varying tile configuration on Cora dataset. Values are normalized to Tile-4 configuration.}
	\label{fig:tile_analysis}
% \vspace{-1.7em}
\end{figure}

\textbf{Tile Size Variation}: We introduce three distinct configurations of NeuraChip, named Tile-4, Tile-16, and Tile-64, derived from experimenting with various workloads. The detailed configurations of NeuraCore and NeuraMem components are provided in \tabref{tab:comp_config}, while the overall accelerator configurations for these tile sizes are listed in \tabref{tab:chip_configuration}. We focus on six key parameters to assess the architectural impact of these configurations, as shown in \figref{fig:tile_analysis}.

Key observations include:
\begin{itemize}
    
    \item \textbf{Register File Size}: Expanding the register file size allows more \mmhf instructions to be in-flight and increases the number of read memory instructions that can be issued to HBM. Beyond $8$ registers per pipeline ($1024$ bits per pipeline), we noticed that the DRAM channels are unable to keep up with the high memory demands.
    This bottleneck is evident in the rise in the cycles per instruction (CPI) and the number of stall cycles, as shown in \figref{fig:tile_analysis}.

    \item \textbf{HashPad Size}: Choosing between smaller HashPads with a larger number of NeuraMems versus larger HashPads with fewer NeuraMems, the former proves advantageous for handling extremely sparse matrices. This configuration benefits from high accumulation throughput as the number of accumulators increases with the number of NeuraMems. This can be seen in the larger number of in-flight HBM memory instructions in \figref{fig:tile_analysis}.

    \item \textbf{Component Counts}: With 32, 128, and 512 NeuraCores and NeuraMems in Tile-4, Tile-16, and Tile-64, respectively, while more components enhance peak compute throughput, the configuration is bound by a peak DRAM bandwidth of $128\ GB/s$. Additionally, workloads do not require a 12 MB on-chip memory  HashPad (of tile-64 configuration).

\end{itemize}

\begin{table}
\centering
\caption{NeuraChip Configuration}
% \vspace*{-2mm}
\label{tab:chip_configuration}
\begin{tabular}{p{2.7in} | x{0.634in} x{0.6395in} x{0.6395in}}
\arrayrulecolor{black}\toprule
   \textbf{Parameter} & \textbf{Tile-4$^*$} & \textbf{Tile-16$^*$} & \textbf{Tile-64$^*$} \\ [0.5ex] 
\arrayrulecolor{black}\toprule
\arrayrulecolor{black}\toprule

     Tile Count & $8$ & $8$ & $8$ \\
    \arrayrulecolor{black!20}\midrule
    NeuraCores per tile & $4$ & $16$ & $64$ \\
    \arrayrulecolor{black!20}\midrule
    Total NeuraCores & $32$ & $128$ & $512$ \\
    \arrayrulecolor{black!20}\midrule
    NeuraMems per tile & $4$ & $16$ & $64$ \\
    \arrayrulecolor{black!20}\midrule
    Total NeuraMems & $32$ & $128$ & $512$ \\
    \arrayrulecolor{black!20}\midrule
    Memory Controller Count & $8$ & $8$ & $8$ \\
    \arrayrulecolor{black!20}\midrule
    Routers per tile & $8$ & $32$ & $128$ \\
    \arrayrulecolor{black!20}\midrule
    Total Routers & $64$ & $256$ & $1024$ \\
    \arrayrulecolor{black!20}\midrule
    Total Pipelines & $64$ & $512$ & $4096$ \\
    \arrayrulecolor{black!20}\midrule
    Register File Size per pipeline (bits) & $512$ & $1024$ & $2048$ \\
    \arrayrulecolor{black!20}\midrule
    Total Hash-Engines & $64$ & $512$ & $4096$ \\
    \arrayrulecolor{black!20}\midrule
    \mtag comparators per Hash-Engine & $2$ & $4$ & $8$ \\
    \arrayrulecolor{black!20}\midrule
    Total \mtag comparators & $128$ & $2048$ & $32768$ \\
    \arrayrulecolor{black!20}\midrule
    Total HashPad Size (MB) & $1.5$ & $3$ & $12$ \\
    \arrayrulecolor{black!20}\midrule
    Max operating frequency ($GHz$) & $1$ & $1$ & $1$ \\

     \arrayrulecolor{black}\bottomrule
% \vspace{-2.7em}
\end{tabular}
% \vspace{0.2mm}

\scriptsize

\centering
$^{*}$Values represent the count of components/elements across the entire NeuraChip accelerators for three different tile configurations.

% \vspace{-1.9em}
\end{table}

%     Tile Count & 8 & 8 & 8 \\
%     \arrayrulecolor{black!20}\midrule
%     NeuraCores per tile & 4 & 16 & 64 \\
%     \arrayrulecolor{black!20}\midrule
%     Total NeuraCores & 32 & 128 & 512 \\
%     \arrayrulecolor{black!20}\midrule
%     NeuraMems per tile & 4 & 16 & 64 \\
%     \arrayrulecolor{black!20}\midrule
%     Total NeuraMems & 32 & 128 & 512 \\
%     \arrayrulecolor{black!20}\midrule
%     Memory Controller Count & 8 & 8 & 8 \\
%     \arrayrulecolor{black!20}\midrule
%     Routers per tile & 8 & 32 & 128 \\
%     \arrayrulecolor{black!20}\midrule
%     Total Routers & 64 & 256 & 1024 \\
%     \arrayrulecolor{black!20}\midrule
%     Total Pipelines & 64 & 512 & 4096 \\
%     \arrayrulecolor{black!20}\midrule
%     Register File Size per pipeline (bits) & 512 & 1024 & 2048 \\
%     \arrayrulecolor{black!20}\midrule
%     Total Hash-Engines & 64 & 512 & 4096 \\
%     \arrayrulecolor{black!20}\midrule
%     \mtag comparators per Hash-Engine & 2 & 4 & 8 \\
%     \arrayrulecolor{black!20}\midrule
%     Total \mtag comparators & 128 & 2048 & 32768 \\
%     \arrayrulecolor{black!20}\midrule
%     Total HashPad Size (MB) & 1.5 & 3 & 12 \\
%     \arrayrulecolor{black!20}\midrule
%     Max operating frequency ($GHz$) & 1 & 1 & 1 \\
% \arrayrulecolor{black!20}\midrule

\begin{figure}[htbp]
	\centering
	\includegraphics[width=0.78\textwidth]{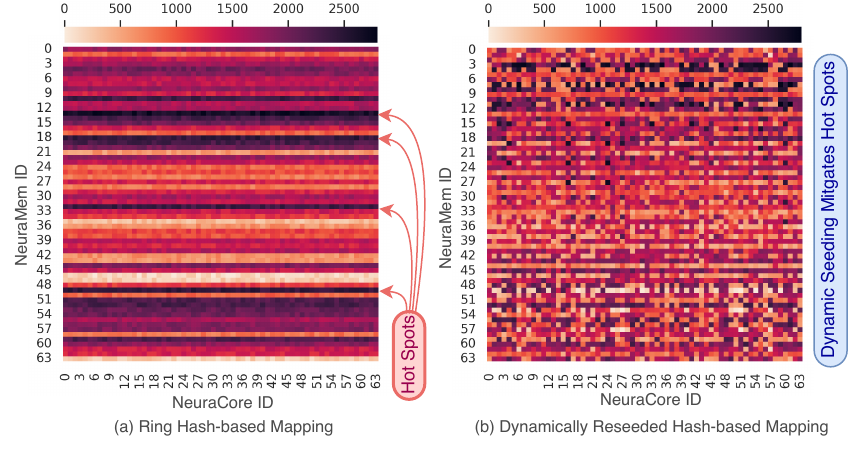}
	% \vspace{-1.2em}
	\caption{Compute mapping heat map, where the X-axis represents multiplications mapped to NeuraCores and Y-axis represents accumulations mapped to NeuraMem.}
	\label{fig:hop_dynamic}
% \vspace{-1.5em}
\end{figure}

\begin{figure*}[htbp]
	\centering
	\includegraphics[width=0.98\textwidth]{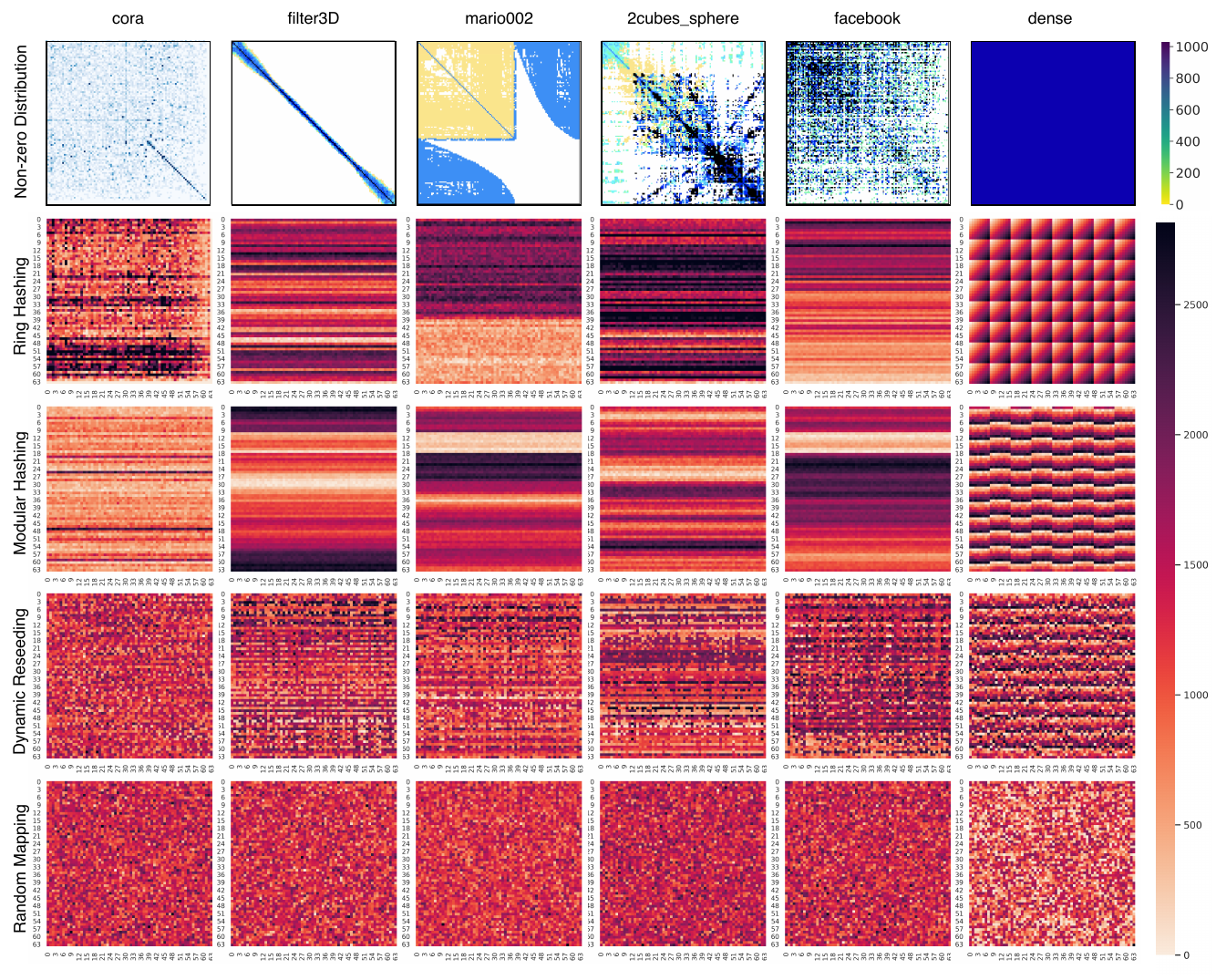}
	% \vspace{-1.2em}
	\caption{Computation mapping heat maps for four distinct hash-based mapping methods, evaluated across five sparse matrices and one dense matrix multiplication. The dynamic reseeding mapping technique is insensitive to sparsity patterns and effectively addresses hot spots in dense matrix computations.}
	\label{fig:dynamic_hashing}
% \vspace{-1.7em}
\end{figure*}

\textbf{Hash-based Mapping Algorithm Variations}: We tested four hash-based mapping schemes. The first, a ring-based mapping (see \figref{fig:hop_dynamic}), follows round-robin resource allocation, though encounters hot spots in workload distribution. The second, a modular hash-based mapping, uses prime numbers for workload mapping, proposed in previous studies~\cite{zhang2018fast, gou2018single, ng2015two, song2009design}. DRHM, shown in \figref{fig:hop_dynamic}, addresses hot spots in modular and ring-based mappings by reseeding the hash function after each row of computations. Lastly, we evaluate a random mapping that maintains a lookup table for each entry. All four techniques are compared in \figref{fig:dynamic_hashing} for varying sparsity patterns.

\begin{figure}[hbp]
	\centering
 % \vspace{-1.7em}
	\includegraphics[width=0.98\textwidth]{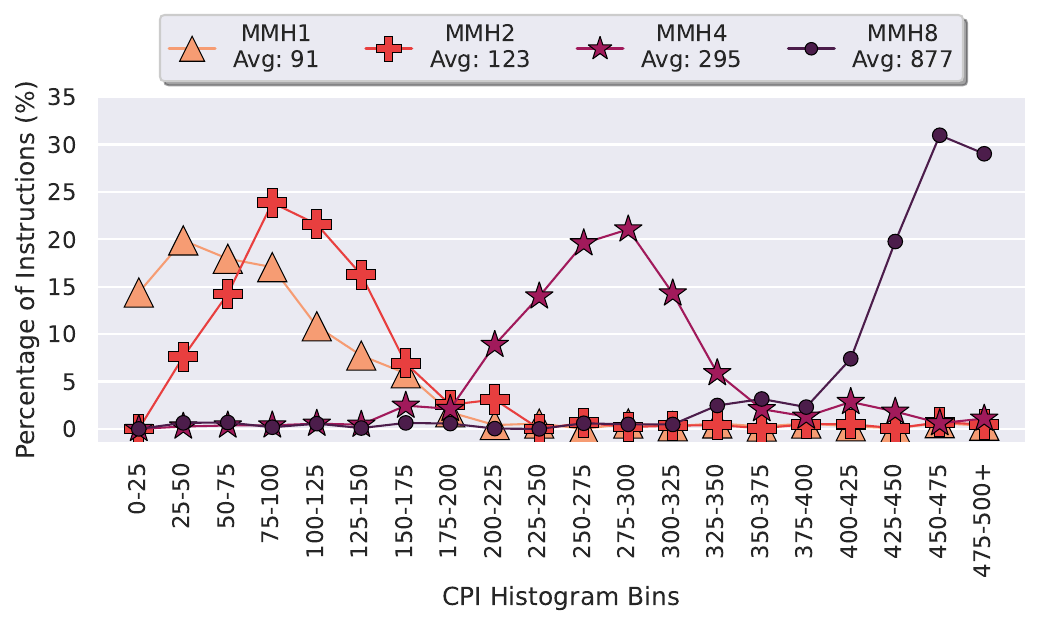}
	% \vspace{-1.2em}
	\caption{Cycles Per Instruction (CPI) histogram plot for four \texttt{MMH} instructions with varying tile sizes.}
	\label{fig:mmh_cpi}
% \vspace{-1.7em}
\end{figure}

\textbf{Variations in \texttt{MMH} and \texttt{HACC} Instructions}: 
NeuraChip introduces \texttt{MMH} and \texttt{HACC} instructions (bit layout of these instructions is illustrated in~\figref{fig:mmh4} and~\figref{fig:hacc}), supporting its decoupled architecture. We analyze the cycle count of various \texttt{MMH} instruction tile sizes, presented in a CPI histogram in \figref{fig:mmh_cpi}. \texttt{MMH4} emerges as the top choice, balancing temporal locality benefits and cycle count.

\begin{figure}[hbp]
	\centering
 % \vspace{-1.7em}
	\includegraphics[width=0.98\textwidth]{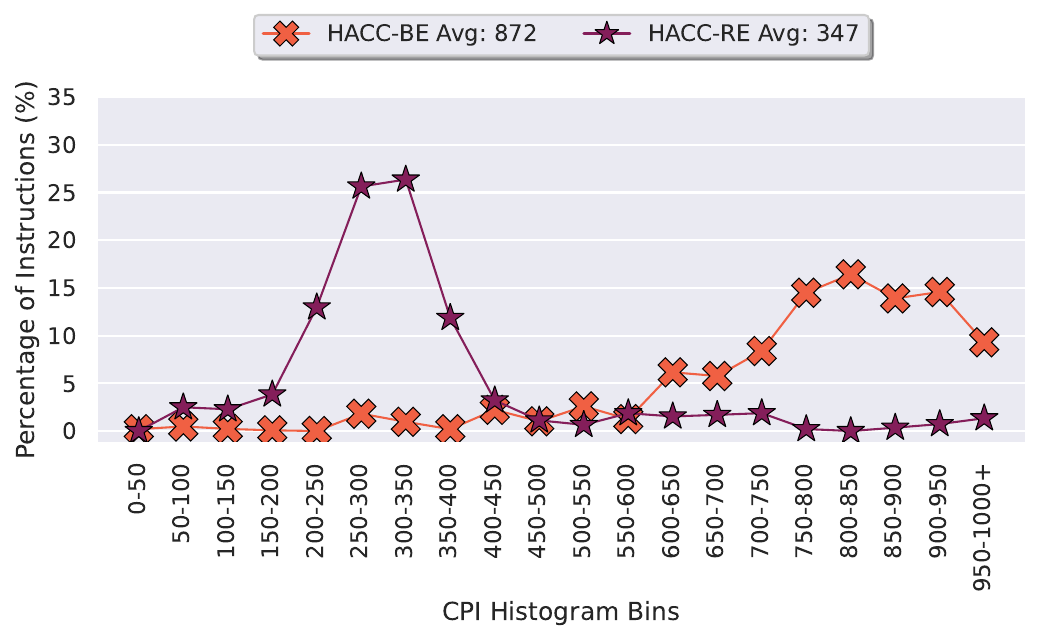}
	% \vspace{-1.2em}
	\caption{Cycles Per Instruction (CPI) histogram plot for \hacc instructions with barrier based evictions \texttt{HACC-BE} and rolling evictions \texttt{HACC-RE}.}
	\label{fig:hacc_cpi}
% \vspace{-1.7em}
\end{figure}

We compare the \texttt{HACC} instruction's efficiency using two eviction schemes: barrier-based eviction (\texttt{HACC-BE}) and our rolling eviction approach (\texttt{HACC-RE}). The latter's superiority in reducing average cycle completion is seen in \figref{fig:hacc_cpi}.

\section{NeuraChip Evaluation}

\subsection{Experimental Setup}

% Our evaluation of NeuraChip involves benchmarking against two distinct workloads. Firstly, we assess the performance of NeuraChip on sparse matrix multiplication tasks using a standard collection of sparse matrices from the Stanford SNAP sparse matrix suite~\cite{XX}. The specific sparse matrix datasets employed for testing our accelerator are detailed in \tabref{tab:sparse_datasets}.
% This evaluation is compared with previous state-of-the-art sparse matrix accelerators and current off-the-shelf hardware platforms.
% We compare our accelerator against Intel MKL using a Xeon E5-2630 CPU, to cuSPARSE and CUSP NVIDIA libraries using a Hopper architecture H100 GPU, and finally to AMD's MI100 GPU using the hipSPARSE library with a rocSPARSE backend.

To evaluate the benefits of NeuraChip, we perform benchmarking across two distinct categories of workloads.
The first category involves examining NeuraChip's efficiency in handling sparse matrix multiplication tasks.
This evaluation uses a standard array of sparse matrices obtained from the Stanford SNAP sparse matrix collection~\cite{snapnets}.
Our evaluation includes a comparison with some of the latest state-of-the-art sparse matrix accelerators~\cite{zhang2020sparch, zhang2021gamma} and off-the-shelf mainstream hardware platforms. NeuraChip is benchmarked against the Intel MKL library~\cite{wang2014intel} with an Intel Xeon E5-2630 CPU. We also compare against cuSPARSE~\cite{naumov2010cusparse} and CUSP~\cite{Cusp} NVIDIA libraries, as run on a Hopper architecture H100 GPU, and we also consider for comparison an AMD's MI100 GPU using the hipSPARSE library with a rocSPARSE backend~\cite{ROCmSoftwarePlatform, bunn2019student}.
For accelerator comparisons, we compare NeuraChip against OuterSPACE~\cite{pal2018outerspace} SpArch~\cite{zhang2020sparch}, and Gamma~\cite{zhang2021gamma}.
Additionally, as to the second category of workloads, our evaluation targets a Graph Convolutional Network (GCN)~\cite{kipf2016semi} layer using various datasets, allowing us to compare NeuraChip against existing Graph Neural Network (GNN) accelerators EnGN~\cite{liang2020engn}, GROW~\cite{hwang2023grow}, HyGCN~\cite{yan2020hygcn}, and FlowGNN~\cite{sarkar2023flowgnn}.
% \todo{Explain how we have simulated these accelerators (see my overleaf's comments in this paragraph)}.

\subsection{Simulator Framework}

In this study, we present NeuraSim, a cycle-accurate, multi-threaded, modular simulation engine inspired by the Structural Simulation Toolkit (SST)~\cite{rodrigues2011structural}. NeuraSim's modular framework allows for flexible integration of new architectural features, without the need for an entire overhaul of the simulation engine. Developed using POSIX threads (\texttt{pthreads}), NeuraSim facilitates parallel simulation. Its dispatcher unit recognizes independent tasks and concurrently executes them on different threads. Additionally, NeuraSim employs MongoDB for backend data storage. NeuraSim also incorporates HBM2 memory simulation, integrating with DRAMsim3~\cite{li2020dramsim3}, a cycle-accurate and validated DRAM simulator.

Regarding simulation efficiency, NeuraSim achieves \mbox{$112$ Kilocycles per second (KCPS)}, $48$ KCPS, and $11$ KCPS on average for the Tile-4, Tile-16, and Tile-64 configurations, respectively. NeuraSim is open-source and faithfully simulates the extended NeuraChip ISA. The NeuraSim source code is accessible on our GitHub repository\footnote{\url{https://github.com/NeuraChip/neurachip}}\footnote{\url{https://neurachip.us/}}.

\subsection{Comparative Analysis with Sparse Matrix Accelerators}

\begin{figure*}[htbp]
	\centering
	\includegraphics[width=0.98\textwidth]{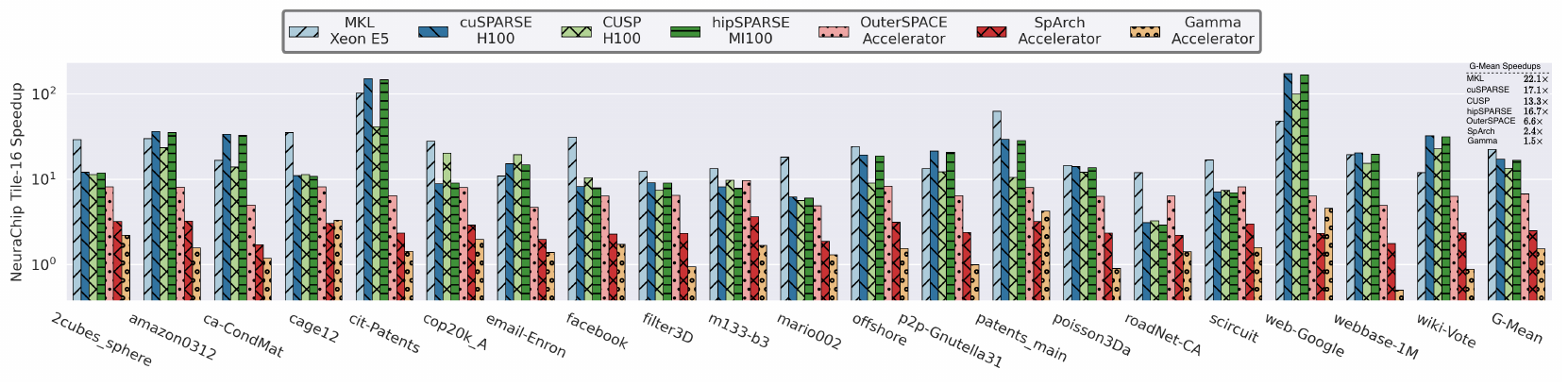}
	% \vspace{-1.2em}
	\caption{Speedup of NeuraChip Tile-16 configuration compared against CPU, GPUs, and SpGEMM accelerators benchmarking sparse matrix multiplication.}
	\label{fig:speedup}
% \vspace{-1.7em}
\end{figure*}

In \figref{fig:speedup}, the performance of the NeuraChip in sparse matrix multiplication tasks is compared against various off-the-shelf high-end CPU and GPU platforms, as well as against state-of-the-art SpGEMM accelerators.

NeuraChip provides benefits when compared to Intel's MKL running on an Xeon CPU, surpassing it by a factor of $22.1\times$. Additionally, when compared against NVIDIA's H100 GPU using the CUSP library, NeuraChip achieves a performance boost of $13.3\times$. In comparison to the prior leading sparse matrix multiplication accelerator, Gamma, NeuraChip achieves an average performance improvement of $1.5\times$.
%NeuraChip is compared with a server-grade CPU, the Intel Xeon E5-2630, utilizing the Intel MKL library, where it exhibits a $22.2\times$ performance improvement. Subsequently, it is compared with the NVIDIA H100 GPU, based on the Hopper architecture, using the cuSPARSE and CUSP libraries for sparse matrix computations. Here, NeuraChip achieves a $17.1\times$ and $13.3\times$ speedup over cuSPARSE and CUSP, respectively.

As we can see, NeuraChip outperforms the CPU and GPU computing platforms in all cases. In particular, average performance improvements of $22.2\times$ over the CPU, a $17.1\times$ and $13.3\times$ average speedup over the NVIDIA Hopper GPU using the cuSPARSE and CUSP libraries, respectively, and $16.7\times$ speedup on average over the AMD's MI100 GPU using the hipSPARSE library.

Further, the performance of NeuraChip is evaluated against two outer-product-based sparse matrix accelerators: OuterSPACE~\cite{pal2018outerspace} and SpArch~\cite{zhang2020sparch}. While OuterSPACE leverages input data reuse, it encounters excessive generation of partial products (the memory bloat issue), leading to degraded performance. SpArch addresses this with on-chip merger trees; however, these trees require large comparator arrays, occupying about $60\%$ of the chip area. NeuraChip counters the memory bloat through an on-chip cache organization with rolling counters, effectively managing the eviction of accumulating partial products and alleviating the bloat issue. In comparison, NeuraChip surpasses OuterSPACE and SpArch by factors of $6.6\times$ and $2.4\times$, respectively.

Additionally, the performance of NeuraChip is compared with a row-wise product-based SpGEMM accelerator, Gamma~\cite{zhang2021gamma}, which is based on Gustavson's algorithm. Gamma employs a resource-intensive storage mechanism, FiberCache, to prefetch data, aiming to reduce data fetch latency and prevent compute stalls. However, this approach results in data remaining idle in the caches prior to being accessed by the processing elements. NeuraChip, in contrast, optimizes on-chip storage through a rolling-eviction strategy, enabling automatic eviction of partial products after the reduce operation is complete. Against Gamma, NeuraChip demonstrates a performance superiority of $1.5\times$ average speedup.

% On average, the Tile-16 configuration of NeuraChip is $1.55\times$, $2.42\times$, $6.65\times$, $16.7\times$, $13.3\times$, $17.1\times$, and $22.1\times$, faster than MKL, cuSPARSE, CUSP, hipSPARSE, OuterSPACE, SpArch, and Gamma respectively.

% NeuraChip outperforms the baselines on every matrix. This mainly comes from the reduction of redundant memory access of partial matrices, with 2.8× less DRAM access compared to OuterSPACE. We also achieve higher memory bandwidth utilization compared to OuterSPACE, by virtue of hiding the DRAM latency with row prefetcher and the regular write pattern of streaming merge tree.

%% Matraptor has a different HBM model
%% At the time of writing ...

\subsection{Comparative Analysis of GNN Accelerators}

\begin{figure}[htbp]
	\centering
 % \vspace{-1.6em}
	\includegraphics[width=0.78\textwidth]{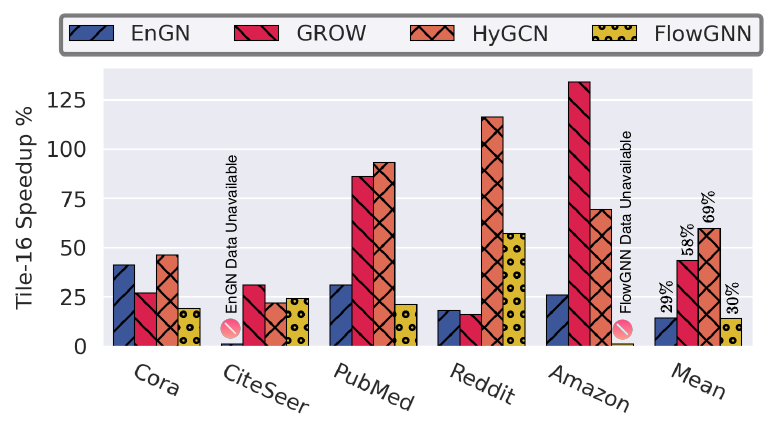}
	\vspace{-1.2em}
	\caption{Percentage speedup of Tile-16 configuration over prior GNN accelerators with GCN workload over different graph datasets.}
	\label{fig:gnn_speedup}
\vspace{-1.0em}
\end{figure}

In \figref{fig:gnn_speedup}, we compare the GNN performance of NeuraChip against various state-of-the-art GNN accelerators.
The NeuraChip configuration used for GNN assessment differs from that used to compare to SpGEMM accelerators in Table~\ref{tab:chip_configuration}.
Specifically, for the Tile-16 configuration in the GNN accelerator analysis, an architecture comprising $8$ tiles is used. Each tile includes a $16\times16$ grid of NeuraCores, with each core featuring a quad-pipeline design. We have significantly reduced the number of TAG comparators and port buffers, while retaining the hashpad sizes. This particular configuration is capable of delivering a peak performance of $8192\ \mathrm{GFLOPs}$, with an average power consumption of $4.3$W.

First, we consider EnGN, a hash-based GNN accelerator~\cite{liang2020engn}, and GROW~\cite{hwang2023grow}. EnGN employs a unique ring-based edge reducer to efficiently map vertex IDs. However, it encounters challenges in achieving a uniform distribution of computational tasks among its processing elements. In comparison, NeuraChip demonstrates superior performance, outperforming EnGN by $29\%$ on average. This improvement is primarily attributed to the dynamic reseed hashing function within NeuraChip, which ensures balanced task distribution across its computational resources, namely NeuraCore and NeuraMem, thus minimizing processing delays.

GROW utilizes a row-wise multiplication method, incorporating hardware and software co-design elements. A notable aspect of GROW’s software strategy is its reliance on graph partitioning, which significantly increases the computational overhead for GNN processing. From a hardware perspective, GROW is equipped with vector processors and employs streaming buffers for handling input and output matrix data. Despite these features, GROW encounters issues similar to those seen in Gamma's prefetcher system, where data idling results in suboptimal usage of on-chip memory resources. Comparative performance metrics indicate that NeuraChip surpasses GROW’s performance by an average of $58\%$.

% Next, we compare NeuraChip to a hybrid GNN accelerator called HyGCN~\cite{yan2020hygcn}. HyGCN implements dedicated engines for the aggregation and combination phases, which allows it to pipeline the computation for multiple GNN layers that have alternating aggregation and combination phases.
% NeuraChip, on the other hand, contains dedicated components for multiplication and accumulation stages, which are integral to both aggregation and combination phases, thereby circumventing the issue of phase computation time disparity. NeuraChip demonstrates a performance improvement of $XX\%$ over HyGCN.

Next, we evaluate our accelerator compared to HyGCN, a hybrid Graph Neural Network (GNN) accelerator, which has specialized engines for aggregation and combination phases~\cite{yan2020hygcn}. The primary advantage of HyGCN's architecture is its ability to pipeline computations, which is particularly beneficial for GNN layers that typically alternate between aggregation and combination phases. However, a significant limitation arises when the compute duration for one phase substantially exceeds the other, leading to a pipeline stall due to the uneven execution duration of each pipeline stage.

Instead, NeuraChip incorporates distinct components specifically for multiplication and accumulation operations, utilized in both the aggregation and combination phases. This design choice renders NeuraChip impervious to the inefficiencies caused by varying computational times between aggregation and combination phases. On average, NeuraChip outperforms HyGCN's performance by $69\%$.

Our final comparison is with FlowGNN~\cite{sarkar2023flowgnn}, a reconfigurable dataflow GNN accelerator comprising Node Transformation Units (NTs) and Message Passing Units (MPs). FlowGNN employs queues for real-time task buffering and relies on dynamic pull-based mapping for task distribution to NTs and MPs. In contrast, NeuraChip adopts a push-based mapping strategy for multiplication tasks and a hash-based approach for accumulation tasks. The Dispatcher in NeuraChip assigns \mmhf instructions to NeuraCores, optimizing input data temporal locality (reuse in NeuraCore register files). The dynamic reseeding hash-based mapping, as detailed in \secref{ssec:drhm}, ensures uniform workload distribution regardless of sparsity patterns. Consequently, NeuraChip achieves an average speedup of $30\%$ over GCN workloads tested on the FlowGNN architecture.

\subsection{Power Consumption and Area Analysis}

We assess our accelerator's area and power overheads by implementing its design in Register Transfer Level (RTL). Using Cadence Genus Synthesis Solutions, we synthesize these RTL components targeting an ASAP7 technology library~\cite{clark2016asap7}, allowing us to determine the area and power consumption for each proposed microarchitectural element. The synthesized chip area requirements for NeuraChip amount to $2.37mm^2$, $10.2mm^2$, and $35.26mm^2$ for the Tile-$4$, Tile-$16$, and Tile-$64$ configurations, respectively.

\begin{table}
\centering
\caption{NeuraChip Power and Area Breakdown}
% \vspace*{-2mm}
\label{tab:power}
% \resizebox{0.98\textwidth}{!}{%
\begin{tabular}{p{1.25in} | x{0.60in} x{0.60in} x{0.60in} | x{0.60in} x{0.60in} x{0.60in}}
\arrayrulecolor{black}\toprule
     & \multicolumn{3}{c|}{Area ($mm^2$)} & \multicolumn{3}{c}{Average Power (W)}  \\
    \textbf{Unit} &   \textbf{Tile-4} & \textbf{Tile-16} & \textbf{Tile-64} &   \textbf{Tile-4} & \textbf{Tile-16} & \textbf{Tile-64} \\ [0.5ex] 
\arrayrulecolor{black}\toprule
\arrayrulecolor{black}\toprule
      NeuraCore & 0.28 & 2.74 & 9.36 & 1.05 & 1.86 & 5.76  \\
    \arrayrulecolor{black!20}\midrule
        NeuraMem & 1.22 & 5.10 & 18.64 & 6.85 & 7.36 & 11.19  \\
    \arrayrulecolor{black!20}\midrule
        Router & 0.49 & 1.98 & 6.88 & 2.15 & 4.88 & 4.43  \\
    \arrayrulecolor{black!20}\midrule
        Memory Controller & 0.38 & 0.38 & 0.38 & 1.41 & 1.96 & 2.84  \\
    \arrayrulecolor{black}\toprule
        \textbf{Total} & 2.37 & 10.2 & 35.26 & 11.46 & 16.06 & 24.22  \\

     \arrayrulecolor{black}\bottomrule
% \vspace{-2.7em}
\end{tabular}
% }%end resizebox
% \vspace{-6mm}
\end{table}

The breakdown of NeuraChip's area and power is shown in Table~\ref{tab:power}. The majority of the area requirement for NeuraChip is allocated to the NeuraMem unit, as it includes the tag comparator array and the hash-pad (on-chip storage).

\section{Comparison with Prior Custom Accelerators}
\label{sec:related_work}

Next, we discuss previous studies on sparse matrix multiplication (spGEMM) and Graph Neural Networks (GNN).

\textbf{SpGEMM Accelerators}: 
InnerSP~\cite{baek2021innersp} introduces an accelerator that applies the inner-product method for matrix multiplication. This method offers advantages, such as eliminating the need for on-chip memory for accumulation. However, it suffers from limited input data reuse from both matrices, leading to performance issues when the sparsity patterns do not align with their task mapping algorithm.
MatRaptor~\cite{srivastava2020matraptor} employs a row-wise multiplication strategy and a round-robin greedy algorithm for allocating input rows to processing elements (PEs). Although this approach enhances input data reuse, it struggles with skewed sparsity patterns. The simplistic round-robin distribution may result in computational hot spots (as elaborated in \secref{sec:dse}).
SIGMA~\cite{qin2020sigma} offers an SpGEMM accelerator equipped with adaptable interconnects. Utilizing a smart global controller, SIGMA dynamically assigns each non-zero pair to PEs via a Benes network. Despite its efficiency in general SpGEMM tasks, SIGMA is less effective with large sparse matrix computations due to the substantial overhead introduced by its bitmap compression format.

\textbf{GNN Accelerators}: 
LISA~\cite{li2022lisa} performs GNN computations on Coarse-Grained Reconfigurable Arrays (CGRAs). LISA generates a dataflow graph and utilizes a simulated annealing method for mapping. 
I-GCN~\cite{geng2021gcn} aims to enhance data locality through an {\em islandization} strategy, clustering densely connected nodes to reduce off-chip memory accesses. 
However, both the simulated annealing and graph clustering methods introduce considerable computational overheads.

\begin{table*}
\centering
\caption{Performance comparison of SpGEMM workload accelerators across various off-the-shelf hardwares.}

\label{tab:spgemm_compare_1}

\begin{tabular}{x{1.90in} | x{1.10in} x{0.97in} x{0.97in} } 
    % \hline
    % \hline
    % \toprule[1pt]
    \arrayrulecolor{black}\toprule
   \textbf{Architectural Parameters} & Xeon E5 & NVIDIA H100 & AMD MI100  \\
   \arrayrulecolor{black}\toprule
   \arrayrulecolor{black}\toprule
   Compute Units & 8 Cores AVX2 & 7296 FP64 & 7680 FP64 \\
   \arrayrulecolor{black!20}\midrule
   Frequency $(\mathrm{GHz})$ & $2.9$ & $1.6$ & $1.5$ \\
   \arrayrulecolor{black!20}\midrule
   Peak Performance & $186$ $\mathrm{GFLOPs}$ & $26$ $\mathrm{TFLOPs}$ & $11.5$ \\
   \arrayrulecolor{black!20}\midrule
   SpGEMM Perf.$^{\Phi}$ $(\mathrm{GOP}/s)$ & $1.12$ & $1.86$ & $1.48$ \\
   \arrayrulecolor{black!20}\midrule
   On-chip Memory & $15~\mathrm{MB}^{\tau}$ & $50~\mathrm{MB}^{\dagger}$ & $8~\mathrm{MB}^{\dagger}$ \\
   \arrayrulecolor{black!20}\midrule
   Off-chip Memory & DDR4 $136\mathrm{GB}/\mathrm{s}$ & HBM $2\mathrm{TB}/\mathrm{s}$ & HBM $1.2\mathrm{TB}/\mathrm{s}$ \\
   \arrayrulecolor{black!20}\midrule
   Technology $(nm)$ & $32$ & $4$ & $7$ \\
   \arrayrulecolor{black!20}\midrule
   Area $(mm^2)$ & $356$ & $814$ & $750$ \\
   \arrayrulecolor{black!20}\midrule
   Power $(W)$ & $85^{\diamond}$ & $300^{\diamond}$ & $300^{\diamond}$ \\
   \arrayrulecolor{black!20}\midrule
   Tile-16 Speedup & $22.1\times$ & $13.3\times$ & $16.7\times$ \\
   \arrayrulecolor{black}\bottomrule

\end{tabular}

\vspace*{0.5mm}

\scriptsize

% \centering

$\diamond$Max thermal dissipation power from datasheet

$^{\dagger}$ L2 cache size
$^{\tau}$ L3 cache size
$^{\Phi}$Computed on common set of matrices as shown in Table~\ref{tab:bloat}.

\end{table*}

\begin{table*}
\centering
\caption{Performance comparison of state-of-the-art SpGEMM accelerators across various NeuraChip (\textbf{NC}) system configurations.}
\vspace{0.7em}
\label{tab:spgemm_compare_2}

\begin{tabular}{x{1.00in} | x{0.50in} x{0.71in} x{0.65in} x{0.63in} x{0.63in} x{0.63in} } 
    % \hline
    % \hline
    % \toprule[1pt]
    \arrayrulecolor{black}\toprule
   \textbf{Architectural Parameters} & Outer SPACE & SpArch & Gamma & \textbf{NC \ Tile-4} & \textbf{NC Tile-16} & \textbf{NC Tile-64} \\
   \arrayrulecolor{black}\toprule
   \arrayrulecolor{black}\toprule
   Compute Units & $256$~PEs & $2\times8$~Mults $16~\times~16$ Merger  & $32$~PEs Radix-$64$ & $2\times4$ NeuraCores & $2\times16$ NeuraCores & $2\times64$ NeuraCores \\
   \arrayrulecolor{black!20}\midrule
   Frequency $(\mathrm{GHz})$ & $1.5$ & $1$ & $1$ & $1$ & $1$ & $1$ \\
   \arrayrulecolor{black!20}\midrule
   Peak Performance & $384$ $\mathrm{GFLOPs}$ & $32$ $\mathrm{GFLOPs}$ & $32$ $\mathrm{GFLOPs}$ & $8$ $\mathrm{GFLOPs}$ & $32$ $\mathrm{GFLOPs}$ & $128$ $\mathrm{GFLOPs}$ \\
   \arrayrulecolor{black!20}\midrule
   SpGEMM Perf.$^{\Phi}$ $(\mathrm{GOP}/s)$ & $2.9$ & $10.4$ & $16.5$ & $5.15$ & $24.75$ & $30.69$ $93.17^{\alpha}$  \\
   \arrayrulecolor{black!20}\midrule
   On-chip Memory & $4~\mathrm{MB}$ & $15~\mathrm{MB}^{\star}$ & $3~\mathrm{MB}^{\ast}$ & $0.75~\mathrm{MB}^{\delta}$ & $3~\mathrm{MB}^{\delta}$ & $12~\mathrm{MB}^{\delta}$ \\
   \arrayrulecolor{black!20}\midrule
   Off-chip Memory & HBM $128\mathrm{GB}/\mathrm{s}$ & HBM $128\mathrm{GB}/\mathrm{s}$ & HBM $128\mathrm{GB}/\mathrm{s}$ & HBM $128\mathrm{GB}/\mathrm{s}$ & HBM $128\mathrm{GB}/\mathrm{s}$ & HBM $128\mathrm{GB}/\mathrm{s}$ \\
   \arrayrulecolor{black!20}\midrule
   Technology $(nm)$ & $32$ & $40$ & $45$ & $7$ & $7$ & $7$ \\
   \arrayrulecolor{black!20}\midrule
   Area $(mm^2)$ & $86.74$ & $28.49$ & $30.6^{\ddagger}$ $20.44^{\ddagger}$ & $2.37$ & $10.2$ & $35.26$ \\
   \arrayrulecolor{black!20}\midrule
   Power $(W)$ & $24$ & $9.26$ & \ding{118} & $11.46$ & $16.06$ & $24.22$ \\
   \arrayrulecolor{black!20}\midrule
   Energy Efficiency $(\mathrm{GOPS}/W)$ & $0.120$ & $1.123$ & \ding{118} & $0.449$ & $1.541$ & $1.266$ \\
   \arrayrulecolor{black!20}\midrule
   Area Efficiency $(\mathrm{GOPS}/mm^2)$ & $0.034$ & $0.365$ & $0.539$ & $2.171$ & $2.426$ & $0.870$ \\
   \arrayrulecolor{black!20}\midrule
   Tile-16 Speedup & $6.6\times$ & $2.4\times$ & $1.5\times$ & $4.8\times$ & $1\times$ & $0.807\times$ \\
   \arrayrulecolor{black}\bottomrule

\end{tabular}

\vspace*{0.5mm}

\scriptsize

% \centering

\ding{118}Gamma does not provide a power performance model
% $^{\dagger}$ L2 cache size
% $^{\tau}$ L3 cache size
$^{\delta}$HashPad Size
$^{\ast}$FiberCache Size

% \ding{117}Power and area metrics were sourced from the vendor's datasheets, thus derived metrics have not been included.
$^{\alpha}$ Simulated using dual stacked HBM providing peak bandwidth of $256$ GB/s
${\ddagger}$Gamma synthesizes accelerator using both $45$ nm and $40$ nm processes, resulting in computing areas of $30.6$ $mm^2$ and $20.44$ $mm^2$, respectively

$^{\star}$Represents column fetchers, row prefetchers, and partial matrix fetchers and writers.
$^{\Phi}$Computed on common set of matrices as shown in Table~\ref{tab:bloat}.

\end{table*}

\section{NeuraChip Summary}

% The timed eviction feature (Hot potato caching) ensures that each partial product spends the least amount of cycles as possible on the on-chip SRAM. Thus not wasting the precious on-chip storage.

% The timed eviction functionality, also known as Hot Potato caching, ensures that each partial product resides in the on-chip SRAM for the minimum number of cycles necessary, thereby maximizing the efficiency of the valuable on-chip storage capacity.

% In this work we recognise that workloads involving sparse matrix multiplication can benefit from a decoupled architecture design, where multiplication and aggregation phases can be individually optimized on two distinct components. 
% We present a cycle-accurate simulator called NeuraSim that forms a platform for our simulations.
% We accelerate GNN workloads with a mix of high-level optimizations along with microarchitectural features.
% Finally we implement our design using RTL to compute power and area requirements for different NeuraChip Tile sizes.

In this thesis we presented, NeuraChip, that demonstrates the potential advantages that sparse matrix multiplication workloads can gain from a decoupled architectural design. NeuraChip optimizes multiplication and aggregation phases separately using two distinct components.
We have presented an open-source\footnote{\url{https://github.com/NeuraChip/neurachip}}, cycle-accurate simulator called NeuraSim, used to demonstrate the effectiveness of our design. 
The acceleration of GNN workloads is achieved through a blend of high-level optimizations and  microarchitectural features.
We have synthesized our design using RTL, thereby allowing us to calculate power and area requirements for various NeuraChip Tile sizes.  NeuraChip is able to outperform state-of-the-art SpGEMM accelerator by a factor of $1.5\times$ and prior GNN accelerator by $1.46\times$ on average.

\chapter{Conclusion}
\label{chp7}

This dissertation presents the design and implementation of accelerators for graph computing. Accelerating graph computations poses significant challenges due to the irregular structure of graphs. Through this research, we contribute to the field in three major aspects. Firstly, we analyze the architectural implications of Graph Neural Networks (GNNs) on GPUs and state-of-the-art accelerators. Secondly, we develop algorithmic strategies to enhance the Sparse Generalized Matrix Multiply (SpGEMM) kernel, which is crucial for GNN workloads. Lastly, we introduce a custom Coarse-Grained Reconfigurable Array (CGRA)-based accelerator for GNNs, designed to overcome the bottlenecks identified during GNN profiling.

\section{GNN Workload Characterization}

This dissertation introduces GNNMark, a comprehensive benchmark suite specifically designed to evaluate the performance of Graph Neural Networks (GNNs) on GPUs. This initiative marks the first attempt within the architectural research community to focus on a GNN training-oriented benchmark suite. Utilizing GNNMark, we conduct an in-depth analysis of GNN workloads to identify the architectural challenges associated with training GNN models on GPU platforms. Our findings contribute novel insights into the primary architectural bottlenecks encountered during GNN training, alongside strategies for their potential mitigation.

The performance characteristics of a single GNN model can vary significantly based on the input graph's structure. Contrary to Deep Neural Networks (DNNs), where General Matrix Multiply (GEMM) and convolution operations predominate, our analysis reveals that graph processing within GNNs predominantly requires integer operations. This observation highlights the need for enhancing integer computation performance to improve overall GNN execution efficiency. Moreover, our study highlights a substantial impact of instruction fetch stalls on GNN performance, indicating that GPU instruction cache limitations could serve as a significant bottleneck. Additionally, our research presents findings on the sparsity observed during GNN training and the efficacy of strong scaling, thereby offering a comprehensive overview of the performance dynamics of GNN training on GPU systems.

\section{SMASH: GNN Algorithmic Acceleration}

Further in this \thesis, we describe the advancements made in the domain of SpGEMM algorithmic acceleration, encapsulating a series of pivotal contributions towards optimizing SpGEMM kernels. The cornerstone of this effort is characterized by a multifaceted analysis aimed at identifying the inherent challenges in developing efficient SpGEMM kernels. This investigation covers a critical evaluation of various sparse matrix multiplication techniques, scrutinizing the merits and drawbacks inherent to each method. 

A significant development in this research is the development of a novel sparse matrix storage format, termed as MAPCSR (Memory Aligned Parallel CSR). This innovative approach facilitates parallel computation of each sparse matrix row, significantly enhancing memory access efficiency through memory-aligned storage. The implementation of MAPCSR has demonstrably bolstered SpGEMM performance, yielding a remarkable $1.58\times$ improvement.

Building upon these foundational advancements, we introduce SMASH (Sparse Matrix Atomic Scratchpad Hashing), an optimized SpGEMM kernel tailored for distributed memory architectures. SMASH is implemented in three distinct versions, each iteration exploiting specific architectural features to incrementally enhance performance. This tiered approach not only exemplifies the adaptability and scalability of SMASH but also underpins its efficacy in leveraging the unique capabilities of custom accelerators. The iterative development of SMASH in this \thesis significantly propels forward the state-of-the-art in SpGEMM algorithmic acceleration.

\section{NeuraChip: GNN Hardware Acceleration}

This segment of the dissertation introduces NeuraChip, a state-of-the-art spatial accelerator tailored for Graph Neural Network workloads, marking a significant stride in hardware acceleration for graph computing. The design and implementation of NeuraChip embodies a series of innovative contributions that collectively address critical challenges in GNN acceleration.

\textbf{Heterogeneous Processing Approach}: Central to NeuraChip's architecture is a heterogeneous processing strategy that divides computation tasks into distinct multiplication and accumulation phases. This decoupled computation pipeline is meticulously designed to improve data reuse, optimizing the efficiency of operations through strategic component mapping.

\textbf{Adaptive Hash-Based Compute Mapping}: NeuraChip incorporates an adaptive, dynamic reseeding hash-based compute mapping (DRHM) mechanism, specifically engineered for GNN computations. Leveraging the consistent lookup times provided by hash functions, DRHM distributes computing tasks across the accelerator's resources. By dynamically updating the seed at regular computation intervals, it ensures an uniform workload distribution, effectively neutralizing the challenges posed by the varying sparsity patterns inherent in graph data.

\textbf{Mechanism for Rolling Evictions}: To address the bottleneck of memory congestion, a consequence of accumulating partial products, NeuraChip introduces a novel rolling eviction strategy. This approach facilitates the timely eviction of partial products, thereby mitigating memory bloat and ensuring sustained high throughput.

NeuraChip establishes itself as a significant development in the domain of hardware acceleration of GNNs, showcasing an architecture that is not only highly efficient but also adaptable to the diverse and dynamic nature of graph-based computations. The NeuraChip segment of this dissertation underscores the potential of specialized hardware designs in overcoming the unique challenges of accelerating GNN workloads, paving the way for future innovations in the field.

\section{Contributions of this Dissertation}

This dissertation has been dedicated to advancing the field of Graph Neural Network acceleration through a comprehensive exploration of both hardware and software aspects. It has established a new benchmark in the study of GNNs, providing a suite of contributions that lay a solid foundation for future research in this domain. These contributions span from the development of benchmark suites for GNN evaluation to novel algorithmic strategies for sparse matrix multiplication, and from the introduction of an innovative hardware accelerator to the creation of a cycle-accurate simulator for Coarse-Grained Reconfigurable Arrays (CGRA). Specifically, this work includes:

\begin{enumerate}

    \item The development of a benchmark suite specifically designed for assessing GNN performance, addressing both single and multi-GPU environments. This suite facilitates a nuanced understanding of GNN workloads, enabling targeted improvements in GPU-based graph processing.

    \item The presentation of GPU and multi-GPU benchmarking results for GNNs, which shed light on the architectural implications of GNN workloads. These insights contribute to the optimization of GPU resources for enhanced GNN processing efficiency.

    \item The introduction of a novel sparse matrix storage format that significantly advances the state of SpGEMM (Sparse Generalized Matrix Multiply) operations. This format underpins the algorithmic optimizations for SpGEMM workloads, demonstrating substantial performance improvements.

    \item An in-depth evaluation of SpGEMM optimizations on a custom accelerator, showcasing the potential of hardware-specific adaptations to elevate GNN processing speeds.

    \item The proposal of a dynamically reseeding hash-based mapping algorithm tailored for GNN workloads, which optimizes computation distribution and efficiency in hardware accelerators designed for GNNs.

    \item The creation of NeuraSim, a cycle-accurate simulator for CGRA architectures, facilitating precise analysis and optimization of GNN accelerator designs.

    \item A comprehensive chip power and area analysis of the NeuraChip GNN accelerator, providing valuable metrics for assessing the viability and efficiency of GNN-specific hardware solutions.

\end{enumerate}

Together, these contributions not only provide a path towards optimized GNN processing but also equip the research community with the tools and methodologies necessary for continued innovation in the acceleration of graph neural networks.

\section{Future Work}

The primary objective of this dissertation was to advance the field of Graph Neural Network acceleration through a comprehensive exploration of benchmark suites, algorithmic optimizations, and hardware solutions. Despite the significant strides made in this research, the domain of GNN acceleration remains vast, with ample opportunities for further exploration and innovation. The following outlines potential directions for future research, building upon the foundational work presented in this thesis:

\begin{enumerate}

    \item \textbf{Enhanced Integration of GNN Workloads with Emerging Hardware Accelerators}: The current landscape of hardware accelerators, including but not limited to NVIDIA’s Tensor Cores~\cite{markidis2018nvidia} and Google’s TPUs~\cite{kimm2021performance, kumar2019scale}, presents a promising avenue for improving GNN performance and energy efficiency. Future work could explore deeper integrations with these accelerators, adapting GNN algorithms to leverage their specific architectural advantages more effectively.

    \item \textbf{Advanced Memory Technologies for GNN Acceleration}: The evolution of memory technologies, such as in-memory processing~\cite{hur2016memory} and near-memory processing~\cite{ke2020recnmp}, offers new possibilities for enhancing GNN execution. Investigating holistic designs that incorporate these advanced memory solutions could lead to significant improvements in GNN processing speed, energy efficiency, and overall system performance.

    \item \textbf{Scalability and Efficiency in Multi-GPU Systems}: While this dissertation addressed multi-GPU benchmarking and architectural implications for GNNs, scaling performance linearly with the addition of GPUs remains a challenge. Future research could focus on developing new techniques for GPU-level cooperative thread array (CTA)~\cite{jog2013owl} scheduling and thread migration\cite{constantinou2005performance} to minimize performance overhead in multi-GPU systems.

    \item \textbf{Dynamic Scheduling and Control Flow Optimization for GNN Workloads}: Addressing the limitations in control flow transition between CPUs and GPUs is crucial for optimizing the performance of GNN workloads~\cite{kim2021minimizing}. Investigating hardware-based GPU schedulers integrated into CPU cores could reduce kernel-launch overhead and memory synchronization challenges, enabling more efficient CPU-GPU collaboration.

    \item \textbf{Computing Capabilities in Network Devices for Distributed GNN Processing}: As network switches evolve to incorporate computing capabilities, there exists an opportunity to offload certain GNN processing tasks to these devices, potentially reducing data movement costs and improving the efficiency of distributed GNN training and inference~\cite{tokusashi2019case}.

    \item \textbf{Exploration of Novel Computing Devices for GNN Acceleration}: The advent of novel computing devices extends the horizon for GNN acceleration. Future research should consider incorporating a broader array of devices, including dedicated accelerators, non-traditional memory devices, and smart network components, to achieve comprehensive performance, energy, reliability, and security enhancements.

\end{enumerate}

Building on the contributions of this dissertation, these future research directions aim to further push the boundaries of GNN acceleration, addressing the complex and evolving challenges inherent in graph-based computations and fostering the development of more efficient, scalable, and adaptable GNN processing systems.

\bibliographystyle{plain}

% include bibliography definition
% \bibliography{bib/thesis, bib/zotero}
\bibliography{bib/thesis}

%% Maybe make it an appendix
\chapter*{Biography}

Kaustubh Shivdikar was born in Mumbai, India, on December 5, 1994. He obtained his Bachelor of Science degree in Electrical Engineering from the Veermata Jijabai Technological Institute, University of Mumbai, in 2016. He went on to receive his Master of Science and Doctor of Philosophy degrees in Electrical and Computer Engineering from Northeastern University, Boston, USA, in May 2020 and May 2024, respectively. His Ph.D. research was supervised by Dr. David Kaeli at the Northeastern University Computer Architecture Research (NUCAR) Laboratory. Kaustubh is a member of IEEE and ACM. His research fields encompass Computer Architecture Simulator Design, Graph Neural Network Accelerators, Sparse Matrix Accelerators, and Homomorphic Encryption Accelerators.\footnote{\url{https://wiki.kaustubh.us}}

% \printglossary[type=\acronymtype,title=List of Acronyms]

% --- Index ----
\printindex

% --- that's it ---
\end{document}